\documentclass[a4paper, notoc]{JHEP3}
\usepackage[dvips]{graphicx}
\usepackage{amsfonts}
\usepackage{amssymb}
\usepackage{epsfig}
\usepackage{cite}

\newcommand{\be}{\begin{equation}}
\newcommand{\ee}{\end{equation}}
\newcommand{\bea}{\begin{eqnarray}}
\newcommand{\eea}{\end{eqnarray}}
\newcommand{\mbb}{\mathbb}
\newcommand{\ti}{\times}

\newcommand{\mc}{\mathcal}

\newcommand{\beqa}{\begin{eqnarray}}
\newcommand{\eeqa}{\end{eqnarray}}

\newcommand{\zo}{z_{1}}
\newcommand{\zt}{z_{2}}
\newcommand{\bzo}{\bar{z}_{1}}
\newcommand{\bzt}{\bar{z}_{2}}
\newcommand{\tile}{\tilde{e}}
\newcommand{\del}{\partial}

\newcommand{\tgamma}{{\tilde{\gamma}}}
\newcommand{\ph}{\phantom}

 \title{Wave Functions and Yukawa Couplings in Local String
Compactifications}
\author{Joseph P. Conlon$^{1,2}$, Anshuman Maharana$^{2}$, and Fernando Quevedo$^{2}$
 \\$^{1}$Cavendish Laboratory, J J Thomson Avenue \\
 Cambridge CB3 0HE, UK \\
 \\$^{2}$DAMTP,
  Centre for Mathematical
Sciences,\\
  Wilberforce Road, Cambridge, CB3 0WA, United Kingdom
}

\abstract{We consider local models of magnetised D7 branes in IIB string
 compactifications, focussing on cases where an explicit metric can be
 written for the local 4-cycle. The presence of an explicit metric
 allows analytic expressions for the gauge bundle and for the chiral
 matter wavefunctions through solving the
Dirac and Laplace equations. The triple overlap of the normalised matter
 wavefunctions generates the physical Yukawa couplings.
Our main examples are the cases of D7 branes on $\mbb{P}^1 \ti
 \mbb{P}^1$ and $\mbb{P}^2$. We consider both supersymmetric and
 non-supersymmetric gauge backgrounds and both
Abelian and non-Abelian gauge bundles.
We briefly outline potential phenomenological
applications of our results.}

\preprint{DAMTP-2008-33}

\begin{document}

\tableofcontents

\section{Introduction}

Understanding the structure of the Standard Model - the gauge groups, matter content and
Yukawa couplings - represents one of the principal problems of theoretical physics.
The Standard Model is not self-justifying and does not motivate a reason for its parameter values.
Any deeper explanation of the Standard Model will therefore likely involve new physical ideas
and concepts, possibly of a very different nature to those used in the Standard Model itself.
In this respect string theory stands out as an attractive and powerful complex of ideas.

One attractive feature of string theory is that it
naturally gives rise to chiral matter and non-Abelian gauge groups,
thereby reproducing the gross features of the Standard Model.
One of the main tasks of string phenomenology consists in finding vacuum configurations
resembling, as closely as possible, the gauge group and matter content of the Standard Model.
In the heterotic string this is achieved through appropriate gauge
bundles and Wilson lines, while for the type II theories this requires an appropriate model of intersecting D-branes.
Such constructions should be regarded as proofs of principle: no current construction reproduces all details of the
Standard Model and, in any case, it is unclear
what the desired low-energy matter content actually is, as nothing precludes
the existence of extra massive vector-like matter.
Recent reviews of string theoretic model building can be found in 
\cite{hepth0502005, hepth0610327, hepth0702094, 07112451, 08063905}.

Such string constructions can be classified as either global or local.
Global constructions rely on the topological features of the totality of a
compact space, with the classic example being the weakly-coupled heterotic string where the spectrum and gauge group
are determined by a bundle over the entire Calabi-Yau. In this case the gauge coupling is given by the volume of the entire
Calabi-Yau. In local models, the principal avatar of which is models of branes at singularities \cite{hepth0005067}, 
the gauge group and matter content depend only on local physics and are independent of the details of the bulk geometry.
The distinction between global and local models is that in a global model the
Standard Model gauge couplings always vanish in the limit that the bulk volume is taken to infinity;
in local models such a limit leaves the gauge couplings finite. Recent
years have seen a renewal of interest in local models, and examples
of work in this direction include \cite{hepth0203129, hepth0508089, hepth0604208,
  hepth0610007, hepth0703047, 08023391, 08052943, 08060102, 08060634}.

This is one of the most attractive features of local models.
It is well-known that for global models the two control requirements of large volume and
weak coupling can never be parametrically satisfied: the known running of the
Standard Model gauge couplings implies that at the compactification
scale
\be
\label{eq1}
\alpha_{SM}^{-1} \sim 25 \sim \frac{\mc{V}}{l_s^6 g_s}.
\ee
A weak string coupling $g_s \sim 0.1$ therefore implies $\mc{V} \sim 2 l_s^6$, making
it difficult to control the $\alpha'$ expansion.
However in local models eq. (\ref{eq1}) no longer holds,
allowing both large volume and weak coupling to be simultaneously realised.
In some cases the use of local models can even be forced upon us by the moduli stabilisation procedure.
An example is the LARGE volume scenario of \cite{hepth0502058, hepth0505076}, where the volume is
stabilised exponentially large, thereby generating hierarchically small supersymmetry breaking.
As $m_{3/2} \sim M_P/\mc{V}$, a TeV gravitino mass requires a volume $\mc{V} \sim 10^{15} l_s^6$, which implies that any
realisation of the Standard Model in this scenario is 
necessarily a local realisation.

A second attractive feature of local models is that they drastically simplify
the geometric complexity of model building.
The local geometry is non-compact and typically involves far fewer moduli than the
hundreds present in typical Calabi-Yaus. Local models can also be constructed on very simple geometries such as
$\mbb{C}^3/\mbb{Z}_n$ singularities or their resolutions. In some cases
the local Calabi-Yau metric is known exactly, in contrast to the cases of global compact Calabi-Yaus where no such explicit metrics are known.
In the limit that the bulk volume is very large - which is the case 
for the LARGE volume models -
the exact local metric is a parametrically good approximation to the true Calabi-Yau metric.

The knowledge of an explicit local metric has several consequences.
The Laplace and Dirac equations can be solved directly, allowing the
exact normalised wavefunctions of the chiral matter fields to be determined. Such wavefunctions give the extra-dimensional profile
of the fields. In local models of magnetised D7 branes, the Yukawa couplings schematically descend from the 
triple overlap integral
\be
\label{eq2}
\mc{L}_{YUK} = \int d^{8}x \, \bar{\psi} \Gamma^M [A_M, \psi].
\ee
Using the form of the wavefunctions the triple overlap integrals of eq. (\ref{eq2}) can be computed and the physical Yukawa couplings,
including the non-holomorphic parts, can be
evaluated. 
In contrast, algebro-geometric techniques, while
very powerful in determining the holomorphic superpotential, are unable to determine the physical couplings
which require knowledge of non-holomorphic functions such as the K\"ahler metrics.
Furthermore, 
the presence of explicit metrics in principle also allows the spectrum and wavefunctions of Kaluza-Klein modes to be computed.
While these will not contribute to the renormalisable Lagrangian, integrating out these modes
will generate highly suppressed non-renormalisable operators. Kaluza-Klein modes are generically gauge singlets and
thus in the R-parity MSSM count as right-handed neutrinos.
Knowledge of the explicit form of such KK wavefunctions may then have important consequences for string theory models of neutrino masses
and mixing matrices \cite{hepph0611144}.

The use of explicit metrics to study wavefunctions and Yukawa
couplings through direct dimensional reduction has been carried out in detail
for toroidal models of magnetised D9 branes \cite{hepth0404229} (for a
recent discussion for orbifolds of toroidal models see \cite{kobayashi}).
The purpose of this paper is to perform a similar analysis for models of magnetised D7 branes.
We shall study in detail local models of magnetised D7 branes wrapping curved spaces.
We aim to compute the chiral massless spectrum and wavefunctions, and use these to analyse the structure of
the resulting overlap integrals and Yukawa couplings. Our two principal examples will be
$\mbb{P}^1 \ti \mbb{P}^1$ and $\mbb{P}^2$. The Dirac and Laplace
equations on such geometries have also been studied in \cite{Pope, hepth0203264,
  hepth0207078, hepth0304037, hepth0505107}.
The local Calabi-Yau geometries these correspond to are
$\mc{O}_{\mbb{P}^1 \ti \mbb{P}^1}(-2, -2)$ and
$\mc{O}_{\mbb{P}^2}(-3)$.
While the former admits chiral supersymmetric D7 brane configurations in the
geometric regime, the latter does not. The $\mc{O}_{\mbb{P}^2}(-3)$ geometry is however of great interest
as the resolution of the
$\mbb{C}^3 / \mbb{Z}_3$ singularity, which has played a central role
in phenomenologically attractive models of branes at singularities \cite{hepth0005067}.

This paper is organised as follows. In section \ref{sec2} we describe
the dimensional reduction of the D7 brane action, the classification of the four-dimensional 
fields that arise, the structure of the Yukawa couplings and
the equations that need to be solved to determine the wavefunctions and Yukawa
couplings. In section \ref{sec3} we review the solutions of the Dirac and Laplace equations on $\mbb{P}^1$, while in sections
\ref{sec4} and \ref{sec5} we study the cases of $\mbb{P}^1 \ti \mbb{P}^1$ and $\mbb{P}^2$ respectively. In these sections
we also discuss the twisting of the Dirac and Laplace equations that is necessary to account for the curved nature of the
D-brane embedding.

\section{Dimensional Reduction and the Low Energy Action}
\label{sec2}

In this section we aim to collect the equations of motion that are satisfied by the various fields,
and to describe how the solutions of these equations can be used to compute the Yukawa couplings.
We focus first on deriving the equations of motion that need to be satisfied to obtain scalar, fermion and
vector zero modes, and subsequently describe the origin of the Yukawa couplings. The presence of a nontrivial brane embedding will
cause the equations determining the zero modes to be twisted.

We start with the ten-dimensional $U(N)$ super-Yang Mills action, which will be dimensionally reduced to eight dimensions.
The ten-dimensional action is
\begin{equation}
     S = \frac{1}{g_{10}^2} \int d^{10} x  \bigg(- { 1 \over 4} F_{MN} F^{MN}  + {1 \over 2}\bar{\psi} \Gamma^{M} \mc{D}_{M} \psi \bigg),
\end{equation}
where $\psi$ is a ten dimensional Weyl-Majorana spinor and $\mc{D}_M
\psi = \nabla_M \psi - i [A_M,\psi]$.
On reduction to eight dimensions the field content becomes one eight-dimensional gauge boson, ($A_M, M= 0 \dots 7$),
one complex scalar ($\phi_z = \phi_8 + i \phi_9$) and a single physical fermion $\lambda$.
The action is
\begin{equation}
     S =  { 1 \over g^{2} }\int d^{8}x \bigg(- {1 \over 4}{\rm{Tr}}\{ F_{MN} F^{MN}\} -
     {1 \over 2} {\rm{Tr}}\{{\cal{D}}_{M} \phi^{r}  {\cal{D}}^{M} \phi^{r} \}
     + i {\rm{Tr}} \{ \bar{\lambda} \gamma^{M} {\cal{D}}_{M} \lambda
     \} + \frac{i}{2}\bar{\lambda} \Gamma^r [\phi_r, \lambda]  \bigg).
\label{eqsss}
\end{equation}
Here we retain the indices $M,N$ for the 8d quantities and use the
index $r$ for the $8-9$ directions. There are also additional quartic
scalar interactions in eq. (\ref{eqsss}) that we will neglect as not
relevant to our purposes. 
There is a $U(1)_R$ symmetry transverse to the brane, under which a field
with $R$ charge $q$ is multiplied by
$e^{{ iq \theta}}$ under rotation by an angle $\theta$ in the 89 plane.
The $U(1)_R$ charges of the fields are
$$
Q_{A_M} = 0, \qquad
Q_{\lambda, \bar{\lambda}} = \pm 1/2 , \qquad
Q_{\phi^{8} \pm i\phi^{9}} = \pm 1.
$$
where in writing $U(1)_R(\lambda) = 1/2$ we have treated $\lambda$ as a Weyl fermion of positive chirality.

We want to decompose the eight dimensional fields into four-dimensional ones. For the bosonic degrees of freedom this is
straightforward. The transverse scalar $\phi_z = \phi_8 + i \phi_9$ becomes
$$
\phi_z(x,y) = \sum_i \phi^i(x) \phi^i_z(y),
$$
where $x$ refers to the $0,1,2,3$ directions 
and $y$ to the $4,5,6,7$ directions. 
The eight dimensional vector decomposes into a 4 dimensional vector $A_\mu$ and
4 real scalars
associated to the internal vector degrees of freedom $A_m$, $m=4,5,6,7$.
$$
A_{\mu}(x,y) = \sum_i A^i_{\mu}(x) A^i(y), \qquad A_m(x,y) = \sum_i \Phi^i(x) A^i_M(y).
$$
To provide an explicit decomposition of the ten-dimensional Majorana-Weyl fermion, we start
by taking the ten dimensional gamma matrices to be in a product representation,
\begin{equation}
\Gamma^{\mu} = \gamma^{\mu} \otimes \mathbb{I} \otimes \mathbb{I}  \ \ \  \
\Gamma^{m} = \gamma^{5} \otimes \tgamma^{m-3} \otimes \mathbb{I}  \ \ \ \
\Gamma^{r} = \gamma^{5} \otimes \tgamma^{5} \otimes \tau^{r},  \ \ \ \
\end{equation}
%
where $\mu =  0 \ldots 3 $, $m = 4 \ldots 7$, and $r=8,9$.
$\gamma^{\mu}$ are the four dimensional Minkowski gamma matrices
\begin{equation}
  \gamma^{0} = \left( \begin{array}{cc}
0 & - \mathbb{I} \\
 \mathbb{I} & 0  \\ \end{array} \right) \ \ \ \gamma^{1} = \left( \begin{array}{cc}
0 & \sigma_{x} \\
\sigma_{x} & 0  \\ \end{array} \right) \ \ \ \gamma^{2} = \left( \begin{array}{cc}
0 & \sigma_{y} \\
\sigma_{y} & 0  \\ \end{array} \right) \ \ \
\gamma^{3} = \left( \begin{array}{cc}
0 & \sigma_{z} \\
\sigma_{z} & 0  \\ \end{array} \right),
\end{equation}
while $\tgamma^{m}$ are four dimensional Euclidean gamma matrices
\begin{equation}
  \tgamma^{1} = \left( \begin{array}{cc}
0 & - i\mathbb{I} \\
 i\mathbb{I} & 0  \\ \end{array} \right) \ \ \ \tgamma^{2} = \left( \begin{array}{cc}
0 & \sigma_{z} \\
\sigma_{z} & 0  \\ \end{array} \right) \ \ \ \tgamma^{3} = \left( \begin{array}{cc}
0 & \sigma_{x} \\
\sigma_{x} & 0  \\ \end{array} \right) \ \ \
\tgamma^{4} = \left( \begin{array}{cc}
0 & \sigma_{y} \\
\sigma_{y} & 0  \\ \end{array} \right).
\label{conve}
\end{equation}
The $\tau^{a}$ are Pauli matrices with
$\tau^{8} = \sigma^{x}$ and $\tau^{9} = \sigma^{y}$.
$\gamma^{5}$ and $\tilde{\gamma}^{5}$ denote the chirality matrices
\begin{equation}
\gamma^{5} = \left( \begin{array}{cc}
\mathbb{I} & 0 \\
 0 & -\mathbb{I}  \\ \end{array} \right), \ \ \  \tgamma^{5} = \left( \begin{array}{cc}
\mathbb{I} & 0 \\
0 & -\mathbb{I} \\ \end{array} \right).
\end{equation}
The ten dimensional chirality matrix is
$\Gamma = \gamma^{5} \otimes \tgamma^{5} \otimes \sigma^{z}$, and a Weyl fermion $\lambda$ is defined by
$\lambda = \Gamma \lambda$. We can also define a Majorana matrix
\be
\label{majorana}
B = \Gamma^2 \Gamma^4 \Gamma^7 \Gamma^9 = \left( \begin{array}{cc} 0 & - \sigma_y \\ \sigma_y & 0 \end{array} \right)
\otimes \left( \begin{array}{cc} -i \sigma_y & 0 \\ 0 & -i \sigma_y \end{array} \right) \otimes \left( \begin{array}{cc} 0 & -i \\
i & 0 \end{array} \right).
\ee
which satisfies $B B^{*} = \mbb{I}$. The Majorana condition is $\lambda^{*} = B \lambda$, and we will require that
$\lambda$ be both Majorana and Weyl.

In writing the spinor it will be convenient to use the following notation. Spinors can be labelled by their chirality in each
of the $\mu$, $m$ and $r$ directions. We will use superscript to indicate positive chirality and subscript to indicate negative chirality,
so the spinor $\lambda^{ab\alpha}$ has positive chirality
($a,b=1,2,3,4; \alpha =1,2$) in each of the $\mu$, $m$ and $r$
directions. 
The Weyl condition restricts a general
10-dimensional spinor to
$$
\{ \lambda_1, \lambda_2, \lambda_3, \lambda_4 \} = \{
\lambda^{ab\alpha}, \lambda^{a}_{\ph{a}b\alpha},
\lambda_{ab}^{\phantom{ab}\alpha},
\lambda_{a\phantom{b}\alpha}^{\phantom{a}b}
\}.
$$
From the form of the Majorana matrix (\ref{majorana}), we see that its action corresponds to a chirality flip in both $\mu$ and $89$ directions.
The Majorana condition $\lambda^{*} = B \lambda$ therefore imposes relations between $\lambda_1$ and $\lambda_4$, and $\lambda_2$ and $\lambda_3$.
A general Majorana-Weyl spinor can be schematically written as
$$
\lambda_{MW} = (\lambda_1 + \lambda_4) \oplus (\lambda_2 + \lambda_3).
$$
To be more explicit, we write the $\lambda_i$ as
\bea
\lambda_1^{ab\alpha} & = & \xi_1^a(x) \psi_1^b(y) \theta_1^{\alpha} (z), \nonumber \\
\lambda^{\ph{2,}a}_{2,\ph{a}b\alpha} & = & \xi_2^a(x) \psi_{2,b}(y)
\theta_{2,\alpha} (z), \nonumber \\
\lambda_{3,ab}^{\ph{4,ab}\alpha} & = & \xi_{3,a}(x) \psi_{3,b}(y) 
\theta_3^{\alpha} (z), \nonumber \\
\lambda^{\ph{3,} \ph{a}b\ph{\alpha}}_{4,a\ph{b}\alpha} & = & \xi_{4,a}(x) \psi_4^{b}(y) \theta_{4,\alpha} (z).
\eea
The Majorana condition $B^{*} \lambda^{*} = \lambda$ then imposes the constraints
\be
\label{majconstraints}
\begin{array}{ccc} \xi_4 = - \sigma_y \xi_1^{*}, & \psi_4 = - i \sigma_y \psi_1^{*}, & \theta_4 = -i \theta_1^{*}. \\
\xi_3 = - \sigma_y \xi_2^{*}, & \psi_3 = -i \sigma_y \psi_2^{*}, & \theta_3 = i \theta_2^{*}.
\end{array}
\ee
From a four dimensional viewpoint there are two distinct types of left-handed spinors, distinguished by
their extra-dimensional chirality. These can be written as
\bea
\lambda_1 + \lambda_4 & = &
\left( \begin{array}{c} \xi_1 \\ 0 \end{array} \right) \otimes \left( \begin{array}{c} \psi_1 \\ 0 \end{array} \right) \otimes
\left( \begin{array}{c} \theta_1 \\ 0 \end{array} \right)  +
\left( \begin{array}{c} 0 \\ - \sigma_y \xi_1^{*} \end{array} \right) \otimes \left( \begin{array}{c} -i \sigma_y \psi_1^{*} \\ 0 \end{array} \right) \otimes
\left( \begin{array}{c} 0 \\ -i \theta_1^{*} \end{array} \right), \nonumber \\
\lambda_2 + \lambda_3 & = &
\left( \begin{array}{c} \xi_2 \\ 0 \end{array} \right) \otimes \left( \begin{array}{c} 0 \\ \psi_2 \end{array} \right) \otimes
\left( \begin{array}{c} 0 \\ \theta_2 \end{array} \right)  +
\left( \begin{array}{c} 0 \\ - \sigma_y \xi_2^{*} \end{array} \right) \otimes \left( \begin{array}{c} 0 \\ -i \sigma_y \psi_2^{*} \end{array} \right) \otimes
\left( \begin{array}{c} i \theta_2^{*} \\ 0 \end{array} \right).
\label{lambdas}
\eea
In terms of representation content under the $SU(2)_L\times
SU(2)_R\sim SO(3,1)$
of 4D Minkowski space,  the $SU(2)_L\times SU(2)_R\sim SO(4)$ of the
internal 4D space and the $U(1)_R$ we can write
\bea
\lambda_1 + \lambda_4 & = &
[ (\textbf{2},\textbf{1})\otimes (\textbf{2},\textbf{1})]_{1/2} 
\oplus\,
[ (\textbf{1},\textbf{2})\otimes
(\textbf{2},\textbf{1})]_{-1/2} \nonumber \\ 
\lambda_2 + \lambda_3 & = & [
 (\textbf{2},\textbf{1})\otimes (\textbf{1},\textbf{2}) ]_{-1/2}
\oplus\,
[ (\textbf{1},\textbf{2})\otimes (\textbf{1},\textbf{2})
]_{1/2} .  
\eea
The subscript denotes the $U(1)_R$ charge.
We will tend to use left-handed spinors in Minkowski space and will treat $\lambda_1$ and $\lambda_2$ as the two
independent dynamical degrees of freedom, with $\lambda_3$ and $\lambda_4$ determined as above.

The geometric background we visualise is a stack of $(P+Q)$ D7 branes
wrapped on a 4-cycle, with a magnetic flux background turned on $P$ of
the branes. The flux background breaks the original $U(P+Q)$ gauge
group. In the case that a $U(1)$ bundle is turned on, the low energy
gauge group is $U(P) \ti U(Q)$. All fields start off valued in
$\textrm{Adj}({\bf P + Q})$, but decompose and give bifundamentals
under the flux background.
An arbitrary field $\Phi$ can be written as
\be
\Phi = \left( \begin{array}{cc} W & X \\
Y & Z \end{array} \right) = \left( \begin{array}{cc}
(\textrm{Adj}(\bf{P}), \bf{1}) & (\bf{P}, \bar{\bf{Q}}) \\ (\bar{\bf{P}}, \bf{Q}) & (\bf{1}, \textrm{Adj}(\bf{Q})) \end{array} \right).
\ee
Chiral bifundamental matter arises from zero modes descending from
either the $X$ or $Y$ sectors. The $X$ sector gives $({\bf P},{\bf
  \bar{Q}})$ modes and the $Y$ sector $({\bf \bar{P}}, {\bf Q})$ modes.

We suppose that $M$ units of flux have been turned on in the $W$
sector. 
When writing the Dirac or Laplace
equations for the $X$ or $Y$, the covariant derivative term $[A,
  \Phi]$ will imply that modes in the $X$ ($Y$) sector are effectively
charged under a $U(1)$ field with 
$M$ (-$M$) 
units of flux. The number
of fields in fundamental and anti-fundamental representations is
determined by the number of zero modes with $M$ (-$M$) units of flux,
and the net chirality is given by $\mc{N}_{+M} - \mc{N}_{-M}$.

\subsection{Equations of Motion}

In describing the equations of motion for the fields, we shall go into considerable detail for the vector modes, and then
be more concise in our description of the scalar and fermion equations of motion.

An important general feature here is the fact that the brane is
wrapping a cycle with a nontrivial normal bundle. This implies that
many of the equations of motion will need to be twisted
\cite{hepth9511222}, reflecting the fact that the field is not
scalar-valued but rather bundle-valued over the cycle. This does not
however hold for the internal degrees of freedom valued in the tangent
bundle, such as the (internal) vector modes. We first describe
the equations of motion for scalar fields that come from vector
degrees of freedom in the internal space.

\subsubsection{Vectors}

The equations of motion for vectors start from the Yang-Mills action
\be
\mc{S}_{YM} = \int_{\mc{M}^4 \ti \Sigma} \mc{L}_{YM} = - {1 \over 4g^2} \int_{\mc{M}^4 \ti \Sigma} {\rm{Tr}}\left[ F_{MN} F^{MN} \right]
\ee
with
$$
F_{MN} = \nabla_M A_N - \nabla_N A_M  - i[A_M, A_N].
$$
$A_M$ is valued in the adjoint representation of $U(N)$ and the action is invariant under the gauge transformation
$$
A_M \to A_M + \partial_M \theta + i [\theta, A_M].
$$
On dimensional reduction, the vector degrees of freedom give rise to a 4-dimensional vector ($A_{\mu}$) and 4-dimensional
scalars $(A_i)$ arising from vectors in the internal space.
As the vectors $A_M$ are all valued in the tangent bundle, they are uncharged under the
R-symmetry transverse to the brane. Their equations of motion are therefore not subject to twisting and
can be found by a direct dimensional reduction of the action $\mc{L}_{YM}$ on the surface $\Sigma$
in the presence of a background magnetic field.

To establish notation and conventions, we start with the Lie algebra of $U(N)$. Our
discussion will follow that in the appendix of
\cite{hepth0404229}. The elements of the Lie algebra can be taken to 
be\footnote{Although strictly $e_{ab}$ is defined for $a \neq b$, 
for convenience of notation we will allow ourselves to
write $e_{ab}$ even when potentially $a=b$, an example being the last
expression of eq. (\ref{abc}).}
$$
(U_a)_{ij} = \delta_{ai} \delta_{aj}, \qquad (e_{ab})_{ij} =
\delta_{ai} \delta_{bj}, \quad (a\neq b).
$$
The gauge field $A_M$ is expanded as
\be
A_M = B_M + W_M = B_M^a U_a + W_M^{ab} e_{ab}.
\ee
Requiring that $A_M^{\dagger} = A_M$ implies that $B_M^a$ is real and $(W_M^{ab})^{*} = W_{\bar{M}}^{ba}$.
Let us state some useful relations for the $U_a$ and $e_{ab}$:
\be
(U_a)_{ij} (U_b)_{jk} = \delta_{ab} (U_a)_{ik}, \quad
[U_a, U_b]_{ik} = (\delta_{ab}) \left((U_a)_{ik} - (U_b)_{ik} \right) = 0.
\ee
\be
(U_a)_{ij} (e_{bc})_{jk}  = \delta_{ab} (e_{bc})_{ik}, \quad
(e_{bc})_{ij}(U_a)_{jk} = \delta_{ca} (e_{bc})_{ik}, \quad
[U_a, e_{bc}] = (\delta_{ab} - \delta_{ac}) \, e_{bc}.
\ee
\be
\label{abc}
(e_{ab} e_{cd})_{ik} = \delta_{bc} (e_{ad})_{ik}, \quad
[e_{ab}, e_{cd}] = \delta_{bc} (e_{ad}) - \delta_{ad} \, (e_{cb}).
\ee
Also, ${\rm{Tr}}(U_a) = 1$, ${\rm{Tr}}(e_{bc}) = \delta_{bc}, {\rm{Tr}}(U_a e_{bc}) = \delta_{abc}$, where $\delta_{abc} = 1$
if and only if $a$, $b$ and $c$ are all identical.

If we now write $A_M = B_M + W_M$, we can expand
\bea
\mc{L}_{YM} & = & -\frac{1}{4g^2} \textrm{Tr} \bigg[ \left( G^{MN} + D^M W^N - D^N W^M -i [W^M, W^N] \right) \ti \nonumber \\
& & \left( G_{MN} + D_M W_N - D_N W_M -i [W_M, W_N] \right)  \bigg] .
\eea
Here $G_{MN} = \nabla_M B_N - \nabla_N B_M$ and $D_M W_N = \nabla_M W_N - i[B_M, W_N]$.
Further expanding we obtain
\bea
\label{expansion}
\mc{L}_B & = & -\frac{1}{4g^2} \textrm{Tr} \left[ G_{MN} G^{MN} \right] - \frac{1}{2g^2}
\textrm{Tr}\bigg[ D_M W_N D^M W^N - D_M W_N D^N W^M - i G_{MN} [W^M, W^N] \bigg] \nonumber \\
& & + \frac{1}{4g^2} \textrm{Tr}\bigg[ [W_M, W_N][W^M, W^N] \bigg] + \frac{i}{2g^2} \textrm{Tr} \bigg[
(D_M W_N - D_N W_M)[W^M, W^N] \bigg].
\eea
In backgrounds with Abelian magnetic flux, $G_{MN}$ has a non-zero vev
but $W_M$ does not.\footnote{For a careful study of vector modes in
  backgrounds with non-Abelian magnetic fluxes, such as used in
  section (\ref{sec5}), we would need to construct analogous equations
  in which $\langle W_M^{ab} \rangle \neq 0$.}
The fluctuations of $W_M$ will correspond to zero modes of the low energy theory. We focus on the 2-point interactions, and neglect the
3- and 4-point interactions present in equation (\ref{expansion}). Expanding the terms of (\ref{expansion}) we obtain
\be
\frac{i}{2g^2} \textrm{Tr} \bigg[G_{MN} [W^M, W^N]\bigg] = \frac{i}{4g^2} \left( G^a_{MN} - G^b_{MN} \right) \left( W^{Mab} W^{Nba} - W^{Nab} W^{Mba}\right).
\ee
We similarly obtain
\bea
\textrm{Tr}( D_M W_N D^M W^N) & = & (\tilde{D}_M W_N^{ba}) (\tilde{D}^M W^{Nab}), \\
\textrm{Tr}( D_M W_N D^N W^M) & = & (\tilde{D}_M W_N^{ba}) (\tilde{D}^N W^{Mab}).
\eea
where $\tilde{D}_M W_N^{ab} = \nabla_M W_N^{ab} -i (B_M^a - B_M^b) W_N^{ab}$. We write $\langle B \rangle_M^{ab} =
B_M^a - B_M^b$.

We now expand about the background fields, writing
\bea
B_M^{a}(y) & = & \langle B_M^{a}(y) \rangle + \delta B_M^{a}(y), \\
W_M^{ab}(y) & = & 0 + \Phi_M^{ab}(y).
\eea
Then
\bea
\textrm{Tr}\bigg[ D_M W_N D^M W^N - D_M W_N D^N W^M \bigg]
& = & D_{\mu} \Phi_i^{ba} D^{\mu} \Phi^{i,ab} + \tilde{D}_j \Phi_i^{ba} \tilde{D}^j \Phi^{i,ab} \nonumber \\
& & - \tilde{D}_j \Phi_i^{ba} \tilde{D}^i \Phi^{j,ab} - 2 (D_{\mu} \Phi_i^{ba} )(\tilde{D}^i W^{\mu,ab}) \nonumber
\eea
The action becomes\footnote{This differs in the first term by a factor of $-1/2$ from the expression in (A.18) of \cite{hepth0404229}.}
\bea
\label{expand}
\mc{L}_{YM}  & = & \frac{i}{4g^2} \left( G^a_{ij} - G^b_{ij} \right)
\left( \Phi^{i,ab} \Phi^{j,ba} - \Phi^{j,ab} \Phi^{i,ba} \right) - \frac{1}{2g^2} \left[ (D_{\mu} \Phi_i^{ba} D^\mu \Phi^{i,ab}) \right. \\
& & + \left. (\tilde{D}_i \Phi_j^{ba} \tilde{D}^i \Phi^{j,ab})
- 2(\tilde{D}_i W_{\mu}^{ba})(D^{\mu} \Phi^{i,ab})
- (\tilde{D}_i \Phi_j^{ba})(\tilde{D}^j \Phi^{i,ab}) \right] \nonumber
\eea
There are extra interactions not included in (\ref{expand}), for example 3- and 4-point interactions,
but these are less relevant for our purposes.
Note that the covariant derivatives $\tilde{D}^i$ reduce to ordinary derivatives in the absence of flux, $G^a_{ij} =0$ -
in this case the gauge connection generates only 3-point (or higher) interactions.

We want to examine this action and
work out the mass eigenstates. We will do this term by term to work
out the contributing parts.

\begin{itemize}

\item{}
 First,
$$
\frac{i}{4g^2} \left(G^a_{ij} - G^b_{ij} \right) \left( \Phi^{i,ab} \Phi^{j,ba}
- \Phi^{j,ab} \Phi^{i,ba} \right) = \frac{i}{2g^2} \Phi^{j,ba} \left( G^a_{ij} - G^b_{ij} \right) \Phi^{i,ab},
$$
where we have used $G_{ij} = - G_{ji}$. We denote $G_{ij}^a - G^b_{ij}$ by $\langle G \rangle_{ij}^{ab}$.
$\langle G \rangle_{ij}^{ab}$ represents the flux difference seen by the $a$ and $b$ sectors. In this case we can then write
\be
\label{put1}
\frac{i}{4g^2} \left(G^a_{ij} - G^b_{ij} \right) \left( \Phi^{i,ab} \Phi^{j,ba}
- \Phi^{j,ab} \Phi^{i,ba} \right) = \frac{i}{2g^2} \Phi^{j,ba} \langle G \rangle_{ij}^{ab} \Phi^{i,ab}.
\ee
This generates a quadratic flux-dependent mass term.

\item{}
The next term we consider is the term
$$
- 2 (D_\mu \Phi_i^{ba})(\tilde{D}^i W^{\mu ab}).
$$
On integration by parts, these will both give rise to terms of the form
$$
(\tilde{D}^i \Phi_i^{ba})(D_\mu W^{\mu ab}).
$$
As we will impose the gauge-fixing condition $(\tilde{D}^i \Phi^{ab}_i = 0)$, these will vanish and will not generate mass terms.

\item{}
The next non-trivial term is
\bea
\frac{1}{2g^2} \left( \tilde{D}_i \Phi_j^{ba} \right) \left( \tilde{D}^j \Phi^{i,ab} \right) & = &
 -\frac{1}{2g^2} \Phi_j^{ba} \left( \tilde{D}_i \tilde{D}^j \Phi^{i,ab} \right) \nonumber \\
& = &  -\frac{1}{2g^2} \Phi_j^{ba} \left( [\tilde{D}_i, \tilde{D}^j ] + \tilde{D}^j \tilde{D}_i \right) \Phi^{i,ab}. \nonumber
\eea
We have integrated by parts here.
As we have gauge-fixed $\tilde{D}_i \Phi^{i,ab} = 0$, we obtain
$$
\frac{1}{2g^2} \left( \tilde{D}_i \Phi_j^{ba} \right) \left( \tilde{D}^j \Phi^{i,ab} \right)
= - \frac{1}{2g^2} \Phi_j^{ba} [\tilde{D}_i, \tilde{D}^j] \Phi^{i,ab}
$$
We need the action of $[\tilde{D}_i, \tilde{D}^j]$ on a vector field. Now,
\bea
[\tilde{D}_i, \tilde{D}_j] & = & [\nabla_i - i \langle B \rangle_i^{ab}, \nabla_j - i \langle B \rangle_j^{ab}] \nonumber \\
& = & [\nabla_i, \nabla_j] - i \left( \nabla_i \langle B \rangle_j^{ab} - \nabla_j \langle B \rangle_i^{ab} \right) \nonumber \\
& = & [\nabla_i, \nabla_j] - i \langle G \rangle_{ij}^{ab}.
\eea
We therefore obtain
\be
\label{put2}
\frac{1}{2g^2} \left( \tilde{D}_i \Phi_j^{ba} \right) \left( \tilde{D}^j \Phi^{i,ab} \right)
= -\frac{1}{2g^2} \Phi_j^{ba} [\nabla_i, \nabla_j] \Phi^{i,ab} +
\frac{i}{2g^2} \Phi^{j,ba} \langle G \rangle_{ij}^{ab} \Phi^{i,ab}.
\ee
\end{itemize}
Putting together the terms (\ref{put1}) and (\ref{put2}) considered
so far, we have
$$
\frac{i}{4g^2} \left(G^a_{ij} - G^b_{ij} \right) \left( \Phi^{iab} \Phi^{jba}
- \Phi^{jab} \Phi^{iba} \right) +
\frac{1}{2g^2} \left( \tilde{D}_i \Phi_j^{ba} \right) \left( \tilde{D}^j \Phi^{i,ab} \right)
$$
\be
\label{extra}
 = \frac{2i}{2g^2} \Phi^{j,ba} \langle G \rangle_{ij}^{ab} \Phi^{i,ab}
- \frac{1}{2g^2} \Phi^{j,ba} [\nabla_i, \nabla_j] \Phi^{i,ab}.
\ee
These represent the `extra' terms that contribute to the vector action
in addition to the naive $\tilde{D}_i \tilde{D}^i$ term,
which gives
\be
\label{put3}
- \frac{1}{2g^2} \tilde{D}_i \Phi_j^{ba} \tilde{D}^i \Phi^{j,ab} = \frac{1}{2g^2} \Phi_j^{ba} \left( \tilde{D}_i \tilde{D}^i \Phi^{j,ab} \right)
\ee
We can finally combine equations (\ref{extra}) and (\ref{put3}) to 
write down the equation of motion satisfied by the vector modes,
\be
\label{veceqn}
\tilde{D}_i \tilde{D}^i \Phi_j^{ab} + 2i \langle G
\rangle^{ab,i}_{\phantom{ab,i}j} \Phi_i^{ab} - [\nabla^i, \nabla_j]
\Phi_i^{ab} 
= -m^2 \Phi_j^{ab}.
\ee
Eigenmodes of the vector fields are obtained by finding solutions of 
eq. (\ref{veceqn}) for $\Phi_i^{ab}$. Zero modes correspond to
solutions with vanishing $m^2$.
For intrinsically massive modes (such as KK modes), we show in the appendix that the masses and profiles of vector modes can be
derived from those of the scalar modes.

We note that, regarded as a adjoint-valued vector in real coordinates, $\Phi_m^{ab}$
must satisfy $(\Phi_m^{ab})^{*} = \Phi_m^{ba}$ and so the he $ab$
modes determine the $ba$ modes. Working with complex coordinates, we require
$(\Phi_m^{ab})^{*} = \Phi_{\bar{m}}^{ba}$. In this notation, $\Phi_m^{ab}$ zero modes
correspond to $({\bf P}, \bar{\bf Q})$ complex scalars and
$\Phi_{\bar{m}}^{ab} \equiv \Phi_m^{ba}$
zero modes correspond to $(\bar{\bf P}, {\bf Q})$ scalars.

\subsubsection{Scalars}

There are two degrees of freedom transforming as scalars in the extra dimensions. One corresponds to the
$A_{\mu} \otimes 1$ mode that transforms as a vector in Minkowksi
space, and the 
other to the transverse scalar mode $\phi_M$ that is valued in the
normal bundle. As for the vector, the scalar can be written as
$\phi^{ab}$. In the presence of an Abelian magnetic flux background, 
$({\bf P}, \bar{\bf Q})$ 
and $(\bar{\bf P}, {\bf Q})$ representations come from $\phi^{ab}$
modes with $ab$ in the upper-right or lower-left blocks.

On dimensional reduction the basic equation determinining
4-dimensional scalar modes is the Laplace equation on the compact space,
\be
\label{scalarlaplace}
-\tilde{D}_i \tilde{D}^i \phi^{ab} = m^2 \phi^{ab},
\ee
where $\tilde{D}_i$ is the gauge-covariant derivative, $\tilde{D}_i
\phi = \nabla_i \phi -i [A_i, \phi]$. 
For the  $A_{\mu} \ti 1$ mode, eq. (\ref{scalarlaplace}) is
sufficient. This mode has all degrees of freedom in the tangent bundle
of the brane and so is untwisted. All eigenmodes can be found directly
from solving (\ref{scalarlaplace}).
 
In contrast, the transverse scalar mode is valued in the normal
 bundle and so must satisfy a twisted version of the Laplace
 equation. As the normal bundle has non-trivial curvature, the
 covariant derivatives must be modified to account for the curvature
 of the bundle. This modification is equivalent to assuming the
 existence of an additional flux background, proportional to the
 K\"ahler form, in the equations of motion.
For cases where the cycle is rigid (i.e. the normal
 bundle has no holomorphic sections), in the absence of flux there are no solutions to
 $\tilde{D}_i \phi = 0$, and all eigenmodes are massive.

In this paper we shall never encounter cases where transverse scalars
have zero modes in a supersymmetric flux background (according to
\cite{08023391}, this can never occur in the geometric regime). For
non-supersymmetric flux configurations, transverse scalars can have
`zero modes', although due to lack of supersymmetry these modes are
not massless. The form of these `zero modes' can be found by solving
$\tilde{D}_i \phi = 0$, where $\tilde{D}_i$ incorporates both the
gauge connection and that due to twisting,
 to obtain the holomorphic section of the bundle.

\subsubsection{Fermions}

For the fermionic degrees of freedom the basic equation of motion is
the twisted Dirac equation \cite{hepth9511222},
\be
\Gamma^M \tilde{D}_M \psi^{ab} = 0.
\ee
$\tilde{D}_M$ incorporates both the spin connection and the gauge
connection due to the fluxes. 
We can write the fermion field $\psi$ as
\be
\psi = \left( \begin{array}{c} \left( \lambda_1 \right) \\ \left(
 \lambda_2 \right) \end{array} \right).
\ee
Here $\lambda_1$ and $\lambda_2$ are as in eq. (\ref{lambdas}). Both
modes are left-handed in four dimensions but have opposite
extra-dimensional chirality and opposite R-charges. The twisting
consists in a shift in the effective magnetic flux felt by the
fermions: as $\lambda_1$ and $\lambda_2$ have different R-charges the
shift takes a different sign for the two modes.

The effect of the twisting is that the gauge connection is 
shifted by an amount equivalent to the normal bundle, 
in such a way that for a pure stack of wrapped branes 
a single constant zero mode
exists in the absence of any magnetic flux.
This zero mode, which comes from the $\lambda_2$ sector, 
corresponds to the gaugino of 4d super Yang-Mills.

In supersymmetric configurations, the $\lambda_1$ modes are fermionic
partners of the scalar modes that come from internal $A_M$ vector
degrees of freedom. The $\lambda_2$ modes are fermionic partners of
either the transverse scalars $\phi_i$ or the 4-dimensional vector
bosons, $A_{\mu} \ti 1$. The equations of motion for the CPT partners
$\lambda_4$ and $\lambda_3$ follow from those for $\lambda_1$ and
$\lambda_2$: the zero modes of these
 are determined by the $\lambda_1, \lambda_2$ modes 
as in eq. (\ref{majconstraints}).

\subsection{Yukawa Interactions}

Four dimensional scalar fields can 
arise either from the $\phi_z$ or $A_M$ degrees of freedom, 
while fermions arise from $\lambda$ either in the
$(\lambda_1 + \lambda_4)$ or $(\lambda_2 + \lambda_3)$ structure. Yukawa couplings will combine two of the fermions with a scalar.
From a higher dimensional perspective
Yukawa couplings originate from the ten-dimensional fermion kinetic term term
$$
\int d^{10}x \textrm{Tr}(\bar{\lambda} \Gamma^m \mc{D}_m \lambda) \to \int d^{10} x \textrm{Tr} (\bar{\lambda}
\Gamma^m [ A_m, \lambda]).
$$
To extract the gauge indices we write $\bar{\lambda} = \bar{\lambda}_I^{ab} e_{ab}$, $A_M = A_{M,J}^{cd} e_{cd}$, and $\lambda = \lambda_K^{ef} e_{ef}$,
where $I,J,K$ are all species indices.
Using $\textrm{Tr}(e_{ab} e_{cd} e_{ef}) = \delta_{af} \delta_{bc} \delta_{de}$, we get
\be
\textrm{Tr} (\bar{\lambda} \Gamma^M [A_M, \lambda] ) =
\left( \bar{\lambda}_I^{ab} \Gamma^M A_{M,J}^{bc} \lambda_K^{ca} - \bar{\lambda}^{ab}_I \Gamma^M A_{M,J}^{ca} \lambda_K^{bc} \right).
\ee
Once the species $I,J,K$ are specified, the Yukawa couplings can be directly evaluated. As $\bar{\lambda} = \lambda^{\dagger} \Gamma^0$,
the basic integral is
\be
\label{fullyukawa}
\int d^{10} x
\left( \lambda_I^{\dagger, ab} \Gamma^0 \Gamma^M A_{M,J}^{bc} \lambda_K^{ca} - \lambda^{\dagger,ab}_I \Gamma^0 \Gamma^M A_{M,J}^{ca} \lambda_K^{bc} \right).
\ee
Here both $\lambda_I$ and $\lambda_K$ are 10-dimensional Majorana-Weyl spinors and so will either be of $(\lambda_1 + \lambda_4)$ or of
$(\lambda_2 + \lambda_3)$ type.

Neglecting the gauge indices in (\ref{fullyukawa}), we can now focus on the spinor structure.
The basic integral we need to evaluate is
\be
\int d^{10} x \left( \lambda_I^{\dagger} \Gamma^0 \Gamma^M A_{M,J} \lambda_K \right).
\ee
This integral takes a different form depending on whether $M = 4,5,6,7$ or $M=8,9$,
correspond to the 4-dimensional scalar arising either from a vector mode valued in the
D7 tangent bundle or a scalar valued in the normal bundle.

\subsubsection*{Transverse Scalars}

We first consider $M=8,9$, where the scalar mode corresponds to the transverse scalar. In this case
\be
\Gamma^0 \Gamma^M = \left( \begin{array}{cc} 0 & \mbb{I} \\ \mbb{I} & 0 \end{array} \right)
\otimes \left( \begin{array}{cc} \mbb{I} & 0 \\ 0 & -\mbb{I} \end{array} \right) \otimes \left( \tau^a \right),
\ee
which gives a chirality flip in both the $0,1,2,3$ and $8,9$ directions.
To obtain a non-vanishing integral we therefore need 
$\lambda_I$ and $\lambda_K$ to be either both of the form $(\lambda_1 + \lambda_4)$ or both of the form
$(\lambda_2 + \lambda_3)$.

We first assume the form $\lambda_I, \lambda_K = (\lambda_1 + \lambda_4)$, when the total Yukawa interaction is
\bea
\mc{L}_{YUK} & = & \int d^{10} x \, \, \left(
\lambda_{4,I}^{\dagger} \Gamma^0 \Gamma^M A_{M,J} \lambda_{1,K} +
\lambda_{1,I}^{\dagger} \Gamma^0 \Gamma^M A_{M,J} \lambda_{4,K} \right) \nonumber \\
& = &
\int d^{10} x \, \,
\left( \xi_{4I}^{\dagger}(x) \xi_{1,K}(x) \right) \left( \psi_{4,I}^{\dagger} (y) \psi_{1,K}(y) \right)
\begin{array}{c} \left( \begin{array}{cc} 0 & \theta_{4I}^{\dagger}(z) \end{array} \right) \\ \ph{0} \end{array}
\Bigg( \tau^a \phi_{aJ} \Bigg) \left( \begin{array}{c} \theta_{1K} \\ 0 \end{array} \right)
+ \nonumber \\
& & \left( \xi_{1I}^{\dagger}(x) \xi_{4,K}(x) \right) \left( \psi_{1,I}^{\dagger} (y) \psi_{4,K}(y) \right)
\begin{array}{c} \left( \begin{array}{cc} 0 & \theta_{1I}^{\dagger}(z) \end{array} \right) \\ \ph{0} \end{array}
\Bigg( \tau^a \phi_{aJ} \Bigg) \left( \begin{array}{c} \theta_{4K} \\ 0 \end{array} \right).
\eea
Using the relations (\ref{majconstraints}), we can express everything in terms of $\lambda_1$ alone, eliminating all
$\lambda_4$ dependence. We take $(\begin{array}{cc} \theta_1(z) & 0 \end{array}) = (\begin{array}{cc} 1 & 0 \end{array})$ as fields only have a trivial dependence
on the transverse coordinates. We obtain
\bea
\mc{L}_{YUK} & = &
\int
-d^4 x \left( \xi_{1I}^T(x) \sigma_y \xi_{1K}(x) \phi_J(x) \right) \int d^4 y \left( \psi_{1I}^T(y) \sigma_y \psi_{1K}(y) \right)
\phi_{z,J}(y) \quad + \nonumber \\
& & \int d^4 x \left( \xi_{1I}^{\dagger}(x) \sigma_y \xi^{*}_{1K}(x) \phi_J(x) \right)
\int d^4 y \left( \psi_{1I}^{\dagger}(y) \sigma_y \psi_{1K}^{*}(y)
\right) \phi_{\bar{z},J}(y).
\label{yukky}
\eea
Here $\phi_{z(\bar{z})}(y) = \phi_8(y) +(-) i \phi_9(y)$. In four dimensional language these interactions
correspond to the Yukawa interactions $\psi_I \psi_K \phi_J +
\psi_I^{*} \psi_K^{*} \phi_J^{*}$, and the structure of this term
is $\lambda_1 \lambda_1 \phi_i$. In terms of chiral superfields this
 corresponds to the interaction between two internal vector modes
and one transverse scalar mode, consistent with the superpotential
$W = h_{ijk} {\bf A_i} {\bf A_j} {\bf \Phi_k}$
given in \cite{08023391}.

Therefore we will need to evaluate integrals of the form
\begin{equation}
  \int d^4 y \left( \psi_{1I}^T(y) \sigma_y \psi_{1K}(y) \right)
\phi_{z,J}(y)
\end{equation}
in order to determine the physical Yukawa couplings once the normalised
wave functions for fermions and scalars are known.

We can perform a similar computation for fermions of the form $(\lambda_2 + \lambda_3)$. In this case similar manipluations show that
the `Yukawa interaction' is
\bea
\mc{L}_{YUK} & = & \int d^{10} x \, \, \left(
\lambda_{2,I}^{\dagger} \Gamma^0 \Gamma^M A_{M,J} \lambda_{3,K} +
\lambda_{3,I}^{\dagger} \Gamma^0 \Gamma^M A_{M,J} \lambda_{2,K}\right) , \nonumber \\
& = & \int
d^4 x \left( \xi_{2I}^T(x) \sigma_y \xi_{2K}(x) \phi_J(x) \right) \int d^4 y \left( \psi_{2I}^T(y) \sigma_y \psi_{2K}(y) \right)
\phi_{\bar{z},J}(y) \quad + \nonumber \\
& & \int d^4 x \left( \xi_{2I}^{\dagger}(x) \sigma_y \xi^{*}_{2K}(x) \phi_J(x) \right)
\int d^4 y \left( \psi_{2I}^{\dagger}(y) \sigma_y \psi_{2K}^{*}(y) \right) \phi_{z,J}(y).
\eea
However this case should not be interpreted as a Yukawa
interaction. The structure of this term is $\lambda_2 \lambda_2
\phi$. The $\sigma_y$ giving a chirality flip in the extra dimensions
implies that, in the language of superfields, one of the $\lambda_2$
fermions is the fermionic part 
of a transverse scalar $\phi_i$ superfield whereas the other is
necessarily the fermionic partner of a
gauge boson vector superfield. This interaction is therefore a gauge
interaction in 4-dimensions of the schematic form $\phi^{*}
\tilde{A}_{\mu} \tilde{\phi}$, where the tildes denote fermions, and
is the supersymmetrisation of the $\phi^{*} A_{\mu} \partial^\mu \phi$
gauge interaction.

\subsubsection*{Internal Vector Modes}

We next consider $M=4,5,6,7$, when the 4d scalar arises from an internal vector mode
valued in the D7 brane tangent bundle. In this case
\be
\Gamma^0 \Gamma^M = \left( \begin{array}{cc} 0 & \mbb{I} \\ \mbb{I} & 0 \end{array} \right)
\otimes \Bigg( \, \, \, \, \tilde{\gamma}^m \, \, \, \, \Bigg) \otimes \mbb{I},
\ee
which generates a chirality flip in both the $0,1,2,3$ and $4,5,6,7$ directions.
In this case to obtain a non-vanishing
integral $\lambda_I$ and $\lambda_K$ must take different forms.
We first consider the case $\lambda_K = (\lambda_1 + \lambda_4)$, and $\lambda_I =
(\lambda_2 + \lambda_3)$. The total Yukawa interaction is
\bea
\mc{L}_{YUK} & = & \int d^{10} x \,
\lambda_{3,I}^{\dagger} \Gamma^0 \Gamma^M A_{M,J} \lambda_{1,K} +
\lambda_{2,I}^{\dagger} \Gamma^0 \Gamma^M A_{M,J} \lambda_{4,K} \nonumber \\
& = & \int d^{10} x \quad
\left( \xi_{3I}^{\dagger}(x) \xi_{1,K}(x) \right) \left(
\begin{array}{c} \left( \begin{array}{cc} 0 & \psi_{3,I}^{\dagger} (y) \end{array} \right) \\ \ph{0} \end{array}
\Bigg( \tilde{\gamma}^M A_{MJ} \Bigg) \left( \begin{array}{c} \psi_{1,K}(y) \\ 0 \end{array} \right) \right)
\left( \theta_{3I}^{\dagger}(z) \theta_{1K} \right)
+ \nonumber \\
& & \left( \xi_{2I}^{\dagger}(x) \xi_{4,K}(x) \right) \left(
\begin{array}{c}
\left(  \begin{array}{cc} 0 & \psi_{2,I}^{\dagger} (y) \end{array} \right) \\ \ph{0} \end{array}
\Bigg( \tilde{\gamma}^M A_{MJ} \Bigg) \left( \begin{array}{c} \psi_{4,K}(y) \\ 0 \end{array} \right) \right)
\left( \theta_{4I}^{\dagger}(z) \theta_{2K} \right).
\eea
We again use the relations (\ref{majconstraints}) to write everything in terms of $\lambda_1$ and
$\lambda_2$. We also use $(\begin{array}{cc} \theta_1(z) & 0 \end{array}) = (\begin{array}{cc} 1 & 0 \end{array})$,
as fields will be brane-valued and so will have trivial dependence
on the transverse coordinates. The Yukawa interactions then become
\bea
\mc{L}_{YUK} & = & - \int
d^4 x \left( \xi_{2I}^T(x) \sigma_y \xi_{1K}(x) \phi_J(x) \right)
\int d^4 y
\left( \begin{array}{cc} 0 & \psi_{2,I}^{T} \sigma_y \end{array} \right)
\Bigg( \tilde{\gamma}^M A_{MJ} \Bigg) \left( \begin{array}{c} \psi_{1,K}(y) \\ 0 \end{array} \right)
\nonumber \\
& + & \int d^4 x \left( \xi_{2I}^{\dagger}(x) \sigma_y \xi^{*}_{1K}(x) \phi_J(x) \right)
\int d^4 y
\left( \begin{array}{cc} 0 & \psi_{2,I}^{\dagger} \sigma_y \end{array} \right)
\Bigg( \tilde{\gamma}^M A_{MJ} \Bigg) \left( \begin{array}{c} \sigma_y \psi^{*}_{1,K}(y) \\ 0 \end{array} \right). \nonumber
\eea
Here $\phi(x)$ represents the 4-dimensional scalar field that is partnered to the vector fluctuation $A_M(y)$ in the low
energy theory.

We can perform a similar computation for the case that $\lambda_K = (\lambda_2 + \lambda_3)$
and $\lambda_I = \lambda_1 + \lambda_4$, where we end up with
\bea
\mc{L}_{YUK} & = &
\int d^{10} x
\lambda_{4,I}^{\dagger} \Gamma^0 \Gamma^M A_{M,J} \lambda_{2,K} +
\lambda_{1,I}^{\dagger} \Gamma^0 \Gamma^M A_{M,J} \lambda_{3,K} , \\
& = & \int
d^4 x \left( \xi_{1I}^T(x) \sigma_y \xi_{2K}(x) \phi_J(x) \right)
\int d^4 y
\left( \begin{array}{cc} 0 & \psi_{1,I}^{T} \sigma_y \end{array} \right)
\Bigg( \tilde{\gamma}^M A_{MJ} \Bigg) \left( \begin{array}{c} \psi_{2,K}(y) \\ 0 \end{array} \right)
 \nonumber \\
& & -\int d^4 x \left( \xi_{1I}^{\dagger}(x) \sigma_y \xi^{*}_{2K}(x) \phi_J(x) \right)
\int d^4 y
\left( \begin{array}{cc} 0 & \psi_{1,I}^{\dagger} \sigma_y \end{array} \right)
\Bigg( \tilde{\gamma}^M A_{MJ} \Bigg) \left( \begin{array}{c} \sigma_y \psi^{*}_{2,K}(y) \\ 0 \end{array} \right). \nonumber
\eea
In both cases the 4-dimensional interaction takes the form $\psi_I \psi_K \phi_J + \psi_I^{*}
\psi_K^{*} \phi_J^{*}$, where the scalar comes from an internal vector
mode. These interactions take the form $\lambda_1
\lambda_2 A_M$. In the language of chiral superfields, two of the
fields ($\lambda_1$ and $A_M$) come from internal-vector superfields,
whereas one ($\lambda_2$) comes from a transverse scalar
superfield. The 4-dimensional interaction comes from the same
superpotential $h_{ijk} {\bf A_i} {\bf A_j} {\bf \Phi_k}$ that generated the interaction (\ref{yukky}).

\section{Compactification on $\mbb{P}^1$}
\label{sec3}

Having described the formalism we now want to apply it local models of
wrapped D7 branes. However as a warm-up example we start by studying the spectrum of fermions,
scalars and vectors 
on $\mbb{P}^1$ with a non-trivial magnetic flux background.
In sections \ref{sec4} and \ref{sec5} we move on to the more interesting cases of $\mbb{P}^1 \ti \mbb{P}^1$ and
$\mbb{P}^2$. We shall compare our results for $\mbb{P}^1$ with those
of \cite{hepth0203264, hepth0304037, hepth0505107}, by finding explicit solutions of the Dirac equation.

$\mbb{P}^1$ is the geometry of the Riemann sphere or the compactified
complex plane. 
In complex coordinates the canonical metric on $\mbb{P}^1$ is
the Fubini-Study metric, which is equivalent to the round sphere
metric with radius $R$. The metric is given by
\begin{equation}
          ds^2 =  { 4R^2 dz d\bar{z} \over ( 1 + z \bar{z} )^2 }.
\end{equation}

\subsection{Fermions on $\mbb{P}^1$}

To write the Dirac equation we need the
spin connection. We choose an orthonormal set of tangent vectors
($e^1, e^2$) for the zweibein, with
\be
e^{1}_{z} = {R \over ( 1 + z \bar z )}, \quad e^{1}_{\bar{z}} = {R \over (1 + z \bar z)}, \quad
e^{2}_{z} = { iR \over (1 + z \bar z)}, \quad e^{2}_{\bar{z}} = { -iR \over (1 + z \bar z)}.
\ee
Raising the indices, we obtain
\be
e^{1 z} = {( 1 + z \bar z ) \over 2R}, \quad
e^{1 \bar{z}} = {(1 + z \bar z) \over 2R}, \quad
e^{2 z} = {  -i(1 + z \bar z) \over 2R}, \quad
e^{2 \bar{z}} = {  i(1 + z \bar z) \over 2R}.
\ee
The spin connection is given by
\be
w_{\mu}^{ab} = {1 \over 2}e^{a \nu} \big( \partial_{\mu} e^{b}_{\nu} -
\partial_{\nu} e^{b}_{\mu} \big) - 
{ 1 \over 2} e^{b \nu} \big( \partial_{\mu} e^{a}_{\nu} - \partial_{\mu} e^{a}_{\nu} \big) - 
{1 \over 2 } e^{\psi a} e^{\sigma b} \big( \partial_{\psi} e_{\sigma c}  - \partial_{\sigma} e_{\psi c} \big) e^{c}_{\mu},
\ee
which evaluates to
\be
w_{z}^{12} = { i \bar{z} \over ( 1 + z \bar{z} ) }, \qquad
w_{\bar{z}}^{12} = { -i z \over ( 1 + z \bar{z} ) }.
\ee
In two dimensions the flat space gamma matrices can be chosen to be
$$
\gamma^{1} = \left( \begin{array}{cc}
0 & 1  \\
1 & 0  \\ \end{array} \right), \qquad
\gamma^{2} = \left( \begin{array}{cc}
0 & -i  \\
i & 0  \\ \end{array} \right).
$$
For future reference, we note that
${1 \over 2}[ \gamma^{1}, \gamma^{2}] = \gamma^{12} = i \left( \begin{array}{cc}
1 & 0  \\
0 & -1  \\ \end{array} \right) = i \sigma^{3}$.
The curved space gamma matrices are
\be
\gamma^{z} = e^{za} \gamma_a = \frac{1}{R}\left( \begin{array}{cc}
0 & 0  \\
(1 + z \bar{z}) & 0  \\ \end{array} \right), \qquad
\gamma^{\bar{z}} = e^{\bar{z} a} \gamma_a = \frac{1}{R} \left( \begin{array}{cc}
0 & (1 +  z \bar z)  \\
0 & 0  \\ \end{array} \right).
\ee
We also want to turn on constant magnetic flux on the two sphere.
The gauge field and field strength are
\be
A_{z}  = { i M \bar{z} \over 2 ( 1 + \bar{z} z )  }, \quad
A_{\bar{z}} = { -iM z \over 2 ( 1 + \bar{z} z ) },
\quad F_{z \bar{z}} = { -i M \over ( 1 + \bar{z} z )^2 }.
\ee
with $M$ integer. 
This field strength is quantised as $\int_{S^{2}} F = -2 \pi M$. We can now write down the Dirac equation
\begin{equation}
  {\bf{\gamma}}^{m} {\cal{D}}_{m} \psi = 0,
\end{equation}
where the covariant derivative $\mc{D}_m$ is
${\cal{D}}_{m} = \partial_{m} + { 1 \over 4 } w_{m \alpha \beta} \gamma^{\alpha \beta} - i A_{m}$,
with $ \gamma^{\alpha \beta} = { 1 \over 2 } [ \gamma^{\alpha} , \gamma^{b} ]$. Using the above results,
the Dirac equation can be written as
\[   \frac{1}{R} \left( \begin{array}{cc}
0 &  (1 + z \bar{z}) \partial_{\bar{z}} - z \left( \frac{M+1}{2} \right) \\
(1 + z \bar{z}) \partial_{{z}} - \bar{z}\left( \frac{-M+1}{2} \right)  & 0
\\ \end{array} \right) \left( \begin{array}{c} \psi_{1} \\ \psi_{2} \\ \end{array} \right) = 0  \]
The equations for different chiralities decouple and the general solutions for zero modes $\psi_{1}$ and $\psi_{2}$ are
\be
 \left( \begin{array}{c} \psi_{1} \\ \psi_{2} \end{array} \right) 
= \left( \begin{array}{c} f(\bar{z})( 1 + z \bar{z} )^{\left(
    \frac{1-M}{2}\right)} \\
g({z})( 1 + z \bar{z} )^{\left( \frac{1+M}{2}\right)} \end{array} \right),
\ee
where $f(\bar{z})$ and $g({z})$ are anti-holomorphic and holomorphic
functions, constrained by normalisability.

In order to obtain physically relevant solutions we must impose that solutions
are normalisable and well-defined. This first implies
that only positive integral powers of
$z$ and $\bar{z}$ be present in $f(\bar{z})$ and $g(z)$.
We also demand that the solutions are square integrable, namely that
\begin{equation}
            \int d^2 z \sqrt{g} \psi^{\dagger} \psi \ \ \ \ \ \textrm{is finite.}
\end{equation}
In order to simplify our discussion we begin by focussing on $\psi^{1}$.
\begin{equation}
            \int d^2 z \sqrt{g} \psi^{1 \dagger} \psi^{1} = 2 \int d^2 z { |f(\bar{z})|^2  \over ( 1 + z \bar z )^{(M+1)} }
\end{equation}
Let us now
examine convergence properties at $z \to \infty$. For $M \leq 0$ there are no solutions for which
the integral is convergent. However for $M > 0$, the integral is convergent if
$f(\bar{z})$ is taken to be a polynomial of degree $M-1$, thus giving
$M$ linearly independent solutions. For $M < 0$, similar arguments imply that
$\psi_2$ has normalisable solutions with $g({z})$ a polynomial of degree
$|M|-1$.

Let us summarise these results:
\begin{itemize}

 \item For $M > 0$, there are normalisable solutions for $\psi_{1}$ only. These are given by
\begin{equation}
 { f_{|M|-1}(\bar{z}) \over ( 1 + z \bar{z} )^{\left(\frac{M-1}{2}\right)} }
\end{equation}
where $f_{M-1}(\bar{z})$ is a polynomial of degree $M-1$, giving
$\vert M \vert$ linearly independent solutions.
\item For $M < 0$, there are normalisable solutions for $\psi_{2}$ only. These are given by
\begin{equation}
 \psi_2 = {  g_{|M|-1}({z}) \over ( 1 + z \bar{z} )^{\left(\frac{|M|-1}{2}\right)} }
\end{equation}
where $g_{|M|-1|}({z})$ is a polynomial of degree $|M|-1$, giving
$|M|$ linearly independent solutions.
\item For $M=0$, there are no normalisable zero modes.
 \end{itemize}

\subsubsection*{Twisting}

If the $\mbb{P}^1$ is embedded in a Calabi-Yau, the fermionic
equations of motion will be twisted to account for the nontrivial
normal bundle. The twist must incorporate the effect of the normal bundle
and corresponds to a shift in the flux value $M
\to M +1$.
A reason to see why this must occur is because a
unfluxed stack of D-branes wrapped on a $\mbb{P}^1$ embedded in a
Calabi-Yau is supersymmetric. The twisting is necessary to ensure that
the constant gaugino zero mode exists. After the twisting procedure,
the fermionic solutions become
\be
 \left( \begin{array}{c} \psi_{1} \\ \psi_{2} \end{array} \right) 
= \left( \begin{array}{c} f(\bar{z})( 1 + z \bar{z} )^{
    \frac{-M}{2}} \\
g({z})( 1 + z \bar{z} )^{\left( 1+\frac{M}{2}\right)} \end{array} \right).
\ee
The number of zero modes of the twisted Dirac equation is then
$|M-1|$.

\subsection{Scalars on $\mbb{P}^1$}
The scalar spectrum on $\mbb{P}^1$ is determined by solving
\be
- g^{\mu \nu} D_{\mu} D_{\nu} \phi = - \big( g^{z \bar{z}} D_{z} D_{\bar{z}} + g^{\bar{z} z} D_{\bar{z}} D_{z} \big) \phi = m^2 \phi.
\ee
Here $D_{z}$ and $D_{\bar{z}}$ are covariant derivatives, incorporating the gauge and geometric connections.
Acting on a scalar they are given by
\bea
D_{z} \phi = (\partial_{z} - iA_{z}) \phi & = & \left(\partial_{z} + { M \bar{z} \over 2 (1 + z \bar{z} ) }\right) \phi, \nonumber \\
D_{\bar{z}} \phi = (\partial_{\bar{z}} - iA_{\bar{z}}) \phi & = & \left(\partial_{\bar{z}} - { M {z} \over 2 (1 + z \bar{z} )}\right) \phi.
\eea
We note that
\be
[D_{z}, D_{\bar{z}}] \phi = -iF_{z \bar{z}} \phi = -{ M \over ( 1 + z \bar{z} )^{2} } \phi.
\ee
We can then write
\bea
- g^{\mu \nu} D_{\mu} D_{\nu} \phi & = & - \big( g^{z \bar{z}} D_{z} D_{\bar{z}} + g^{\bar{z} z} D_{\bar{z}} D_{z} \big) \phi \nonumber \\
  & = & - 2 g^{z \bar{z}} D_{\bar{z}} D_{z} \phi - g^{\bar{z}z}[D_{z}, D_{\bar{z}}] \phi \nonumber \\
  & = & - 2 g^{z \bar{z}} D_{\bar{z}} D_{z} \phi + \frac{M}{2R^2}\phi.
\eea
The spectrum of $- 2 g^{z \bar{z}} D_{\bar{z}} D_{z}$ is positive semi-definite,
which follows from arguments similar to those used to show
the spectrum of the scalar Laplacian on a compact manifold is positive semi-definite.

We can now analyse the spectrum. The simplest case is $M=0$, for which the
lowest mode has eigenvalue zero and is constant over the two sphere.
For $M>0$.
we know from our analysis of the Dirac equation that
$D_{z}$ has zero modes, and thus
the lowest mass modes can be obtained by choosing wavefunctions $\phi$, such
that $D_{z} \phi = 0$. These modes have mass ${M \over 2}$ and degeneracy $|M|+1$.
Note that $D_{z} \phi = 0$ are the same equations solved in the fermionic analysis,
subject to a shift in the effective value of $M$. The shift is due to the fact that fermions couple to the curvature of the
sphere. As the field strength of the Dirac monopole is proportional to the K\"ahler form,
the effect of the curvature precisely corresponds
to a shift in the value of $M$.

An identical argument for $M < 0$ shows that the lowest modes have mass
${|M| \over 2}$ in this case, and are given by the solutions of $D_{\bar{z}} \phi = 0$.
The lowest modes are therefore always massive for $M \neq 0$, with the 
forms of the solutions being
\bea
M > 0: \quad \phi(z, \bar{z}) & = & \frac{f_M(\bar{z})}{( 1 + z \bar{z})^{\vert M \vert/2} }, \quad m^2 = \frac{\vert M \vert}{2R^2},  \nonumber \\
M = 0: \quad \phi(z, \bar{z}) & = & \phi_0, \quad m^2 = 0, \nonumber \\
M < 0: \quad \phi(z, \bar{z}) & = & \frac{f_M(z)}{( 1 + z \bar{z})^{\vert M \vert/2} }, \quad m^2 = \frac{\vert M \vert}{2R^2}.
\eea
where $f_M$ is a holomorphic polynomial of degree $M$. The restriction to degree $M$ is to ensure that the
wavefunctions are square integrable, namely that $\int \sqrt{g} \phi^* \phi$ is finite. The fact that the polynomial
is of degree $M$ implies that the degeneracy of the zero mode solutions is $\vert M \vert + 1$.

\subsubsection*{Twisting}

As the normal bundle is nontrivial, scalars which correspond to
transverse deformations of branes wrapped on $\mbb{P}^1$ will
be twisted. This follows from the fact that, embedded into a
Calabi-Yau,  the normal bundle has no global sections due to the
adjunction formula. Transverse deformations of the brane are therefore
automatically
massive, and so in the zero flux case no massless modes can exist. 
The twisting will correspond to a shift in the flux, $M \to M+1$, such
that in the absence of flux there are no massless scalar
modes.\footnote{The fluxed case is always non-supersymmetric. While in
  the fluxed case `zero modes' can exist, in the sense that
  holomorphic sections of the bundle exist, the scalar `zero modes' are not
  massless due to the lack of supersymmetry.}

\subsection{Vectors on $\mbb{P}^1$}

We finally want to study vector zero modes on $\mbb{P}^1$ in the fluxed case. 
As derived in the general case, we start with the equation
\be
\tilde{D}_i \tilde{D}^i \Phi_{j,ab} + 2i \langle G \rangle^{ab,i}_{\ph{ab,i}j}
\Phi_{i,ab} - [\nabla^i, \nabla_j] \Phi_{i,ab} = -m^2 \Phi_{i,ab}.
\ee
We will assume $a>b$ for definiteness. The fluxes are
$$
G^{z\phantom{z}ab}_{\phantom{z}z} = \frac{iM}{2 R^2}, \qquad G^{\bar{z}\phantom{z}ab}_{\phantom{z}\bar{z}} = \frac{-iM}{2 R^2}.
$$
The gauge-fixing condition that we impose is
$$
\tilde{D}_i \Phi^{i,ab} = g^{z \bar{z}} (\tilde{D}_z \Phi_{\bar{z}}^{ab} + \tilde{D}_{\bar{z}} \Phi_{z}^{ab}) = 0.
$$
For zero-modes (or their equivalent) we expect to find the mode by
solving first-order equations. We will do this 
either through
$\Phi_{\bar{z}}^{ab} = 0$ and $\tilde{D}_{\bar{z}} \Phi_z^{ab} = 0$, or with the opposite $z \to \bar{z}$ replacement.
 The non-zero elements of the mode in 
the $({\bf P}, \bar{\bf Q})$
 representation will therefore be
$\Phi_z^{ab}$ and $\Phi_{\bar{z}}^{ba} =  (\Phi_z^{ab})^*$. 
The $(\bar{\bf P}, {\bf Q})$
 representation will correspond to
$\Phi_z^{ba}$ and $\Phi_{\bar{z}}^{ab} =  (\Phi_z^{ba})^*$. 

Throughout this section we shall assume this ansatz for the vector modes,
focussing on the zero modes rather than on the massive KK states.
Let us evaluate the effect of the curvature term on such vector modes,
first assuming $\Phi_z^{ab} \neq 0$ and $\Phi_{\bar{z}}^{ab} = 0$.
We can write
\bea
[\nabla_i, \nabla_j] \Phi^{i,ab} = g^{ik} [\nabla_i, \nabla_j] \Phi_k^{ab} & = & g^{\bar{z}z} [\nabla_{\bar{z}}, \nabla_z] \Phi_z^{ab} \nonumber \\
& = & g^{\bar{z} z} \left( - \partial_{\bar{z}} \Gamma^z_{zz} \right) \Phi_z^{ab} \nonumber \\
& = & \frac{(1+ z \bar{z})^2}{2 R^2} \left( \frac{2}{(1+ z \bar{z})^2} \Phi_z^{ab} \right) = \frac{\Phi_z^{ab}}{R^2}.
\eea
We obtain a similar expression for $\Phi_{\bar{z}}^{ab}$. This allows us to write
\bea
2i \langle G \rangle^{ab,z}_{\phantom{ab,z}z} \Phi_{z,ab} -[\nabla^z, \nabla_{\bar{z}}] \Phi_{z,ab} & = & \frac{1}{R^2}(-M-1)\Phi_{z,ab}, \nonumber \\
2i \langle G \rangle^{ab,\bar{z}}_{\phantom{ab,z}\bar{z}} \Phi_{\bar{z},ab} -[\nabla^z, \nabla_{\bar{z}}] \Phi_{\bar{z},ab} & = & \frac{1}{R^2}(M-1)\Phi_{\bar{z},ab}.
\eea
We finally want to evaluate the Laplacian acting on $\Phi_z^{ab}$, $\tilde{D}_i \tilde{D}^i \Phi_z^{ab}$.
Expanding this out and using
$\tilde{D}_i \Phi^{i,ab} = 0$, we obtain
\be
\label{eq000}
\tilde{D}_i \tilde{D}^i \Phi_z^{ab} 
= \left[ g^{z \bar{z}}\left( \tilde{D}^s_z \tilde{D}^s_{\bar{z}} + \tilde{D}^s_{\bar{z}} \tilde{D}^s_z \right)
-g^{z \bar{z}} \partial_{\bar{z}} \Gamma^z_{zz} \right] \Phi_z^{ab}
\ee
where $\tilde{D}^s$ denotes the scalar covariant derivative,
$\tilde{D}^s_i \Phi^{ab} = \left( \partial_i -i(\langle A_i \rangle^a -
\langle A_i \rangle^b) \right) \Phi^{ab}$.
We can simplify (\ref{eq000}) by commuting the covariant derivatives through, to obtain
\be
\tilde{D}_i \tilde{D}^i \Phi_z^{ab} 
= g^{z \bar{z}} \tilde{D}^s_z \tilde{D}^s_{\bar{z}} \Phi_z^{ab} + \left( \frac{M}{2} + 1 \right) \frac{\Phi_z^{ab}}{R^2}.
\ee
Combining all terms, we obtain
\be
\label{comb1}
\tilde{D}_i \tilde{D}^i \Phi_{z,ab} + 2i \langle G \rangle^{ab,z}_{\ph{ab,z}z} \Phi_{z,ab} - [\nabla^z, \nabla_z] \Phi_{z,ab} =
\left(g^{z \bar{z}} \tilde{D}^s_z \tilde{D}^s_{\bar{z}} - \frac{M}{2}
\right) \frac{\Phi_z^{ab}}{R^2} = -m^2 \frac{\Phi_z^{ab}}{R^2}.
\ee
Performing a similar calculation for $\Phi_{\bar{z}}$ gives
\be
\label{comb2}
\tilde{D}_i \tilde{D}^i \Phi_{\bar{z},ab} + 
2i \langle G \rangle^{ab,\bar{z}}_{\ph{ab,\bar{z}}\bar{z}} \Phi_{\bar{z},ab} - [\nabla^{\bar{z}}, \nabla_{\bar{z}}] \Phi_{\bar{z},ab} =
\left(g^{z \bar{z}} \tilde{D}^s_{\bar{z}} \tilde{D}^s_{z} +
\frac{M}{2} \right) 
\frac{\Phi_{\bar{z}}^{ab}}{R^2} = -m^2 \frac{\Phi_{\bar{z}}^{ab}}{R^2}.
\ee
The lowest mass state is therefore tachyonic with a mass $-\vert M
\vert/2 R^2$, 
and is obtained by solving $D^s_{\bar{z}} \Phi_z^{ab} = 0$ (for $M < 0$)
or $D^s_z \Phi_{\bar{z}}^{ab} = 0$ (for $M > 0$). Note that solving $D^s_{\bar{z}} \Phi_z^{ab} = 0$ automatically ensures that
$D^s_z \Phi_{\bar{z}}^{ba} = 0$.

For the zero mode solution to exist, we require $\vert M \vert \ge 2$, with a zero mode degeneracy of $\vert M \vert - 1$. This
is due to constraints on normalisability. The solution of the 
zero mode equation with $M < 0$ units of flux gives
$$
D^s_{\bar{z}} \Phi_z^{ab} = 0 \to \Phi_z^{ab} = f(z)(1 + z \bar{z})^{-\frac{|M|}{2}}.
$$
The normalisability condition is that
$$
\int \sqrt{g} g^{ij} \Phi_i^{ab} \Phi_j^{ba}
$$
be finite, which requires that $g^{ij} \Phi_i^{ab} \Phi_j^{ba}$ does not diverge as $z \to \infty$. As $g^{z \bar{z}} = \frac{(1+z \bar{z})^2}{2}$,
this requires $\vert M \vert \ge 2$ for any zero mode solutions to
exist. It also follows easily from this condition that 
$f$ has to be a polynomial of degree $\vert M\vert -2$ and therefore 
the degeneracy of zero modes
is given by $\vert M \vert - 1$. All such zero modes are chiral, in the sense that they correspond to $\vert M \vert -1$
complex scalar fields in the $({\bf P}, {\bf \bar{Q}})$
representation of the gauge group.

\subsection{Normalisation and Overlap Integrals}
The functional form of the lowest lying modes takes the same form independent of whether the mode is scalar, vector or spinorial.
Introducing a normalisation constant $\mc{N}^{K}_{M}$, we can write a generic zero mode wavefunction $\psi^K_M$ as (we take $M>0$)
\begin{equation}
 \psi^{K}_{M} = { 1 \over \mc{N}^{K}_{M} } \frac{z^{K}}{( 1 + z \bar{z})^{\frac{M-1}{2}}}.
\end{equation}
Then
\begin{equation}
  \left\vert \mc{N}_{M}^{K} \right\vert^{2} = \int d^{2} z \sqrt{g} \psi^{\dagger} \psi = 4 R^2 \int  { dx dy (z \bar{z})^K \over ( 1 + z \bar{z} )^{M+1} }
     = 8 \pi R^2 \int  { dr r^{2K+1}  \over ( 1 + r^{2} )^{M+1} }.
\end{equation}
Expressions for wavefunction overlaps will involve a similar integral, as any integral with
differing powers of $z$ and $\bar{z}$ will automatically vanish when the angular integration
is performed. We therefore define the standard integral
\begin{equation}
\label{IKM}
  \int  { dr \, r^{2K+1}  \over ( 1 + r^{2} )^{M+1} } = { \Gamma(K +
  1) \Gamma(M-K) \over 2\Gamma(M+1) } \equiv I^{{ K}}_{M} ,
\end{equation}
where the evaluation can be performed using the substitution $r = \tan \theta$ and the relation%
\begin{equation}
\int_{0}^{\pi/2} d \theta \sin ^{2p-1} (\theta)  \cos ^{2q-1} (\theta) =  { \Gamma(p) \Gamma(q) \over 2 \Gamma(p+q) }.
\end{equation}
The normalisation constant $\mc{N}_M^K$ is therefore given by
\begin{equation}
\left\vert \mc{N}_{M}^{K} \right\vert^{2} = { 4 \pi R^2 \Gamma(K + 1)
  \Gamma(M-K) \over \Gamma(M+1) }= 8\pi R^2 I^K_M.
\end{equation}

The non-vanishing triple overlap integrals relevant for Yukawa
couplings 
can also be computed in terms of these closed form integrals and take the
form:
\bea
Y^{KL}_{MNP} & = &
\frac{1}{\mc{N}_M^K\mc{N}_N^L\mc{N}_P^{K+L}}
\int\frac{ d^2z\, \sqrt{g}\,
  z^Kz^L{\bar{z}}^{K+L}}{(1+z\bar{z})^{\frac{M-1}{2}}(1+z\bar{z})^{\frac{N-1}{2}}(1+z\bar{z})^{\frac{P-1}{2}}} \nonumber \\ 
& =& 
\frac{8\pi R^2}{\mc{N}_M^K\mc{N}_N^L\mc{N}_P^{K+L}}
 \, I^{{K+L}}_\frac{M+N+P-1}{2}
\eea

The canonical $\mbb{P}^1$ metric has the $SO(3)$ symmetry of the
2-sphere. This implies that zero modes fall into $SO(3)$
representations, both in degeneracy and in structure. The $SO(3)$ acts
as a flavour symmetry, as it relates different zero modes with the
same gauge charges.
The $SO(3)$ flavour symmetry will govern the
Yukawa couplings, which are determined by the representation theory of
$SO(3)$. In the limit that the compact space really is compact, the
$SO(3)$ isometry of $\mbb{P}^1$ will be broken, and the Yukawa
couplings will be governed by an approximate $SO(3)$ symmetry.

\section{Compactification on $\mbb{P}^1 \ti \mbb{P}^1$}
\label{sec4}

We now move onto models involving D7 branes, starting with $\mbb{P}^1
\ti \mbb{P}^1 \equiv \mbb{F}^0$ 
as the simplest appropriate geometry.
Our interest is not in compactification on $\mbb{P}^1 \ti \mbb{P}^1$
\emph{per se} , but rather on $\mbb{P}^1 \ti \mbb{P}^1$ embedded in a Calabi-Yau. The tangent bundle of $\mbb{P}^1$
has $c_1(\mbb{P}^1) = 2J$, which is $\mc{O}_{\mbb{P}^1}(2)$. From the adjunction formula and the triviality of the first
Chern class for a Calabi-Yau it therefore follows that the normal bundle to $\mbb{P}^1 \ti \mbb{P}^1$ is
$\mc{O}_{\mbb{P}^1 \ti \mbb{P}^1} (-2, -2)$. The fact that this is nontrivial indicates that the gauge theory will be twisted,
and the fact that the normal bundle has no global sections indicates that we can sensibly consider this as a local model.

We will assume that the Calabi-Yau metric restricted to the minimal
volume $\mbb{P}^1 \ti \mbb{P}^1$ is canonical, $ds^2_{\mbb{P}^1 \ti
  \mbb{P}^1} = R_1^2 ds^2_{\mbb{P}^1} \otimes R_2^2
ds^2_{\mbb{P}^1}$. While we do not know of an analytic expression for
the full Calabi-Yau case, we make this assumption on analogy with the
$\mbb{P}^1$ case, where the non-compact Eguchi-Hanson metric restricts to the
canonical $\mbb{P}^1$ metric on the minimal area $\mbb{P}^1$, and the 
case of the resolution of $\mbb{C}^3/\mbb{Z}_3$, where the 
non-compact Calabi-Yau metric restricts to the canonical Fubini-Study
metric on the resolving $\mbb{P}^2$.

We denote the coordinates on the two $\mbb{P}^1$s by
$z$ and $w$. The $\mbb{P}^1 \ti \mbb{P}^1$ metric is
\be
ds^2 = \frac{4 R_1^2 dz d\bar{z}}{(1+ z \bar{z})^2} + \frac{4 R_2^2 dw d\bar{w}}{(1+ w \bar{w})^2}
\ee
and so
$$
g_{z \bar{z}} = \frac{2 R_1^2}{(1 + z \bar{z})^2}, \quad g_{w \bar{w}} = \frac{2 R_2^2}{(1 + w \bar{w})^2}, \quad
g^{z \bar{z}} = \frac{(1 + z \bar{z})^2}{2 R_1^2}, \quad g^{w \bar{w}} = \frac{(1 + w \bar{w})^2}{2 R_2^2}.
$$

\subsection{Fermions on $\mbb{P}^1 \ti \mbb{P}^1$}

We want to study the massless fermion spectrum on $\mbb{P}^1 \ti \mbb{P}^1$.
We first construct and solve the fluxed Dirac equation on $\mbb{P}^1 \ti \mbb{P}^1$, and then
describe how this is twisted to account for the non-trivial normal bundle.
Using the metric we construct a vielbein
\bea
e^1_z = \frac{R_1}{1 + z \bar{z}}, \quad e^1_{\bar{z}} = \frac{R_1}{1+ z \bar{z}}, \quad & &  e^1_w = 0, \quad e^1_{\bar{w}} = 0. \\
e^2_z = \frac{iR_1}{1 + z \bar{z}}, \quad e^2_{\bar{z}} = \frac{-iR_1}{1+ z \bar{z}}, \quad & &  e^2_w = 0, \quad e^2_{\bar{w}} = 0.
\eea
\bea
e^3_z = 0, \quad e^3_{\bar{z}} = 0, \quad & &  e^3_w = \frac{R_2}{1 + w \bar{w}}, \quad e^3_{\bar{w}} = \frac{R_2}{1 + w \bar{w}}. \\
e^4_z = 0, \quad e^4_{\bar{z}} = 0, \quad & &  e^4_w = \frac{iR_2}{1 + w \bar{w}}, \quad e^4_{\bar{w}} = \frac{-iR_2}{1 + w \bar{w}}.
\eea
As the metric factorises, both the vielbein and the spin connection can be broken down into separate parts. Using the
results of section \ref{sec3} for $\mbb{P}^1$, the spin connection is
$$
\omega_z^{12} = \frac{i \bar{z}}{(1+ z \bar{z})}, \quad \omega_{\bar{z}}^{12} = \frac{-i z}{1+z \bar{z}}, \quad
\omega_w^{34} = \frac{i \bar{w}}{(1+ w \bar{w})}, \quad \omega_{\bar{w}}^{34} = \frac{-i w}{1+ w\bar{w}}.
$$
The fact that the metric is a direct product implies there are no cross-terms such as $13$ or $24$.

The gamma matrices are defined
in a coordinate basis through $\tgamma^\alpha = e^{\alpha i} \tgamma_i$, giving
\bea
\tgamma^z & = & e^{z1} \tgamma_1 + e^{z2} \tgamma_2 = \left( \frac{1 + z \bar{z}}{2 R_1} \right) (\tgamma_1 - i \tgamma_2), \\
\tgamma^{\bar{z}} & = & e^{\bar{z}1} \tgamma_1 + e^{\bar{z}2} \tgamma_2 = \left( \frac{1 + z \bar{z}}{2 R_1} \right) (\tgamma_1 + i \tgamma_2), \\
\tgamma^w & = & e^{w3} \tgamma_3 + e^{w4} \tgamma_4 = \left( \frac{1 + w \bar{w}}{2 R_2} \right) \left( \tgamma_3 - i \tgamma_4\right), \\
\tgamma^{\bar{w}} & = & e^{\bar{w}3} \tgamma_3 + e^{\bar{w}4}
\tgamma_4 = \left( \frac{1 + w \bar{w}}{2 R_2} \right) 
\left( \tgamma_3 + i \tgamma_4\right).
\label{gammazp1p1}
\eea
Using these expressions and the form for the $\gamma$ matrices from eq. (\ref{conve}), we can compute the
spin connection terms appearing in the
Dirac equation:
\be
\frac{1}{8} \tgamma^m \omega_{m}^{\alpha \beta} [\tgamma^\alpha, \tgamma^\beta] = \frac{1}{2} \left( \begin{array}{cccc}
0 & 0 & \frac{i \bar{z}}{R_1} & -\frac{w}{R_2} \\ 0 & 0 & - \frac{\bar{w}}{R_2} & \frac{iz}{R_1} \\
-\frac{iz}{R_1} & -\frac{w}{R_2} & 0 & 0 \\ - \frac{\bar{w}}{R_2} & - \frac{i \bar{z}}{R_1} & 0 & 0 \end{array}
\right) .
\ee
The kinetic term is $\tgamma^m \partial_m$, which evaluates to
\be
\tgamma^m \partial_m = \left( \begin{array}{cccc} 0 & 0 & -\frac{(1+ z \bar{z})}{R_1}i\partial_z & \frac{(1+w\bar{w})}{R_2}\partial_{\bar{w}} \\
0 & 0 & \frac{(1+w \bar{w})}{R_2}\partial_w & -\frac{(1+z \bar{z})}{R_1}i \partial_{\bar{z}} \\
i\frac{(1 + z \bar{z})}{R_1} \partial_{\bar{z}} & \frac{(1 + w \bar{w})}{R_2} \partial_{\bar{w}} & 0 & 0 \\
\frac{(1 + w \bar{w})}{R_2} \partial_w & i\frac{( 1 + z \bar{z})}{R_1} \partial_z & 0 & 0  \end{array} \right).
\ee
Prior to including any effects of magnetic flux, we have
 $$
 \tgamma^m \partial_m + \frac{1}{8} \tgamma^m \omega_m^{\alpha \beta} [\tgamma^\alpha, \tgamma^\beta]
 =
 $$
 $$
 \left( \begin{array}{cccc} 0 & 0 & \frac{i}{R_1} \left(-(1+z \bar{z})\partial_z + \frac{\bar{z}}{2} \right) &
 \frac{1}{R_2}\left( (1+w \bar{w}) \partial_{\bar{w}} - \frac{w}{2} \right) \\
 0 & 0 & \frac{1}{R_2} \left( (1+w \bar{w}) \partial_{w} - \frac{\bar{w}}{2} \right) & \frac{i}{R_1}\left(-(1+z \bar{z})\partial_{\bar{z}} + \frac{z}{2} \right) \\
 \frac{i}{R_1}\left((1+z \bar{z})\partial_{\bar{z}} - \frac{z}{2} \right) & \frac{1}{R_2}\left((1+w \bar{w}) \partial_{\bar{w}} - \frac{w}{2} \right) & 0 & 0 \\
\frac{1}{R_2}\left(  (1+w \bar{w}) \partial_{w} - \frac{\bar{w}}{2} \right)
& \frac{i}{R_1} \left((1+z \bar{z})\partial_{z} - \frac{\bar{z}}{2} \right) & 0 & 0 \end{array} \right)
 $$

To obtain chiral fermions, we also require a gauge field background on $\mbb{P}^1 \ti \mbb{P}^1$. We place a Dirac monopole
 background on each $\mbb{P}^1$, with $M$ units of flux on the first $\mbb{P}^1$ and $N$ units of flux on the
 second $\mbb{P}^1$. The resulting gauge field is
 $$
 A = \frac{i M \bar{z}}{2 (1 + z \bar{z})} dz - \frac{i M z}{2(1 + z \bar{z})} d \bar{z}
 + \frac{i N \bar{w}}{2(1 + w \bar{w})} dw - \frac{i N w}{2( 1 + w \bar{w})} d \bar{w},
 $$
 with
 \be
 F = \frac{-iM}{(1+z\bar{z})^2} dz \wedge d \bar{z} - \frac{iN}{(1 + w \bar{w})^2} dw \wedge d \bar{w},
 \ee
 and
 $$
 \int_{\mbb{P}^1(z)} F = -2 \pi M, \quad \int_{\mbb{P}^1(w)} F = -2 \pi N.
 $$
 We can then evaluate
 \be
 -i \tgamma^m A_{m} = \frac{1}{2} \left( \begin{array}{cccc} 0 & 0 & -\frac{iM \bar{z}}{R_1} & -\frac{N w}{R_2} \\
 0 & 0 & \frac{N \bar{w}}{R_2} & \frac{iMz}{R_1} \\
 -\frac{iMz}{R_1} & -\frac{Nw}{R_2} & 0 & 0 \\
 \frac{N \bar{w}}{R_2} & \frac{i M \bar{z}}{R_1} & 0 & 0 \end{array} \right).
 \ee

 We can now write down the Dirac operator on $\mbb{P}^1 \ti \mbb{P}^1$. This is
 $$
 \tgamma^m D_m = \tgamma^m (\partial_m + \frac{1}{8} \omega_\mu^{\alpha \beta} [\tgamma^\alpha, \tgamma^\beta]
 -i A_\mu) = \left( \begin{array}{cc} 0 & \mc{D}_{+} \\ \mc{D}_{-} & 0 \end{array} \right),
 $$
 where
 \bea
 \mc{D}_{+} & = &
 \left( \begin{array}{cc}  \frac{i}{R_1}\left(-(1+z \bar{z})\partial_z + \frac{(1-M)}{2}\bar{z} \right) &
 \frac{1}{R_2}\left((1+w \bar{w}) \partial_{\bar{w}}
 - \frac{(N+1)}{2}w \right) \\
 \frac{1}{R_2}\left((1+w \bar{w}) \partial_{w} + \frac{(N - 1)}{2}\bar{w}\right) & \frac{i}{R_1}\left(-(1+z \bar{z})\partial_{\bar{z}} + \frac{(M+1)}{2}z \right)
 \end{array} \right), \nonumber \\
 \mc{D}_{-} & = &
 \left( \begin{array}{cc}
 \frac{i}{R_1}\left((1+z \bar{z})\partial_{\bar{z}} - \frac{(M+1)}{2}z \right) &
 \frac{1}{R_2}\left((1+w \bar{w}) \partial_{\bar{w}} - \frac{(N+1)}{2}w \right) \\
 \frac{1}{R_2} \left( (1+w \bar{w}) \partial_{w} + \frac{(N-1)}{2} \bar{w} \right) &
 \frac{i}{R_1}\left((1+z \bar{z})\partial_{z} +\frac{(M-1)}{2}\bar{z} \right)
 \end{array} \right).
 \eea
We write the fermions as a column vector,
$
\psi = \left( \begin{array}{cccc} \psi_1 & \psi_2 & \psi_3 & \psi_4 \end{array} \right)^T
$
 and look for zero modes of the Dirac equation. It is easy to verify that, prior to imposing
 normalisability, zero mode solutions of the Dirac equation take the following form:
 \be
 \psi = \left( \begin{array}{c} (1+ w \bar{w})^{\frac{1-N}{2}} (1+z \bar{z})^{\frac{1+M}{2}} f( \bar{w}, z) \\
 (1+ w \bar{w})^{\frac{1+N}{2}} (1+z \bar{z})^{\frac{1-M}{2}} f( w, \bar{z}) \\
 (1+ w \bar{w})^{\frac{1-N}{2}} (1+z \bar{z})^{\frac{1-M}{2}} f( \bar{w}, \bar{z}) \\
 (1+ w \bar{w})^{\frac{1+N}{2}} (1+z \bar{z})^{\frac{1+M}{2}} f(w, z)
 \end{array} \right),
 \label{pretwisted}
 \ee
where $f$ are holomorphic polynomials.

\subsubsection{Twisting of Fermionic Modes}

To compute the wavefunctions for fermions living on D-brane worldvolumes, it is necessary to twist the
Dirac equation to account for the fact that the cycle lives in a
curved space and has a nontrivial normal bundle
\cite{hepth9511222, 08023391}.
As a K\"ahler manifold, the holonomy group of the surface is $U(2)$, broken to $U(1) \ti U(1)$ by the direct
product metric of $\mbb{P}^1 \ti \mbb{P}^1$. This $U(2)$ can be
written as $SU(2) \ti U(1)_J$, where the $U(1)_J$ 
corresponds to the central $U(1)$ of the tangent bundle. In preparation for the description of the twisting procedure, it is helpful to see
how the central $U(1)$ acts on the four spinor modes $\psi_i$. The action of $U(1)_J$ on
tangent vectors is
\be
\left( \begin{array}{c} z \\ w \end{array} \right) \to \left( \begin{array}{cc} e^{i \theta} & 0 \\ 0 & e^{i \theta} \end{array}
\right) \left( \begin{array}{c} z \\ w \end{array} \right).
\ee
The action on spinors is given by $\tilde{\Lambda} = \frac{1}{4} \Lambda_{\alpha \beta} [\gamma^{\alpha}, \gamma^{\beta}]$. Here
$\Lambda^{z}_{\ph{z}z} = i \theta, \Lambda^{w}_{\ph{w}w} = i \theta$ for infinitesimal transformations. Using the expressions (\ref{gammazp1p1}) for
$\gamma^z, \gamma^{\bar{z}}$,
we obtain
\bea
\tilde{\Lambda} & =  & -\frac{\theta}{4} \left( [\gamma_2, \gamma_1] + [\gamma_4, \gamma_3] \right) \\
& = & \left( \begin{array}{cccc} 0 & 0 & 0 & 0 \\ 0 & 0 & 0 & 0 \\
0 & 0 &  i \theta & 0 \\
0 & 0 &  0 & -i \theta \end{array} \right)
\label{tlam}
\eea
From (\ref{tlam}) it is clear that under $U(1)_J$ $\psi_1$ and $\psi_2$ have charge 0, while $\psi_3$ and $\psi_4$ have charges $\pm 1$ respectively.
Identifying $SU(2)_L$ with the $SU(2)$ of the $U(2)$ holonomy group,
we can then regard ($\psi_1, \psi_2$) and ($\psi_3, \psi_4$) as being in the $(\bf{2},\bf{1})$ and $(\bf{1},\bf{2})$
representations of $SO(4) = SU(2)_L \ti SU(2)_R$. As discussed in section \ref{sec2},
on dimensional reduction all the spinor modes correspond to left-handed spinors in 4-dimensions.
The fact that we start with a positive chirality spinor in ten dimensions implies that the representation content is
$$
[({\bf{2},\bf{1})\otimes(\bf{2},\bf{1})]}_{+1/2} \oplus [({\bf{2}, \bf{1})\otimes(\bf{1},\bf{2})]}_{-1/2},
$$
where the last number designates the R-charge of the mode under the $U(1)$ transverse to the brane world volume.
$(\psi_1, \psi_2$) and ($\psi_3, \psi_4$) therefore have R-charges of $\pm 1/2$ respectively.

The twisting corresponds to a replacement of the central $U(1)$ generator $J$ by $J \pm 2 R$,
where the choice of the sign is arbitrary. As the magnetic flux corresponds
to a Dirac monopole and has the same structure as the K\"ahler form, this twist corresponds to $(M,N) \to (M,N) \pm (1,1)$ in the
Dirac equation. The choice of the sign is arbitrary and for convenience we will take $(M,N) \to (M,N) + (1,1)$ for $(\psi_1, \psi_2)$
and $(M,N) \to (M,N) - (1,1)$ for $(\psi_3, \psi_4)$.
The different signs for $(\psi_1, \psi_2)$ and $(\psi_3, \psi_4)$ correspond to
the different R-charges of these fields.
After this twisting, the fermionic zero-mode wavefunctions of (\ref{pretwisted}) become, prior to imposing normalisability,
\be
\label{twistedspinorp1p1}
\psi = \left( \begin{array}{c} \psi_1 \\ \psi_2 \\ \psi_3 \\ \psi_4 \end{array} \right) =
\left( \begin{array}{c} (1+ w \bar{w})^{\frac{-N}{2}} (1+z \bar{z})^{1+\frac{M}{2}} f( \bar{w}, z) \\
 (1+ w \bar{w})^{1+\frac{N}{2}} (1+z \bar{z})^{\frac{-M}{2}} f( w, \bar{z}) \\
 (1+ w \bar{w})^{1-\frac{N}{2}} (1+z \bar{z})^{1-\frac{M}{2}} f( \bar{w}, \bar{z}) \\
 (1+ w \bar{w})^{\frac{N}{2}} (1+z \bar{z})^{\frac{M}{2}} f(w, z)
 \end{array} \right).
\ee
Requiring the wavefunctions to be well-defined
at the origin implies that $f$ can only contain positive powers of $z$
and $w$. 
As for the $\mbb{P}^1$ case of
section \ref{sec3}, the allowed degree of the polynomial is determined by the flux integers $M$ and $N$ and the requirement that
the wavefunction be square integrable.

The counting of normalisable modes of the Dirac equation follows
straightforwardly. The number of zero modes of each type are
\bea
\textrm{Field} \qquad & \qquad \textrm{Number of zero modes} \qquad & \qquad \textrm{Conditions on
  $M$ and $N$} \nonumber \\
\psi_1: \qquad & (|M|-1)(|N|+1), & \qquad M \le -2, N \ge 0. \nonumber \\
\psi_2: \qquad & (|M|+1)(|N|-1), & \qquad M \ge 0, N \le -2. \nonumber \\
\psi_3: \qquad & (|M|-1)(|N|-1), & \qquad M \ge 2, N \ge 2. \nonumber \\
\psi_4: \qquad & (|M|+1)|N|+1), & \qquad M \le 0, N \le 0.
\eea  
We note that when $M=N=0$, eq. (\ref{twistedspinorp1p1}) gives one constant zero mode, which represents the gaugino of dimensionally reduced
super Yang-Mills. 

\subsubsection{Counting Zero Modes}

The number and form of the zero modes on dimensional reduction is in principle contained in eq. (\ref{twistedspinorp1p1}). However we need to recall that
when we reduce fluxed super $U(P+Q)$ Yang-Mills, 4-dimensional fermions start off valued in the adjoint,
\be
\psi_{8d} = \left( \begin{array}{cc} W & X \\
Y & Z \end{array} \right) = \left( \begin{array}{cc}
(\textrm{Adj}(\bf{P}), \bf{1}) & (\bf{P}, \bar{\bf{Q}}) \\ (\bar{\bf{P}}, \bf{Q}) & (\bf{1}, \textrm{Adj}(\bf{Q})) \end{array} \right),
\ee
where we have written the representations of the various modes.
When we turn on magnetic flux along the diagonal $U(1)$ in the $W$ sector and break the gauge group from $U(P+Q)$ to $U(P) \ti U(Q)$,
chiral bifundamental fermions can appear in the $X$ and $Y$ sectors. 
Fermions in the $X$ and $Y$ sectors are sensitive to the difference in magnetic flux between the $W$ and $Z$ sectors.
Fermions in the $W$ and $Z$ sectors do not feel any net magnetic
flux and only generate the adjoint zero mode that is the SYM gaugino.

We suppose that the net amount of magnetic flux in the $W$ sector is $(M,N)$. From section \ref{sec2}, the charge felt by the bifundamental
fermions comes from the $[A, \lambda]$ term and therefore is $(+M, +N)$ for the $X$ modes and $(-M, -N)$ for the $Y$ modes.
The chiral spectrum consists of $\mc{N}_{(M,N)}$ modes in the $(\bf{P}, \bar{\bf{Q}})$ representation from the $X$ sector,
and $\mc{N}_{(-M,-N)}$ modes in the $(\bar{\bf{P}}, \bf{Q})$ representation from the $Y$ sector. The net number of chiral zero modes
is the difference $\mc{N}_{(M,N)} - \mc{N}_{(-M,-N)}$.

A case of particular interest is when $M$ and $N$ have opposite signs, as we shall see below that this is the case corresponding
to supersymmetric brane configurations. For definiteness we take $M > 0, N < 0$.
From equation (\ref{twistedspinorp1p1}), we see that for this case
$\psi_3$ and $\psi_4$ can never have zero modes, whereas $\psi_1$
and $\psi_2$ can have zero modes. We can also see that
\be
\mc{N}_{(M,N)} = (|M| + 1)(|N|-1), \quad \mc{N}_{(-M,-N)} = (|M| - 1)(|N|+1).
\ee
and so
\be
\label{nchiral}
\mc{N}_{(M,N)} - \mc{N}_{(-M,-N)} = 2 (|N| - |M|).
\ee
For the particular case that $M = -N$, there are no
net chiral zero modes, although vector-like pairs do exist.

When $M$ and $N$ have opposite signs we see that all zero modes are either $\psi_1$ or $\psi_2$ in form. As chiral superfields,
this means they are members of multiplets descending from $A_M$ and there are no fields that descend from the transverse scalar
multiplets. As discussed in section \ref{sec2}, this means that all Yukawa couplings will vanish due to the gamma-matrix structure of
$\psi^{\dagger} \gamma^M \psi$.

When $M$ and $N$ have the same sign it becomes possible to find zero modes from both the $(\psi_1, \psi_2)$ sector and the
$(\psi_3, \psi_4)$ sector. In this case $\psi^{\dagger} \gamma^M \psi$ does not necessarily vanish and the trilinear
Yukawa coupling can be non-zero. However in this case the D-terms are non-vanishing and the brane configuration is not
supersymmetric.

\subsection{Scalars on $\mbb{P}^1 \ti \mbb{P}^1$}

The scalar modes on $\mbb{P}^1 \ti \mbb{P}^1$ are determined by the eigenmodes of
\be
\label{scalarspec}
- g^{m n} D_{m} D_{n} \phi =
\Big[- \left( g^{z \bar{z}} D_{z} D_{\bar{z}} + g^{\bar{z} z} D_{\bar{z}} D_{z} \right)
- \left( g^{w \bar{w}} D_{w} D_{\bar{w}} + g^{\bar{w} w} D_{\bar{w}} D_{w} \right) \Big]\phi
\ee
The product nature of the geometry implies that scalar wavefunctions on $\mbb{P}^1 \ti \mbb{P}^1$ factorise
as the product of two scalar wavefunctions on the individual $\mbb{P}^1$s,
\be
\phi(z,\bar{z},w, \bar{w}) = \phi_M(z, \bar{z}) \phi_N(w, \bar{w}).
\ee
As for the $\mbb{P}^1$ case we can write
\bea
- \big( g^{z \bar{z}} D_{z} D_{\bar{z}} + g^{\bar{z} z} D_{\bar{z}} D_{z} \big) \phi_M
& = & - 2 g^{z \bar{z}} D_{\bar{z}} D_{z} \phi_M + { M \over 2 R_1^2}\phi_M, \nonumber \\
- \big( g^{w \bar{w}} D_{w} D_{\bar{w}} + g^{\bar{w} w} D_{\bar{w}} D_{w} \big) \phi_N
& = & - 2 g^{w \bar{w}} D_{\bar{w}} D_{w} \phi_N + { N \over 2 R_2^2}\phi_N.
\eea
The spectrum is a direct product of that for $\mbb{P}^1$: the lowest lying modes have mass
\be
m^2 = \frac{\vert M \vert}{2 R_1^2} + \frac{ \vert N \vert}{2 R_2^2},
\ee
and have degeneracy $(|M|+1)(|N|+1)$.

As for the $\mbb{P}^1$ case the form of the solutions are
\bea
M > 0: \quad \phi_M(z, \bar{z}) & = & \frac{f_M(\bar{z})}{( 1 + z \bar{z})^{\vert M \vert/2} },   \nonumber \\
M = 0: \quad \phi_M(z, \bar{z}) & = & \phi_0, \nonumber \\
M < 0: \quad \phi_M(z, \bar{z}) & = & \frac{f_M(z)}{( 1 + z \bar{z})^{\vert M \vert/2} }.
\eea
where $f_M$ is a holomorphic polynomial of degree $M$. The functions $\phi_N(w, \bar{w})$ have an identical
behaviour depending on the values of $N$.

\subsubsection*{Twisting of Scalars}

The fact that the brane is embedded in a nontrivial background also implies that the scalar equation of motion should be twisted.
From the extra dimensional point of view there are two scalar degrees of freedom. One of these corresponds to $A_\mu \otimes 1$,
which is a vectorlike degree of freedom in 4 dimensions. The other
corresponds to the transverse scalar $\phi$ that parametrises normal motion
of the brane. In the former case, the degree of freedom is internal
(as with $A_M$) and 
so the equation of motion is not twisted. In the compact space this degree
of freedom therefore satisfies the scalar equation (\ref{scalarspec}), 
with $M$ and $N$ directly given by the flux quantum numbers. This
degree of freedom partners the $\psi_4$ fermionic degree of freedom.

This does not hold for the transverse scalar. This is valued in the normal bundle, which for $\mbb{P}^1 \ti \mbb{P}^1$ is
$\mc{O}_{\mbb{P}^1 \ti \mbb{P}^1} (-2,-2)$. The effective values of
$M$ and $N$ for the transverse scalar equation 
are twisted by
two units of flux. The sign of the twist is determined by our knowledge that there are no holomorphic sections of the normal bundle.
Denoting $\Phi = \phi_8 + i \phi_9$, this implies that in the absence of flux there can be no normalisable solutions
of $D_{\bar{z}} \Phi = D_{\bar{w}} \Phi = 0$, and consequently there are no massless scalars in the absence of flux.

The covariant derivatives are
\bea
D_{z} \phi = \left(\partial_{z} + { (M-2) \bar{z} \over 2 (1 + z \bar{z} ) }\right) \phi, \qquad
D_{\bar{z}} \phi & = & \left(\partial_{\bar{z}} - { (M-2) {z} \over 2 (1 + z \bar{z} )}\right) \phi. \nonumber \\
D_{w} \phi = \left(\partial_{w} + { (N-2) \bar{w} \over 2 (1 + w \bar{w} ) }\right) \phi, \qquad
D_{\bar{w}} \phi & = & \left(\partial_{\bar{w}} - { (N-2) {w} \over 2 (1 + w \bar{w} )}\right) \phi.
\eea
When $M,N \ge 2$ holomorphic sections of the bundle exist and `zero
modes' occur. However, as for $M, N \ge 2$ the spectrum is
non-supersymmetric these modes will not be massless but will
generically be tachyonic. To compute the masses of these twisted scalars will involve the addition of a 
curvature contribution to the naive mass eigenvalue, due to the fact that these scalar are valued in a nontrivial bundle
over $\mbb{P}^1 \ti \mbb{P}^1$.
These modes partner the $\psi_3$ fermionic
mode, for which zero modes exist only when $M$ and $N$ have the same sign.

\subsection{Vectors on $\mbb{P}^1 \ti \mbb{P}^1$}

The study of vectors on $\mbb{P}^1 \ti \mbb{P}^1$ clearly parallels that on $\mbb{P}^1$, but there are some
subtle issues that arise. The vector equation follows from the general analysis,
\be
\label{p1p1vec}
\tilde{D}_i \tilde{D}^i \Phi_{j,ab} + 2i \langle G
\rangle^{ab,i}_{\ph{ab,i}j} \Phi_{i,ab} - [\nabla^i, \nabla_j]
\Phi_{i,ab} = 
-m^2 \Phi_{i,ab}.
\ee
We will assume $a>b$ for definiteness. The fluxes are
\bea
G^{z\phantom{z}ab}_{\phantom{z}z} = - G^{\bar{z}\phantom{z}ab}_{\phantom{z}\bar{z}} & = & \frac{iM}{2 R_1^2}, \nonumber \\
G^{w\phantom{w}ab}_{\phantom{w}w} = - G^{\bar{w}\phantom{w}ab}_{\phantom{w}\bar{w}} & = & \frac{iN}{2 R_2^2}.
\eea
The gauge-fixing condition that we impose is
$$
\tilde{D}_i \Phi^{i,ab} = g^{z \bar{z}} (\tilde{D}_z \Phi_{\bar{z}}^{ab} + \tilde{D}_{\bar{z}} \Phi_{z}^{ab})
+ g^{w \bar{w}} (\tilde{D}_w \Phi_{\bar{w}}^{ab} + \tilde{D}_{\bar{w}} \Phi_{w}^{ab})= 0.
$$
As for the $\mbb{P}^1$ case, we will seek solutions for which $\tilde{D}_{\bar{z}} \Phi_z^{ab} = 0$,
with $\Phi_{\bar{z}}^{ab} = 0$. Separate solutions will occur for $\Phi_i^{ab}$ oriented along the $z$ or $w$ directions.
As for the $\mbb{P}^1$ case, the latter terms of equation
(\ref{p1p1vec}) can be simplified to give
\bea
2i \langle G \rangle^{ab,z}_{\phantom{ab,z}z} \Phi_{z,ab} -[\nabla^z, \nabla_{\bar{z}}] \Phi_{z,ab} & = & (-M-1)\frac{\Phi_{z,ab}}{R_1^2}, \nonumber \\
2i \langle G \rangle^{ab,\bar{z}}_{\phantom{ab,z}\bar{z}} \Phi_{\bar{z},ab} -[\nabla^z, \nabla_{\bar{z}}] \Phi_{\bar{z},ab} & =
& (M-1)\frac{\Phi_{\bar{z},ab}}{R_1^2}. \nonumber \\
2i \langle G \rangle^{ab,w}_{\phantom{ab,w}w} \Phi_{w,ab} -[\nabla^w, \nabla_{\bar{w}}] \Phi_{w,ab} & = & (-N-1)\frac{\Phi_{w,ab}}{R_2^2}, \nonumber \\
2i \langle G \rangle^{ab,\bar{w}}_{\phantom{ab,w}\bar{w}} \Phi_{\bar{w},ab} -[\nabla^w, \nabla_{\bar{w}}] \Phi_{\bar{w},ab} & = &
(N-1)\frac{\Phi_{\bar{w},ab}}{R_2^2}.
\eea
The Laplacian operator can also be simplified in analogy with equation (\ref{eq000}), to give
\bea
\label{eq001}
\tilde{D}_i \tilde{D}^i \Phi_z^{ab} & = & \left( g^{z \bar{z}} \tilde{D}^s_z \tilde{D}^s_{\bar{z}} +
g^{w \bar{w}} \tilde{D}^s_w \tilde{D}^s_{\bar{w}} \right)
\Phi_z^{ab} + \left( \frac{M+2}{2R_1^2} + \frac{N}{2R_2^2}  \right) \Phi_{z,ab}. \nonumber \\
\tilde{D}_i \tilde{D}^i \Phi_{\bar{z}}^{ab} & = & \left( g^{z \bar{z}} \tilde{D}^s_{\bar{z}} \tilde{D}^s_{z} +
g^{w \bar{w}} \tilde{D}^s_w \tilde{D}^s_{\bar{w}} \right)
\Phi_{\bar{z}}^{ab} + \left( \frac{-M+2}{2R_1^2} + \frac{N}{2 R_2^2} \right) \Phi_{\bar{z},ab}. \nonumber \\
\tilde{D}_i \tilde{D}^i \Phi_w^{ab} & = & \left( g^{z \bar{z}} \tilde{D}^s_z \tilde{D}^s_{\bar{z}} +
g^{w \bar{w}} \tilde{D}^s_w \tilde{D}^s_{\bar{w}} \right)
\Phi_w^{ab} + \left( \frac{M}{2R_1^2} + \frac{N+2}{2R_2^2} \right) \Phi_{w,ab}. \nonumber \\
\tilde{D}_i \tilde{D}^i \Phi_{\bar{w}}^{ab} & = & \left( g^{z \bar{z}} \tilde{D}^s_z \tilde{D}^s_{\bar{z}} +
g^{w \bar{w}} \tilde{D}^s_{\bar{w}} \tilde{D}^s_{w} \right)
\Phi_{\bar{w}}^{ab} + \left( \frac{M}{2R_1^2} - \frac{N+2}{2 R_2^2}
\right) \Phi_{\bar{w},ab}.
\eea
The mass equations can then be written as, in analogue with equations
(\ref{comb1}) and (\ref{comb2}),
\bea
\label{eq002}
-m^2 \Phi_z^{ab} & = & \left( g^{z \bar{z}} \tilde{D}^s_z \tilde{D}^s_{\bar{z}} +
g^{w \bar{w}} \tilde{D}^s_w \tilde{D}^s_{\bar{w}} \right)
\Phi_z^{ab} + \left( \frac{-M}{2R_1^2} + \frac{N}{2R_2^2}  \right) \Phi_{z,ab}. \nonumber \\
-m^2 \Phi_{\bar{z}}^{ab} & = & \left( g^{z \bar{z}} \tilde{D}^s_{\bar{z}} \tilde{D}^s_{z} +
g^{w \bar{w}} \tilde{D}^s_w \tilde{D}^s_{\bar{w}} \right)
\Phi_{\bar{z}}^{ab} + \left( \frac{M}{2R_1^2} + \frac{N}{2 R_2^2} \right) \Phi_{\bar{z},ab}. \nonumber \\
-m^2 \Phi_w^{ab} & = & \left( g^{z \bar{z}} \tilde{D}^s_z \tilde{D}^s_{\bar{z}} +
g^{w \bar{w}} \tilde{D}^s_w \tilde{D}^s_{\bar{w}} \right)
\Phi_w^{ab} + \left( \frac{M}{2R_1^2} + \frac{-N}{2R_2^2} \right) \Phi_{w,ab}. \nonumber \\
-m^2 \Phi_{\bar{w}}^{ab} & = & \left( g^{z \bar{z}} \tilde{D}^s_z \tilde{D}^s_{\bar{z}} +
g^{w \bar{w}} \tilde{D}^s_{\bar{w}} \tilde{D}^s_{w} \right)
\Phi_{\bar{w}}^{ab} + \left( \frac{M}{2R_1^2} + \frac{N}{2 R_2^2}
\right) \Phi_{\bar{w},ab}.
\eea
An excitation $\Phi_z^{ab}$ is the product of a vector excitation in the $z$ plane
and a scalar excitation in the $w$ plane. For a $\Phi_z$ excitation,
the $\tilde{D}^s_w$ and $\tilde{D}^s_{\bar{w}}$ terms can be exchanged
with an $N \to -N$ shift in the above equations.

For $M, N > 0$, the lowest lying modes are obtained by solving $D^s_{z} \Phi_{\bar{z}}^{ab} = D^s_{w} \Phi_{\bar{z}}^{ab} = 0$, and
likewise $D^s_{z} \Phi_{\bar{w}}^{ab} = D^s_{w} \Phi_{\bar{w}}^{ab} = 0$. The resulting modes can be written as
\bea
\Phi_{\bar{z}}^{ab} = 
(1+z \bar{z})^{-|M|/2}(1+w \bar{w})^{-|N|/2} 
A_{\bar{z},M}(\bar{z}) \phi_N(\bar{w}), & 
\quad \textrm{ with mass } & m^2 = \left( \frac{\vert N \vert}{2R_2^2} - \frac{\vert M \vert}{2R_1^2} \right), \nonumber \\
\Phi_w^{ab} = 
(1+z \bar{z})^{-|M|/2}(1+w \bar{w})^{-|N|/2} 
\phi_M(z) A_{w,N}(w), & \quad \textrm{ with mass } &
m^2 = \left( \frac{\vert M \vert}{2R_1^2} 
- \frac{\vert N \vert}{2R_2^2} \right). \nonumber
\eea
Here both $A_{M}$ and $\phi_N$ are (anti)holomorphic functions of the
appropriate variables.

For $M>0, N<0$, the zero modes are obtained by solving $D^s_{\bar{w}} \Phi_{\bar{z}}^{ab} = D^s_{z} \Phi_{\bar{z}}^{ab} = 0$, and
likewise $D^s_{z} \Phi_w^{ab} = D^s_{\bar{w}} \Phi_{w}^{ab} = 0$. The resulting modes can be written as
\bea
\Phi_{\bar{z}}^{ab} = 
(1+z \bar{z})^{-|M|/2}(1+w \bar{w})^{-|N|/2} 
A_{\bar{z},M}(\bar{z}) \phi_N(w), & \quad \textrm{ with mass } & m^2 = \left( \frac{\vert N \vert}{2R_2^2} -
\frac{\vert M \vert}{2R_1^2} \right), \nonumber \\
\Phi_{w}^{ab} = 
(1+z \bar{z})^{-|M|/2}(1+w \bar{w})^{-|N|/2} 
\phi_M(\bar{z}) A_{w,N}(w), & \quad \textrm{ with mass } & m^2 = \left( \frac{\vert M \vert}{2R_1^2} -
\frac{\vert N \vert}{2R_2^2} \right), \nonumber
\eea
where $A$ and $\phi$ are (anti)holomorphic functions.
Similar expressions are obtained for the  $M<0, N>0$ 
and $M, N <0$ cases.

The normalisation condition is that
$$
\int \sqrt{g} g^{ij} \Phi_i^{ab} \Phi_j^{ba}
$$
is finite. In a mode such as $\Phi_{\bar{z}}^{ab} =
A_{\bar{z},M}(\bar{z}) \phi_N(\bar{w}) (M,N > 0)$, 
this implies that $A_{\bar{z},M}(\bar{z})$ satisfies vector normalisability on
the first $\mbb{P}^1$, whereas $\phi_N(\bar{w})$ 
only need satisfy scalar normalisability.  The degeneracy is then given by $(|M|-1)(|N|+1)$, with a requirement
that $M \ge 2$ for any normalisable zero modes to exist at all.

In general, the degeneracy of the $\Phi_z^{ab}$ (or $\Phi_{\bar{z}}^{ab}$) modes are given by
$(\vert M \vert - 1) (\vert N \vert +1)$, with a requirement of $\vert M \vert \ge 2$ for any normalisable modes to exist, while the
degeneracy of  $\Phi_w^{ab}$ (or $\Phi_{\bar{w}}^{ab}$) modes is given by $(\vert M \vert + 1)( \vert N \vert - 1)$, with a requirement of
$\vert N \vert \ge 2$ for any normalisable modes to exist.

The most interesting case for our purposes is the case $M>0, N<0$ as
this corresponds to a case where a supersymmetric spectrum can be realised.
The relative sizes of the two $\mbb{P}^1$s will adjust to eliminate tachyons\footnote{When present, the
tachyons correspond to a Nielsen and Olesen  instability \cite{Nielsen} of the flux.} from the spectrum, generating genuinely massless
modes in the spectrum. The relative sizes will be such that $R_2^2/R_1^2 = \vert N/M \vert$.

In this case there are $(\vert M \vert + 1)(\vert N \vert - 1)$ modes
in the $({\bf{P}}, {\bf \bar{Q}})$ 
representation and $(\vert M \vert - 1)(\vert N \vert + 1)$
in the $({\bf \bar{P}}, {\bf Q})$ representation. The net number of chiral modes is
$$
\mc{N}_{({\bf{P}}, {\bf \bar{Q}})} - \mc{N}_{({\bf \bar{P}}, {\bf
      Q})} = 
(|M|+1)(|N|-1) - (|M|-1)(|N|+1) = 2 (|N|-|M|).
$$
This coincides precisely with equation (\ref{nchiral}) and the topological index (see eq (3.51) of \cite{08023391})
\be
I = \int_S c_1(F) c_1(S).
\ee
Defining $e_i$ by $\int_{\mbb{P}^1_i} e_i = 1$, we have $c_1(S) = 2(e_1 + e_2)$ (recall $c_1(\mbb{P}^n) = (n+1) e_1$) and
$c_1(F) = |M| e_1 - |N| e_2$, which gives
\be
I = 2 (\vert M \vert - \vert N \vert).
\ee
Note that it is only when $MN < 0$ that we can perform a direct comparison of the topological index with the counting of zero modes
from $A_M$ alone - when $MN \ge 0$ we should include the scalar and $A_\mu$ sector as well in order to compare with the index.
However $MN < 0$ is the only case that is relevant for supersymmetry.

We note that the structure of the vector zero mode wavefunctions is
the same as that for the fermion zero mode wavefunctions. The apparent
difference - single powers of $(1+ z \bar{z})$ or $(1+ w \bar{w})$ -
goes away once one includes the $g^{z \bar{z}}$ necessary to compare
 $g^{z\bar{z}} \Phi_z \Phi_{\bar{z}}$ and $\psi^{\dagger} \psi$. The
fact that the wavefunctions take the same form is of course necessary
for supersymmetry.

\subsection{Supersymmetry}

For the brane embedding to be supersymmetric requires both F- and D-terms to vanish. The F-term conditions are
associated to holomorphy, and require that the brane wrap a complex cycle with a holomorphic vector bundle.
The D-terms involve non-holomorphic conditions and depend on the locus in moduli space, thereby imposing
constraints on the K\"ahler moduli. As the D-term conditions are not holomorphic in nature, they receive quantum corrections
that will also become important at small values of the moduli.

In the geometric regime (all cycle sizes much larger than the string scale),
the D-term condition is that $F \wedge J = 0$, where $F$ is the relative flux between the two wrapped branes.
For the $\mbb{P}^1 \ti \mbb{P}^1$ case, we write
$F = M e_1 + N e_2$ and $J = R_1^2 e_1 + R_2^2 e_2$. In this case the D-term condition gives
\be
\frac{M}{N} = -\frac{R_1^2}{R_2^2}.
\label{dtermeq}
\ee
Eq. (\ref{dtermeq}) can only be satisfied if $M$ and $N$ have opposite signs, and in this case the ratio $\frac{R_1}{R_2}$ is also fixed.
This case is the one that we have considered in most detail above, where we saw that
in the supersymmetric limit the vector and fermionic wavefunctions match precisely in terms of both number and
representation.

However we also considered above cases where $M$ and $N$ have the same sign. This was motivated by the need to obtain non-vanishing
Yukawas, but a second reason for this is our knowledge that the D-term equation will get quantum corrections, particularly at small volumes.
The supersymmetry requirement $\textrm{sign}(M) = - \textrm{sign}(N)$ that held deep in the geometric regime may no longer hold
in the small volume regime, and it may be possible for the case $\textrm{sign}(M) = \textrm{sign}(N)$ to be 
compatible with supersymmetry, particularly if we also allow for the possibility that the D-term is cancelled by a matter field vev rather
than a vanishing Fayet-Iliopoulos term.

\subsection{Normalisation and  Overlap Integrals}

The normalisation conditions for zero modes wavefunctions on $\mbb{P}^1 \ti \mbb{P}^1$ follow straightforwardly from those
on $\mbb{P}^1$. The generic wavefunction takes the form
\be
\psi^{K_1,K_2}_{M_1,M_2}(z,w) = \frac{1}{\mc{N}^{K_1,K_2}_{M_1,M_2}} \frac{z^{K_1}}{(1+z \bar{z})^{\frac{M_1-1}{2}}} \frac{w^{K_2}}{(1+ w \bar{w})^{\frac{M_2-1}{2}}}.
\ee
From the requirement that $\int \sqrt{g} \psi^{\dagger} \psi = 1$ it follows straightforwardly that
\bea
\left\vert \mc{N}^{K_1,K_2}_{M_1,M_2} \right\vert^2  =  \left\vert \mc{N}_{M_1}^{K_1} \right\vert^2 \left\vert \mc{N}_{M_2}^{K_2} \right\vert^2 \nonumber 
 &=&  16 \pi^2 R_1^2 R_2^2\, \frac{ \Gamma(K_1+1) \Gamma(M_1-K_1)
  \Gamma(K_2+1) \Gamma(M_2-K_2)}{\Gamma(M_1+1) \Gamma(M_2+1)}\\
&  = & 64\pi^2R_1^2 R_2^2\, 
I^{K_1}_{M_1}\,  I^{K_2}_{M_2}.
\eea
where $I^K_M$ was defined in eq. (\ref{IKM}).

Similar to $\mbb{P}^1$,  the triple overlap integrals take the
form:
\be
Y^{K_1L_1,K_2L_2}_{M_1N_1P_1,M_2N_2P_2}= \left[\frac{64\pi^2 R_1^2 R_2^2}{
\mc{N}^{K_1,K_2}_{M_1,M_2}\, \mc{N}^{L_1,L_2}_{N_1,N_2}\, \mc{N}^{K_1+L_1,
  K_2+L_2}_{P_1,P_2}}\right] \,
I^{K_1+L_1}_{\frac{M_1+N_1+P_1-1}{2}}\,\,
  I^{K_2+L_2}_{\frac{M_2+N_2+P_2-1}{2}}.
\ee

As for $SO(3)$ in the $\mbb{P}^1$ case, in the $\mbb{P}^1 \ti
\mbb{P}^1$ case the zero modes fall into representations of $SO(3) \ti
SO(3)$. The degeneracies of the zero modes and the structure of the
Yukawa couplings are determined by this global flavour symmetry.
Once the $\mbb{P}^1 \ti \mbb{P}^1$ is embedded in a compact
Calabi-Yau, the local isometries will be lifted and the flavour
symmetry will become approximate, as discussed at greater detail in \cite{BCHKMQ}.

\section{Compactification on $\mbb{P}^2$}
\label{sec5}

We finally consider branes wrapped on $\mbb{P}^2 \equiv dP^0$.
As $c(\mbb{P}^2) = (1+e)^3$, where $e$ is the fundamental class, $c_1(\mbb{P}^2) = 3e$ and
from the adjunction formula it follows that for $\mbb{P}^2$ embedded in a Calabi-Yau
the normal bundle is $\mc{O}_{\mbb{P}^2} (-3)$. This geometry
is interesting as the local geometry of the resolved $\mbb{C}^3/\mbb{Z}_3$
orbifold, which is the basis of some of the most attractive local realisations of the Standard Model \cite{hepth0005067}.
The resolving 4-cycle is a $\mbb{P}^2$ and the metric on the resolution has been computed in \cite{GibbonsFreedman, Lutken}.
The non-compact metric can be written as 
\be
g_{\mu \bar{\nu}} = \frac{(R^{6} + \lambda^6)^{1/3}}{R^2} \delta_{\mu
  \bar{\nu}}
-\frac{\lambda^6}{R^4 (R^6 + \lambda^6)^{2/3}} w_{\mu} w_{\bar{\nu}},
\ee 
where $R^2 = \sum_{i=1}^3 w_i \bar{w}_i$ and $\lambda$ is a scale
parameter. As $R \to 0$ the metric has a coordinate
singularity. Through a change of coordinates it can be seen that the
space is however regular at $R=0$, at which point there exists a minimal area
$\mbb{P}^2$ with the canonical Fubini-Study metric. The size of the
$\mbb{P}^2$ is set by the parameter $\lambda$, and our interest is in
branes wrapped on this $\mbb{P}^2$.

In the non-compact limit the resolving $\mbb{P}^2$ has the canonical Fubini-Study metric.
Denoting coordinates
by $z_1$,$\zt$, $\bzo$, $\bzt$, the Fubini-Study metric
for $\mbb{P}^n$ comes from the K\"ahler potential
\be
    K = { i \over 2 }\log \left( 1 + \sum_{i=1}^n z_i \bar{z}_i \right) \equiv \frac{i}{2} \ln \left( 1 + \rho^2 \right),
\ee
where $\rho^{2} = \sum_i z_i \bar{z}_i$. The metric is given by
\bea
g_{i \bar{j}} & = & \frac{1}{2} \left( \frac{\delta_{i \bar{j}}}{(1+\rho^2)} - \frac{z_{\bar{i}} z_j}{(1+\rho^2)^2} \right), \\
g^{i \bar{j}} & = & 2(1+\rho^2) \left( \delta^{i \bar{j}} + z^i \bar{z}^{\bar{j}} \right).
\eea
We can write these out explicitly as
\bea
\label{fsmetric}
g = & \frac{1}{2 ( 1  + \zo \bzo + \zt \bzt )^2}  & \left[
\begin{array}{cccc}
0 & 0 & 1 + \zt \bzt & - \bzo \zt \\
0 & 0 & -\zo \bzt  & 1 + \zo \bzo \\
1 + \zt \bzt & - \zo \bzt & 0 & 0 \\
- \bzo \zt & 1+ \zo \bzo & 0 & 0
\end{array}
\right], \nonumber \\
g^{-1} = & {  2 ( 1  + \zo \bzo+ \zt \bzt ) } & \left[
\begin{array}{cccc}
0 & 0 & 1 + \zo \bzo &  \zo \bzt \\
0 & 0 &  \bzo \zt & 1 + \zt \bzt \\
1 + \zo \bzo &  \bzo \zt & 0 & 0 \\
 \zo \bzt & 1+ \zt \bzt & 0 & 0
\end{array}
\right].
\eea
The K\"ahler form is given by
\be
J = i g_{i \bar{j}} dz^i \wedge d \bar{z}^{\bar{j}} =
i \frac{dz^i \wedge d \bar{z}^{\bar{i}}}{2(1+\rho^{2})} -i   \frac{ \bar{z}^{\bar{i}} dz^i \wedge z^j d \bar{z}^{\bar{j}}}{2(1+\rho^{2})^2}.
\ee
We note that we can write
\begin{eqnarray}
\label{kahlerform}
J &=& { i \over { 2 ( 1 + \rho^{2})^{2} } } \bigg[ (1 + \zt \bzt) d \zo \wedge d \bzo - \zt \bzo d \zo \wedge d \bzt
                                                   - \zo \bzt d \zt \wedge d \bzo  + (1 + \zo \bzo) d \zt \wedge d \bzt \bigg] \cr
&=& d \left({ 1 \over 4i (1 + \rho^2) } \big[    (\bzo d \zo + \bzt d \zt ) - (\zo d \bzo + \zt d \bzt )\big] \right) = d \left( \frac{\rho e_2}{2} \right),
\end{eqnarray}
where the one-form $e_2$ will be subsequently used in eq. (\ref{ees}).
This last equality is only valid away from the origin,
in keeping with the fact that the K\"ahler form is not globally exact.
Topologically $\mbb{P}^2$ has a single 2-cycle, which we can take to be parametrised by $\{ z_1 \in \mbb{C}, z_2 = 0 \}$.
Using the parametrisation of the K\"ahler form from eq. (\ref{kahlerform})
we can verify that $\int_{\mbb{P}^1} J = \pi$.

\subsection{Scalars on $\mbb{P}^2$}

We first study the lowest-lying eigenmodes of the scalar Laplacian on $\mbb{P}^2$. This is determined by the
eigenfunctions of
\be
-D_m D^m \phi = - g^{i \bar{j}} \left( D_i D_{\bar{j}} \phi + D_{\bar{j}} D_i \phi \right).
\ee
Acting on a scalar
$[D_i, D_{\bar{j}}] \phi = -i F_{i \bar{j}} \phi$,
and so
\bea
-D^a D_a \phi & = & - 2 g^{i \bar{j}} D_i D_{\bar{j}} \phi - i g^{i \bar{j}} F_{i \bar{j}} \phi, \nonumber \\
& = & - 2 g^{i \bar{j}} D_{\bar{j}} D_i  \phi + i g^{i \bar{j}} F_{i \bar{j}} \phi.
\eea
We can evaluate $g^{i \bar{j}} F_{i \bar{j}} = 4iM$ to obtain
\bea
- D_a D^a \phi & = & -2 g^{i \bar{j}} D_i D_{\bar{j}} \phi + 4M \phi, \nonumber \\
& = & -2 g^{i \bar{j}} D_{\bar{j}} D_i \phi - 4M \phi.
\eea
As $-g^{i \bar{j}} D_i D_{\bar{j}}$ is a positive semi-definite operator the lowest eigenfunctions
are obtained by solving $D_{\bar{i}} \phi = (\partial_{\bar{i}} - i A_{\bar{i}}) \phi = 0$ for $M > 0$ and $D_i \phi = 0$ for $M < 0$.
The form of the solutions are
\bea
M > 0: \quad \phi(z, \bar{z}) & = & (1+\rho^2)^{-|M|/2} f_M(z_1, z_2), \quad m^2 = 4 \vert M \vert,  \nonumber \\
M = 0: \quad \phi(z, \bar{z}) & = & \phi_0, \quad m^2 = 0, \nonumber \\
M < 0: \quad \phi(z, \bar{z}) & = & (1+\rho^2)^{-|M|/2} g_M(\bar{z}_1, \bar{z}_2), \quad m^2 = 4 \vert M \vert.
\eea
where $f_M$ and $g_M$ are holomorphic polynomials of degree $|M|$. The degeneracy
of the lowest modes is set by the number of such polynomials, and
is $(|M|+1)(|M|+2)/2$.

As for the case of $\mbb{P}^1 \ti \mbb{P}^1$ the scalar modes come in two types. One corresponds to the four-dimensional vector
$A_{\mu} \otimes 1$, which is not twisted. The masses and degeneracies of this mode can be directly computed from the 
standard scalar Laplacian. The second type of mode corresponds to transverse scalar degrees of freedom. Such transverse scalars are
valued in the normal bundle, $\mc{O}_{\mbb{P}^2}(-3)$. `Zero modes' - i.e. holomorphic sections of the bundle - can occur only in the presence of 
at least three units of flux. In this case the form of the zero mode is given by solving the equation
\be
D_{\bar{i}} \phi = (\partial_{\bar{i}} - i A_{\bar{i}}) \phi = 0,
\ee
where the flux is twisted by three units, $M \to M - 3$. As non-vanishing flux is not compatible with supersymmetry in the geometric regime,
such modes will not be massless, and the computation of their mass will require a twisting component to be introduced in the naive
eigenvalue equation.

\subsection{Fermions on $\mbb{P}^2$}

To formulate fermions on $\mbb{P}^2$ and the Dirac equation we need to establish a vierbein and compute the spin connection.
We require a set of orthonormal frame vectors $\tilde{e}_{\mu \alpha}$ satisfying
\begin{equation}
              g^{\mu \nu}\tile_{\mu \alpha} \tile_{\nu \beta} = { 1 \over 2 }\delta_{\alpha \bar{\beta}},
\end{equation}
This complex vierbein is given by
\begin{eqnarray}
     \tile_{1} &=& {1 \over 2}{ 1 \over  \rho (1 + \rho^{2}) } (\bzo d \zo + \bzt d \zt ), \nonumber \\
      \tile_{\bar{1}} &=& {1 \over 2}{1  \over \rho (1 + \rho^{2}) }(\zo d \bzo + \zt d \bzt ), \nonumber \\
      \tile_{2} &=& {1 \over 2}{ 1 \over \rho (1 + \rho^{2})^{1/2} }(\zt d \zo - \zo d \zt ), \nonumber \\
      \tile_{\bar{2}} &=& {1 \over 2}{ 1 \over \rho (1 + \rho^{2})^{1/2} }(\bzt d \bzo - \bzo d \bzt ).
\end{eqnarray}
We can write the metric as
\begin{eqnarray}
ds^{2} &=& \left(\tile_1 + \tile_{\bar{1}}\right)^2 + \left({ \tile_1 - \tile_{\bar{1}} \over i } \right)^2  +
\left(\tile_2 + \tile_{\bar{2}}\right)^2 + \left({ \tile_2 - \tile_{\bar{2}} \over i}\right)^2 \cr
&=& {1 \over (1 + \rho^2)^2} d \rho^{2} + { \rho^{2} \over (1 + \rho^2)^2} \sigma^2_{z}  + { \rho^{2} \over (1 + \rho^2)} \sigma^2_{x} + { \rho^{2} \over (1 + \rho^2)} \sigma^2_{y},
\end{eqnarray}
where we have defined
\begin{eqnarray}
d\rho &=& { 1 \over 2 \rho } \big[    (\bzo d \zo + \bzt d \zt ) + (\zo d \bzo + \zt d \bzt )\big], \cr
\sigma_{z} &=& { 1 \over 2i \rho^2 } \big[    (\bzo d \zo + \bzt d \zt ) - (\zo d \bzo + \zt d \bzt )\big], \cr
\sigma_{x} &=& { 1 \over 2 \rho^2 } \big[  (\zt d \zo - \zo d \zt ) + (\bzt d \bzo - \bzo d \bzt ) \big], \cr
\sigma_{y} &=& { 1 \over 2i \rho^2 } \big[  (\zt d \zo - \zo d \zt ) - (\bzt d \bzo - \bzo d \bzt ) \big].
\end{eqnarray}
The one forms $\sigma_{i}$ are the left invariant $SU(2)$ one forms that satisfy
$$
d \sigma_{x} = 2\sigma_{y} \wedge \sigma_{z}, \quad
d \sigma_{y} = 2\sigma_{z} \wedge \sigma_{x}, \quad
d \sigma_{z} = 2\sigma_{x} \wedge \sigma_{y}.
$$
We can then write
\be
ds^2= e_{1}^2 + e_{2}^2 + e_{3}^2 + e_{4}^2,
\ee
with the real vierbein $e_i$ given by
\be
\label{ees}
e_{1} = { 1 \over   (1 + \rho^{2}) } d \rho, \quad
e_{2} = { \rho  \over  (1 + \rho^{2}) } \sigma_{z}, \quad
e_{3} = { \rho \over  (1 + \rho^{2})^{1/2} } \sigma_{x}, \quad
e_{4} = { \rho \over (1 + \rho^{2})^{1/2} } \sigma_{y}.
\ee
If we define
$\hat{e}_{a}$ as the dual basis to $e_{a}$ given by $\hat{e}_{a} = e_{a}^{m} \del_{m}$,
we obtain
\bea
\hat{e}_{1} & = &  { (1 + \rho^{2}) \over \rho } \left[ \zo \del_{\zo} + \zt \del_{\zt} + \bzo \del_{\bzo} + \bzt \del_{\bzt} \right], \nonumber \\
\hat{e}_{2} & = &  { (1 + \rho^{2}) \over i \rho } \left[ -\zo \del_{\zo} - \zt \del_{\zt} + \bzo \del_{\bzo} + \bzt \del_{\bzt} \right]. \nonumber \\
\hat{e}_{3} & = &  { (1 + \rho^{2})^{1/2} \over \rho } \left[ \bzt \del_{\zo} - \bzo \del_{\zt} + \zt \del_{\bzo} - \zo \del_{\bzt} \right], \nonumber \\
\hat{e}_{4} & = &  { (1 + \rho^{2})^{1/2} \over i \rho } \left[ -\bzt \del_{\zo} + \bzo \del_{\zt} + \zt \del_{\bzo} - \zo \del_{\bzt} \right],
\eea

The metric compatible spin-connection is easily computed using the Cartan structure
equations
\begin{equation}
     d e^{\alpha} + \omega^{\alpha}_{\ph{\alpha} \beta} \wedge e^{\beta} = 0.
\end{equation}
Modulo antisymmetry properties the non-vanishing terms are
\begin{equation}
   \begin{array}{cccc}
\omega^{12} = - {(1 - \rho^{2}) \over \rho } e_{2}, &   \omega^{13} = - { 1 \over \rho } e_{3}, \quad & \omega^{14} = - { 1  \over \rho } e_{4}, \quad   \\
 \\
\omega^{23} = { 1  \over \rho } e_{4} & \omega^{24} =  { -1 \over \rho } e_{3}, \quad & \omega^{34} =  {(1 + 2\rho^{2}) \over \rho } e_{2} \quad.   \\
\end{array}
\end{equation}
The Dirac equation for a fermion zero mode is
\begin{equation}
      ie^{m}_{\ph{m} \alpha} \tgamma^{\alpha} \bigg( \del_{m} + {1 \over 8} [\tgamma^\alpha, \tgamma^\beta] \omega_{m \alpha \beta} - i A_{m} \bigg) \psi = 0,
\end{equation}
with the gamma matrices $\tilde{\gamma}^i$ as in eq. (\ref{conve}).
The kinetic part of the Dirac equation is
\begin{equation}
   ie^{m}_{\ph{m} \alpha} \tgamma^{\alpha} \del_m = \left( \begin{array}{cc}
0 & {\cal{D}}_+ \\
{\cal{D}}_- & 0  \\ \end{array} \right),
\end{equation}
where
\bea
{\cal{D}}_+ & = & \left( \begin{array}{cc}
{2 (1+ \rho^2) \over \rho } \big( \bzo \del_{\bzo} +  \bzt \del_{\bzt} \big)  &   {2i(1+ \rho^2)^{1/2} \over \rho }\big( \bzt \del_{\zo} - \bzo \del_{\zt} \big) \\
\\
 {2i (1+ \rho^2)^{1/2} \over \rho }\big( \zt \del_{\bzo} - \zo \del_{\bzt} \big)  & {2 (1+ \rho^2) \over \rho } \big( \zo \del_{\zo} +  \zt \del_{\zt} \big)   \\ \end{array} \right), \nonumber \\
{\cal{D}}_- & = & \left( \begin{array}{cc}
{-2 (1+ \rho^2) \over \rho } \big( \zo \del_{\zo} +  \zt \del_{\zt} \big)  & {2i(1+ \rho^2)^{1/2} \over \rho }\big( \bzt \del_{\zo} - \bzo \del_{\zt} \big) \\
\\
 {2i (1+ \rho^2)^{1/2} \over \rho }\big( \zt \del_{\bzo} - \zo \del_{\bzt} \big) &  {-2 (1+ \rho^2) \over \rho } \big( \bzo \del_{\bzo} +  \bzt \del_{\bzt} \big)  \\ \end{array} \right).
\label{derdef}
\eea
The spin connection term is
\begin{equation}
     ie_{\ph{\mu} \delta}^{\mu} \tgamma^{\delta}  {1 \over 8} \omega_{\alpha \beta m} [\tgamma^{\alpha}, \tgamma^{\beta}] =
     {1 \over 2 \rho } \left( \begin{array}{cc} 0 & -3 \rho^{2} \mathbb{I} \\
\\ -(\rho^2 + 6) \mathbb{I} & 0 \end{array} \right).
\end{equation}

As for the $\mbb{P}^1 \ti \mbb{P}^1$ case we wish to use non-trivial magnetic flux
to generate bifundamental chiral fermions.
We first focus on line bundles, namely Abelian magnetic flux backgrounds.
We shall subsequently discuss the more complicated case of non-Abelian bundles in section \ref{subsecNA} below.

As there is only a single 2-cycle, and as the K\"ahler form itself is topologially non-trivial,
the magnetic flux background satisfies $F = \lambda J$ for some $\lambda$.
To turn on the $U(1)$ gauge bundle, we choose
\be
A_m = \frac{iM}{2} \left( \frac{z^i d \bar{z}^{\bar{i}}}{(1+u)} - \frac{\bar{z}^{\bar{i}} dz^i}{(1+u)} \right) = M \rho e_2,
\ee
giving
\be
F = dA = iM \frac{dz^i \wedge d \bar{z}^{\bar{i}}}{(1+u)} - i M \frac{ \bar{z}^{\bar{i}} dz^i \wedge z^j d \bar{z}^{\bar{j}}}{(1+u)^2} = 2M J,
\ee
with $\int_{\mbb{P}^1} F = 2 \pi M$.

Using the fact that $e_\alpha \cdot e_\beta = \delta_{\alpha \beta}$, it follows that
the gauge coupling term in the Dirac equation, prior to twisting, is
\begin{equation}
    +ie^{m}_{\alpha} \tgamma^{\alpha} (-iA_{m}) = \gamma^{2} { M \rho  } =  { M \rho } \left( \begin{array}{cc}
0 & \sigma_{z} \\
\sigma_{z} & 0  \\ \end{array} \right).
\end{equation}
The Dirac equation
\begin{equation}
      ie^{m}_{\alpha} \tgamma^{\alpha} \bigg( \del_{m} + {1 \over 8} [\tgamma^\alpha, \tgamma^\beta] \omega_{m \alpha \beta} - i A_{m} \bigg) \psi = 0.
\end{equation}
can now be solved. Writing the fermion as $\left( \begin{array}{cccc} \psi_1 & \psi_2 & \psi_3 & \psi_4  \end{array} \right)^T$,
the equations factorise into separate forms for `left'-handed $\left( \begin{array}{cc} \psi_1 & \psi_2 \end{array} \right)^T$
and `right'-handed $\left( \begin{array}{cc}  \psi_3 & \psi_4 \end{array} \right)^T$ modes, where `left' and `right' refer to
chirality in the $\mbb{P}^2$. All modes correspond to left-handed spinors in four dimensions.

The Dirac equation for left handed particles is
\begin{equation}
  \big( {\cal{D}}_{-} + {\cal{B}} \big) \psi_L = 0
\end{equation}
which is
\begin{equation}
\left( \begin{array}{cc}
{-2 (1+ \rho^2) \over \rho } \big( \zo \del_{\zo} +  \zt \del_{\zt} \big)
- \frac{(1-2M)\rho^{2} + 6}{2 \rho}
& {2i(1+ \rho^2)^{1/2} \over \rho }\big( \bzt \del_{\zo} - \bzo \del_{\zt} \big) \\
\\
 {2i (1+ \rho^2)^{1/2} \over \rho }\big( \zt \del_{\bzo} - \zo \del_{\bzt} \big) &
 {-2 (1+ \rho^2) \over \rho } \big( \bzo \del_{\bzo} +  \bzt \del_{\bzt} \big) - \frac{(1+2M) \rho^2 + 6}{2 \rho}
 \\ \end{array} \right) \left( \begin{array}{c} \psi_1 \\ \psi_2 \end{array} \right) = 0. \nonumber
\end{equation}
The Dirac equation for right handed particles is
\be
  \big( {\cal{D}}_{+} + {\cal{A}} \big) \psi_R = 0 \nonumber
\ee
which is
\bea
\left( \begin{array}{cc}
{2 (1+ \rho^2) \over \rho } \big( \bzo \del_{\bzo} +  \bzt \del_{\bzt} \big)  -
 \frac{(3 - 2M)\rho}{2} &  {2i(1+ \rho^2)^{1/2} \over \rho }\big( \bzt \del_{\zo} - \bzo \del_{\zt} \big) \\
 {2i (1+ \rho^2)^{1/2} \over \rho }\big( \zt \del_{\bzo} - \zo \del_{\bzt} \big)  &
 {2 (1+ \rho^2) \over \rho } \big( \zo \del_{\zo} +  \zt \del_{\zt} \big) - \frac{(3 + 2M)\rho}{2}  \\ \end{array} \right)
 \left( \begin{array}{c} \psi_3 \\ \psi_4 \end{array} \right) = 0. \nonumber
\eea

Zero mode solutions are present only for the right handed particles \cite{Pope}.\footnote{The reason for this can be understood
from the form of the left-handed eqautions. The $\frac{-6}{2 \rho}$
term present at small $\rho$ causes 
wavefunctions to have the singular behaviour
$\psi \sim \rho^{-3}$ near $\rho \sim 0$.}
Normalised solutions take the form $\psi_R = (\psi_{3},\psi_{4})$ with
\be
\label{p2wavefunctions}
\left( \begin{array}{c} \psi_{3} \\ \psi_{4} \end{array} \right) =
\left( \begin{array}{c} f(\zo, \zt) ( 1 + \rho^{2} )^{ \frac{3}{4} - \frac{M}{2}} \\
g(\bzo, \bzt)( 1 + \rho^{2} )^{\frac{3}{4} + \frac{M}{2}} \end{array} \right),
\ee
where $f$ and $g$ are holomorphic polynomials of positive degree.
Note that these are local solutions of the Dirac equation, valid only within this patch.
Requiring that the wavefunctions are normalisable and square integrable gives
\begin{itemize}

\item For $\vert M \vert < 3/2$, there are no normalisable zero modes.

\item For $M \ge 3/2$, we requires $\psi_{2} = 0$, and $f(\zo,\zt)$ to be a polynomial in powers
of $\zo$ and $\zt$ of degrees less than or equal to $|M|/2 - 3/4$.

\item or $M \le -3/2$, we requires $\psi_{1} = 0$, and $g(\bzo,\bzt)$ to be a polynomial in powers
of $\bzo$ and $\bzt$ of degrees less than or equal to $|M|/2 -3/4$.

\end{itemize}

\subsubsection*{Twisting}

The above solutions are written in terms of half-integral fluxes. This is because
it is necessary that the fermionic wavefunctions be globally well-defined. It is well known that
$\mbb{P}^2$ is not a spin manifold, as the second Stiefel-Whitney class $H^2(\mbb{P}^2, \mbb{Z}_2)$ is non-zero.
$\mbb{P}^2$ does not admit a globally defined $spin$ structure.
For any integral
$M \in \mbb{Z}$, the wavefunctions of eq. (\ref{p2wavefunctions}) cannot be globally defined:
for appropriate choices of patches $A$, $B$ and $C$ the patch transition functions
satisfy
\be
\label{fw}
\mc{O}_{AB} \mc{O}_{BC} \mc{O}_{CA} = -1.
\ee
However this problem is resolved if $M \in \mbb{Z} + 1/2$ rather than
$M \in \mbb{Z}$. In this case
the transition functions necessarily incorporate both gauge and spin components, and
there is an additional $-1$ in eq. (\ref{fw}) from the half-integral gauge field.
The fermionic wavefunctions are then globally well-defined, and the
fermions transform as sections of a $spin^c$ bundle rather than a $spin$ bundle.
In the context of $\mbb{P}^2$, the necessity of the half-integral
gauge background that allows the fermions to be globally defined
was first realised by Hawking and Pope \cite{HawkingPope}.
In modern language it corresponds to the cancellation of the Freed-Witten
anomaly \cite{hepth9907189}.

For the case of D7-branes wrapping a $\mbb{P}^2$ embedded in a Calabi-Yau,
this half-integral shift in the gauge background
is automatically generated from the twisting necessary to account for the nontrivial normal bundle.
The discussion is very similar to the $\mbb{P}^1 \ti \mbb{P}^1$ case discussed above.
The action of the central $U(1)$ on
tangent vectors is
\be
\left( \begin{array}{c} z \\ w \end{array} \right) \to \left( \begin{array}{cc} e^{i \theta} & 0 \\ 0 & e^{i \theta} \end{array}
\right) \left( \begin{array}{c} z \\ w \end{array} \right),
\ee
which corresponds to an action on spinors of
\bea
\tilde{\Lambda} & =  & -\frac{\theta}{4} \left( [\gamma_2, \gamma_1] + [\gamma_4, \gamma_3] \right) \\
& = & \left( \begin{array}{cccc} 0 & 0 & 0 & 0 \\ 0 & 0 & 0 & 0 \\
0 & 0 &  i \theta & 0 \\
0 & 0 &  0 & -i \theta \end{array} \right)
\eea
As before the twisting corresponds to a replacement of the central $U(1)$ generator $J$ by $J \pm 2 R$,
where the choice of the sign is arbitrary. The difference for $\mbb{P}^2$ is that the
twist corresponds to $M \to M \pm 3/2$ in the
Dirac equation, as the normal bundle is now $\mc{O}_{\mbb{P}^2}(-3)$ rather than $\mc{O}_{\mbb{P}^1 \ti \mbb{P}^1}(-2,-2)$.
For convenience we take $M \to M + 3/2$ for $(\psi_1, \psi_2)$
and $M \to M - 3/2$ for $(\psi_3, \psi_4)$.
The different signs for $(\psi_1, \psi_2)$ and $(\psi_3, \psi_4)$ correspond to
the different R-charges of these spinors.

This twisting procedure implies that the Dirac equation should automatically be solved with a half-integral flux background.
The solutions of the twisted Dirac equation now take the form
\be
\label{twistedp2wavefunctions}
\left( \begin{array}{c} \psi_{3} \\ \psi_{4} \end{array} \right) =
\left( \begin{array}{c} f(\zo, \zt) ( 1 + \rho^{2} )^{ \frac{3}{2} - \frac{M}{2}} \\
g(\bzo, \bzt)( 1 + \rho^{2} )^{\frac{M}{2}} \end{array} \right).
\ee
For the zero flux case $M=0$, there is a constant zero mode, which
corresponds to the gaugino of 4-dimensional super Yang-Mills.
In general (\ref{twistedp2wavefunctions}) has $\frac{(M-1)(M-2)}{2}$ zero modes.

\subsubsection*{Zero Mode Counting}

As for the $\mbb{P}^1 \ti \mbb{P}^1$ case the number and
form of the zero modes on dimensional reduction is in principle contained in eq. (\ref{twistedp2wavefunctions}).
However due to the dimensional reduction structure,
\be
\psi_{8d} = \left( \begin{array}{cc} W & X \\
Y & Z \end{array} \right) = \left( \begin{array}{cc}
(\textrm{Adj}(\bf{P}), \bf{1}) & (\bf{P}, \bar{\bf{Q}}) \\ (\bar{\bf{P}}, \bf{Q}) & (\bf{1}, \textrm{Adj}(\bf{Q})) \end{array} \right),
\ee
fermions in the $X$ and $Y$ sectors are both sensitive to the difference in magnetic flux between the $W$ and $Z$ sectors.
We suppose that the net amount of magnetic flux in the $W$ sector is $M$. From section \ref{sec2} the charge felt by the bifundamental
fermions comes from the $[A, \lambda]$ term and therefore is $+M$ for the $X$ modes and $-M$ for the $Y$ modes.
The chiral spectrum consists of $\mc{N}_{M}$ modes in the $(\bf{P}, \bar{\bf{Q}})$ representation from the $X$ sector,
and $\mc{N}_{-M}$ modes in the $(\bar{\bf{P}}, \bf{Q})$ representation from the $Y$ sector. The net number of chiral zero modes
is the difference $\mc{N}_{M} - \mc{N}_{-M}$.

All zero modes come from the $\psi_3$ and $\psi_4$ sectors. For Abelian bundles, we see that
\be
\mc{N}_{M} = \frac{(M-1)(M-2)}{2}, \quad \mc{N}_{-M} = \frac{(M+1)(M+2)}{2}.
\ee
and so
\be
\label{p2ferct}
\mc{N}_{M} - \mc{N}_{-M} = -3M.
\ee
There exist chiral zero modes whenever the flux is non-vanishing.
As $c(\mbb{P}^2) = (1 + e)^3$ with $e$ the fundamental class, and $c_1(F) = Me$, eq. (\ref{p2ferct}) coincides
with the index $\int c_1(\mbb{P}^2) c_1(F)$.
In all cases the spectrum is non-supersymmetric as $J \wedge F \neq 0$, which follows immediately from the fact that
$F = \lambda J$. Furthermore, the fact that all zero modes lie in the $(\psi_3, \psi_4)$ sector means that no no-vanishing
Yukawa couplings can be generated using only Abelian fluxes.

\subsection*{Vectors on $\mbb{P}^2$}

The fact that with abelian fluxes all zero modes lie in the $(\psi_3, \psi_4)$ sectors and no zero modes can be found in the
$(\psi_1, \psi_2)$ sectors means that there are no vector zero modes for abelian fluxes on $\mbb{P}^2$. Vector zero modes are partners
to the $(\psi_1, \psi_2)$ modes, and the absence of fermionic zero modes means that that there are no bosonic modes that can be considered
as `zero modes'. Of course there are still vector Kaluza-Klein modes which are eigenfunctions of eq. (\ref{veceqn}), 
but these are intrinsically massive.

The structure of the Yukawa interactions means that the absence of any vector zero modes implies that all Yukawa couplings vanish,
even for nonsupersymmetric brane configurations. This motivates the inclusion of non-Abelian bundles, which will allow 
$(\psi_1, \psi_2)$ zero modes to exist and thus generate non-vanishing Yukawas.

\subsection{Normalisation and Overlap Integrals}

The generic form of an Abelian zero-mode wavefunction on $\mbb{P}^2$ is given by (we again take $M>0$)
\begin{equation}
    \psi_{M}^{KL} = { 1 \over \mc{N}^{KL}_{M} } \frac{z_{1}^{K} z_{2}^{L}}{( 1 + z_1 \bar{z}_1 + z_2 \bar{z}_2)^{\frac{M-1}{2}}} \, .
\end{equation}
It then follows from the metric (\ref{fsmetric}) that
\begin{eqnarray}
(\mc{N}^{KL}_{M})^{2} &=& \int d^{2} \zo d^{2} \zt \sqrt{g} (\psi_{M}^{KL})^{\dagger} \psi_{M}^{KL} \cr
              &=& 4 \pi^{2} R^4 \int dr_{1} dr_{2}{ r^{2K+1}_{1} r_{2}^{2L+1}  \over ( 1 + r^{2}_{1} +  r_{2}^{2} )^{M + 2} } \cr
              &\equiv& 4 \pi^{2} R^4 I^{KL}_{M},
\end{eqnarray}
where as for $\mbb{P}^1$ we have defined the standard integral $I^{KL}_{M}$. This is also the integral that will arise when computing
Yukawa couplings and the triple overlap of wavefunctions. Now
\begin{eqnarray}
I^{KL}_{M} &=& \int dr_{1} r^{2K+1}_{1} \int dr_{2} { r_{2}^{2L+1}  \over ( 1 + r^{2}_{1} +  r_{2}^{2} )^{M + 2} } \cr
           &=&  \int  dr_{1} { r_{1}^{2K+1} \over ( 1 + r_{1}^{2} )^{M-L+1} }
                 \int_{r_1 = {\textrm const }}d \alpha { \alpha^{2L+1} \over ( 1 + \alpha^{2} )^{M+2} },
\end{eqnarray}
where we have defined $\alpha = { r_{2} \over \sqrt{ 1  + r_{1}^{2} } }$. Using the
$\mbb{P}^1$ result (\ref{IKM}) for $I^K_M$, we obtain
\begin{equation}
 I^{KL}_{M} = I^{K}_{M-L} I^{L}_{M+1} = { \Gamma(K+1) \Gamma(L+1) \Gamma(M-L-K) \over 4 \Gamma(M+2) }.
\end{equation}
It therefore follows that the normalisation constant $\mc{N}^{KL}_M$ is given by
\begin{equation}
  \left\vert \mc{N}^{KL}_{M} \right\vert^2 =  4 \pi^{2} R^4 I^{KL}_{M} = { \pi^{2} \Gamma(K+1) \Gamma(L+1) \Gamma(M-L-K) \over  \Gamma(M+2) }.
\end{equation}

In this case the triple overlap integrals take the form:
\be
Y^{K_1 L_1, K_2 L_2}_{MNP}= \left[\frac{4\pi R^4}{\mc{N}^{K_1L_1}_M\,
  \mc{N}^{K_2L_2}_N \, \mc{N}^{K_1+K_2, L_1+L_2}_P}\right]\,  I^{K_1+K_2,
  L_1+L_2}_{\frac{M+N+P-1}{2}}
\ee

As for the previously discussed cases of $\mbb{P}^1$ and $\mbb{P}^1
\ti \mbb{P}^1$, the $SU(3)/\mbb{Z}_3$ isometry of the canonical 
$\mbb{P}^2$ metric acts as a flavour symmetry on wavefunction zero
modes, and the possible degeneracies of zero modes are set by the
possible sizes of $SU(3)$ representations.
In particular, in the limit that the bulk is infinitely large
the Yukawa couplings will be ordered by an exact $SU(3)$ family
symmetry. In the limit that the bulk is large but finite, an
approximate $SU(3)$ family symmetry will exist.

\subsection{Non-Abelian Bundles}
\label{subsecNA}

For the case of Abelian bundles, all solutions are right-handed and therefore all Yukawa couplings vanish.
In addition to the intrinsic interest of doing so, this also gives a motivatation
for turning on a non-Abelian bundle on the brane, as this will allow
left-handed zero modes to exist.

 We follow \cite{Charap,GibbonsPope} to obtain an $SU(2)$ bundle on ${\mathbb{P}}^{2}$, deriving
the gauge bundle from the tangent bundle.  Our choice for the $SU(2)$ generators $T^{i}$ i=2,3,4 is\footnote{This rather
unconventional choice is related to our choice of the gamma matrices in (\ref{conve}). }
\begin{equation}
   T^{2} = { \sigma^{z} \over 2} \qquad T^{3} = { \sigma^{x} \over 2 } \qquad T^{4} = { \sigma^{y} \over 2 }
\end{equation}
These satisfy the $SU(2)$ algebra $[ T^{i}, T^{j}] = \epsilon^{ijk} T^{k}$ with $\epsilon^{ijk}$ fully anti-symmetric, $\epsilon^{234}=1$.
The gauge potential $A_{\mu} = A_{\mu}^{i} T^{i}$ is given by
$A_{\mu}^{i} = \omega_{\mu}^{1i} - { 1 \over 2 } \epsilon^{ijk} \omega_{\mu}^{jk}$.
Explicitly this gives
\begin{equation}
  A_{}^{2} = { -2  - \rho^{2} \over \rho}e_{2}, \qquad    A_{}^{3} = { -2 \over \rho} e_{3}, \qquad   A_{}^{4} = { -2 \over \rho}e_{4} .
\end{equation}
The corresponding field strength $F = dA + A \wedge A$ is%
\begin{eqnarray}
 F^{2} &=& 2( e_{1} \wedge e_{2} - e_{3} \wedge e_{4}), \cr
 F^{3} &=& 2(e_1 \wedge e_3 - e_4 \wedge e_2), \cr
 F^{4} &=& 2( e_1 \wedge e_4 - e_2 \wedge e_3).
\end{eqnarray}
This is manifestly anti-selfdual and therefore a solution to the Yang-Mills equations of motion.

We consider backgrounds in which an $SU(2)$ subsector of the $U(Q+2)$ theory obtains a
vev of the above form. The gauge and adjoint fermion fields can be written as
$$
F = \left( \begin{array}{cc}
F_{SU(2)} & 0 \\
0 & 0  \\ \end{array} \right), \qquad
  \psi = \left( \begin{array}{cc}
W & X \\
Y & Z  \\ \end{array} \right).
$$
Here W, X, Y and Z are blocks of size $2 \times 2$, $2 \times Q$, $Q \times 2$ and $Q \times Q$ respectively, and the
instanton is valued in the W block. The instanton background breaks the gauge group down to $U(1) \ti U(Q)$.
The X and Y blocks experience a non-trivial effect from the $SU(2)$ instanton.

The $\psi$ equations of motion are
\begin{equation}
  ie^{m}_{\ph{m}\alpha} \tgamma^{\alpha} \bigg( \nabla_{m} \psi -i[A_{m}, \psi] \bigg) =0
\label{naeom}
\end{equation}
where $\nabla_{m}$ is the covariant derivative, $\nabla_{m} = \del_{m}  + {1 \over 8}[\tilde{\gamma}^{\alpha}, \tilde{\gamma}^{\beta}]w_{m \alpha \beta}$.

For column vectors in block
X (we shall denote such doublets by $(\theta_{1}, \theta_{2})$) this reduces to
\begin{equation}
i\gamma^{M} \nabla_{M} \left( \begin{array}{c}
\theta_{1} \\
\theta_{2} \\ \end{array} \right)
+
\left( \begin{array}{cc}
- {( 2 + \rho^{2}) \over 2 \rho } \gamma^{2} + { M } \rho \gamma^{2} & -{ 1 \over \rho} ( \gamma_{3} - i \gamma_{4} ) \\
 -{1 \over \rho} ( \gamma_{3} + i \gamma_{4}) &   {( 2 + \rho^{2}) \over 2 \rho } \gamma^{2} + { M  } \rho \gamma^{2} \\ \end{array} \right)
\left( \begin{array}{c}
\theta_{1} \\
\theta_{2} \\ \end{array} \right) = 0,
\end{equation}
where we have also included $M$ units of $U(1)$ flux in addition to the $SU(2)$ bundle.\footnote{As in our earlier discussion,
one requires $M \in {\mathbb{Z}} + {1 \over 2} $ for a consistent ${\it{spin}}^{c}$ structure.} The equations of motion for the
the right and left handed modes decouple. For the right handed modes one has
\begin{equation}
  {\cal{D}}_+ \theta_{1}^{R}  +  { 1 \over 2 \rho } \left( \begin{array}{cc}
-2 + (2M-4) \rho^{2} & 0 \\
0  & +2 - (2M+2) \rho^{2}  \\ \end{array} \right) \theta_{1}^{R} - { 1 \over 2 \rho } \left( \begin{array}{cc}
0 & 0 \\
4  & 0  \\ \end{array} \right) \theta_{2}^{R} = 0
\end{equation}
\begin{equation}
  {\cal{D}}_+ \theta_{2}^{R}  +  { 1 \over 2 \rho } \left( \begin{array}{cc}
2 + (2M-2)\rho^{2} & 0 \\
0  & -2 - (2M+4) \rho^{2}  \\ \end{array} \right) \theta_{2}^{R} - { 1 \over 2 \rho } \left( \begin{array}{cc}
0 & 4 \\
0  & 0  \\ \end{array} \right) \theta_{1}^{R} = 0
\end{equation}
and for the left handed modes
\begin{equation}
  {\cal{D}}_- \theta_{1}^{L}  +  { 1 \over 2 \rho } \left( \begin{array}{cc}
-8 + (2M-2)\rho^{2} & 0 \\
0  &  -4 - 2M \rho^{2}  \\ \end{array} \right) \theta_{1}^{L} - { 1 \over 2 \rho } \left( \begin{array}{cc}
0 & 0 \\
4  & 0  \\ \end{array} \right) \theta_{2}^{L} = 0
\end{equation}
\begin{equation}
  {\cal{D}}_- \theta_{2}^{L}  +  { 1 \over 2 \rho } \left( \begin{array}{cc}
-4 + 2M\rho^{2} & 0 \\
0  & -8 - (2M+2) \rho^{2}  \\ \end{array} \right) \theta_{2}^{L} - { 1 \over 2 \rho } \left( \begin{array}{cc}
0 & 4 \\
0  & 0  \\ \end{array} \right) \theta_{1}^{L} = 0
\end{equation}
with ${\cal{D}}_{+}$ and ${\cal{D}}_{-}$ as defined in (\ref{derdef}).

  For the doublets obtained from the rows of the block Y, which we denote by $(\lambda_{1}, \lambda_{2})$
(\ref{naeom}) reduces to
\begin{equation}
i\gamma^{M} \nabla_{M} \left( \begin{array}{c}
\lambda_{1} \\
\lambda_{2} \\ \end{array} \right)
-
\left( \begin{array}{cc}
- {( 2 + \rho^{2}) \over 2 \rho } \gamma^{2} + { M } \rho \gamma^{2} & -{ 1 \over \rho} ( \gamma_{3} + i \gamma_{4} ) \\
 -{1 \over \rho} ( \gamma_{3} - i \gamma_{4}) &   {( 2 + \rho^{2}) \over 2 \rho } \gamma^{2} + { M  } \rho \gamma^{2} \\ \end{array} \right)
\left( \begin{array}{c}
\lambda_{1} \\
\lambda_{2} \\ \end{array} \right) = 0
\end{equation}
which yields
\begin{equation}
  {\cal{D}}_+ \lambda_{1}^{R}  +  { 1 \over 2 \rho } \left( \begin{array}{cc}
2 - (2M+2)\rho^{2} & 0 \\
0  & -2 + (2M-4) \rho^{2}  \\ \end{array} \right) \lambda_{1}^{R} + { 1 \over 2 \rho } \left( \begin{array}{cc}
0 & 4 \\
0  & 0  \\ \end{array} \right) \lambda_{2}^{R} = 0
\end{equation}

\begin{equation}
  {\cal{D}}_+ \lambda_{2}^{R}  +  { 1 \over 2 \rho } \left( \begin{array}{cc}
-2 - (2M+4)\rho^{2} & 0 \\
0  & 2 +  (2M-2)\rho^{2}  \\ \end{array} \right) \lambda_{2}^{R} + { 1 \over 2 \rho } \left( \begin{array}{cc}
0 & 0 \\
4  & 0  \\ \end{array} \right) \lambda_{1}^{R} = 0
\end{equation}
for the right handed modes and
\begin{equation}
  {\cal{D}}_- \lambda_{1}^{L}  +  { 1 \over 2 \rho } \left( \begin{array}{cc}
-4 - 2M\rho^{2}& 0 \\
0  &  -8 + (2M-2) \rho^{2}  \\ \end{array} \right) \lambda_{1}^{L} + { 1 \over 2 \rho } \left( \begin{array}{cc}
0 & 4 \\
0  & 0  \\ \end{array} \right) \lambda_{2}^{L} = 0
\end{equation}
\begin{equation}
  {\cal{D}}_- \lambda_{2}^{L}  +  { 1 \over 2 \rho } \left( \begin{array}{cc}
-8 - (2M+2)\rho^{2} & 0 \\
0  & -4 + 2M\rho^{2}  \\ \end{array} \right) \lambda_{2}^{L} + { 1 \over 2 \rho } \left( \begin{array}{cc}
0 & 0 \\
4  & 0  \\ \end{array} \right) \lambda_{1}^{L} = 0
\end{equation}
for the left handed modes.

We now discuss solutions of these equations. We focus on solutions from the `X' block as the `Y' block case is similar.
We shall also not try to be exhaustive but instead shall focus our search on
`s-wave' solutions that can be written solely as a function of $\rho$. This simplifies the search considerably, as
\bea
\big( \bzo \del_{\bzo} +  \bzt \del_{\bzt} \big) f(\rho) & = & \frac{\rho}{2} f'(\rho), \nonumber \\
\big( \bzt \del_{\zo} - \bzo \del_{\zt} \big) f(\rho) & = & 0.
\eea
We first look to solve the right-handed equations. We recall these equations had many solutions when only U(1) flux was turned on and
we expect this qualitative feature to persist. We note that for the upper component of
$\theta_1^R$ and the lower component of $\theta_2^R$ the equations decouple and we are just let with a single first order equation to solve.
The equations for $\theta_1^{R,u}$ and $\theta_2^{R,d}$ are respectively
\bea
\partial_{\rho} \theta_1^{R,u} & = & \frac{\left( 2 + (4-2M) \rho^2 \right) \theta_1^{R,u}}{2 \rho (1 + \rho^2)}, \nonumber \\
\partial_{\rho} \theta_2^{R,d} & = & \frac{\left( 2 + (4+2M) \rho^2 \right) \theta_2^{R,d}}{2 \rho (1 + \rho^2)}.
\eea
These have solutions
$$
\theta_1^{R,u} = f(z_1, z_2) \rho (1 + \rho^2)^{\frac{1-M}{2}}, \qquad \theta_2^{R,d} = g(\bar{z}_1, \bar{z}_2) \rho (1 + \rho^2)^{\frac{1+M}{2}}.
$$
Normalisability as $\theta \to \infty$ requires that
$\left\vert \frac{\theta}{\rho} \right\vert \to 0 $ as $\rho \to \infty$. This implies that for the $\theta_1^{R,u}$ solutions, $M \ge 3/2$, while for
the $\theta_2^{R,d}$ solutions we need $M \le -3/2$.

In the case of Abelian flux, no left-handed solutions existed at all. One motivation for studying
non-abelian flux backgrounds is that left-handed solutions to the Dirac
equation can now be found. Considering the left-handed equations, as with the right-handed modes the $\theta_1^{L,u}$ and
$\theta_2^{R,d}$ modes are decoupled. In this case we can check that no holomorphic normalisable modes can exist in the vicinity of $\rho = 0$.
However, if we consider the coupled system of equations for $\theta_1^{L,d}$ and $\theta_2^{L,u}$ then it turns out that
finite and normalisable solutions can be found. To see this, note that in the vicinity of $\rho = 0$, these coupled equations become
\be
\partial_\rho \left( \begin{array}{c} \theta_1^{L,d} \\ \theta_2^{L,u} \end{array} \right) = \frac{1}{\rho} \left( \begin{array}{cc} -2 & -2 \\ -2 & -2 \end{array}
\right) \left( \begin{array}{c} \theta_1^{L,d} \\ \theta_2^{L,u} \end{array} \right).
\ee
These equations have a constant mode $ \left( \begin{array}{c} \theta_1^{L,d} \\ \theta_2^{L,u} \end{array} \right) \sim \left( \begin{array}{c}
\lambda \\ -\lambda \end{array} \right)$ near the origin. At large $\rho \to \infty$, the equations take the form
\be
\partial_\rho \left( \begin{array}{c} \theta_1^{L,d} \\ \theta_2^{L,u} \end{array} \right) = \frac{1}{\rho} \left( \begin{array}{cc} -M & -\frac{2}{\rho^2}
\\ -\frac{2}{\rho^2} & M \end{array}
\right) \left( \begin{array}{c} \theta_1^{L,d} \\ \theta_2^{L,u} \end{array} \right).
\ee
These equations admit normalisable solutions as $\rho \to \infty$ for $M = \pm 1/2$, when the $\rho \to \infty$ behaviour is
$\theta_1^{L,d} \sim \rho^M, \theta_2^{L,u} \sim \rho^{-M}$.

We can check numerically that these asymptotic behaviours patch together into a single normalisable solution extending from $\rho =0$
to $\rho = \infty$. This is illustrated in figure \ref{zeromodeplot}. We were not able to obtain an analytic expression for this zero mode.
\FIGURE{\epsfig{file=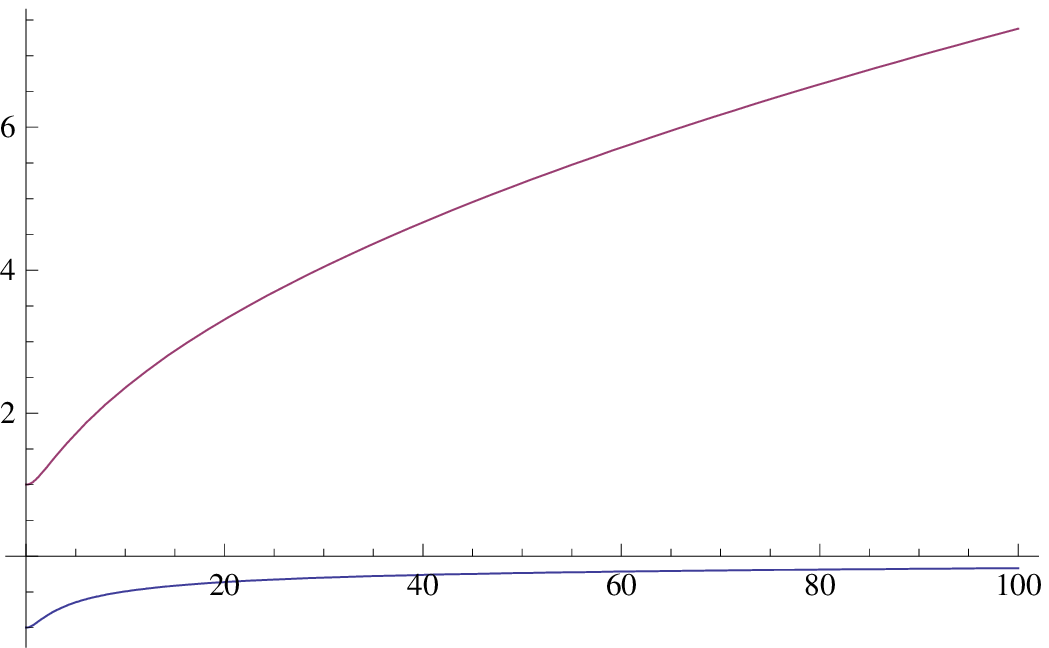,width=0.5\textwidth}
        \caption[p]{\footnotesize{A plot of the numerical
        behaviour of $\theta_1^{L,d}$ and $\theta_2^{L,u}$ for the
        coupled left-handed zero mode, shown for $M=-1/2$. The
        growing (but normalisable) mode is $\theta_1^{L,d}$ and the decaying mode $\theta_2^{L,u}$. }}
    \label{zeromodeplot}}

\section{Conclusions}
\label{sec6}

Local brane realisations of the Standard Model around a (resolved)
singularity have various 
phenomenologically attractive properties. They also drastically reduce the 
geometric complexity associated with  global models. Furthermore, 
such local Standard Model constructions 
are forced on us by certain moduli stabilisation scenarios, such as the LARGE volume models of \cite{hepth0502058, hepth0505076}.

In this paper we have studied certain aspects of such local D7 brane models in great detail. The paper has been devoted
to studying the precise form of the wavefunctions that arise, through solving the appropriate differential equation for the zero modes.
Compared to algebro-geometric approaches, the advantage of having
explicit wavefunctions is that these are not restricted to holomorphic
information and also contains information about the K\"ahler metric - the overlap of the wavefunctions
directly gives the physical Yukawa couplings rather than simply the holomorphic component.
The methods of this paper can be seen as an extension of the approach of \cite{hepth0404229} to local D7 brane models.
Our particular emphasis has been the cases of branes wrapped on $\mbb{P}^1 \ti \mbb{P}^1$ and $\mbb{P}^2$, 
as these spaces represent the simplest examples where we can solve the Dirac and associated equations.
We have dimensionally reduced super Yang-Mills on these surfaces,
and solved the (twisted) equations 
to obtain the normalised zero modes for bifudamental fields that
transform as scalars, spinors and vectors on the internal space.
We have worked with both supersymmetric and non-supersymmetric brane
configurations, and with Abelian and non-Abelian magnetic flux
backgrounds. In $\mbb{P}^2$ even though the Yukawa couplings vanish
for both supersymmetric and non-supersymmetric Abelian flux,
non-Abelian fluxes can lead to non-vanishing Yukawa couplings.

The cycle geometries have isometries which manifest themselves as
flavour symmetries of the low-energy theory, acting on the zero modes.
Where comparison is possible the zero mode degeneracies agree with those computed using index formulae.
The Yukawa couplings vanish if we require a vanishing FI term, $J \wedge F = 0$. If we allow $J \wedge F \neq 0$, and assume either that the 
D-terms are cancelled by quantum corrections or field vevs or that the brane configuration is non-supersymmetric, then the 
Yukawa couplings can be non-vanishing.

Let us close by outlining some directions for future work.
\begin{enumerate}
\item
We would like to extend this work to realistic D-brane  models, where the
gauge group resembles that of the MSSM. This will also require the
construction of globally consistent brane configurations.
This paper has focussed on obtaining the normalised
wavefunctions and we have not imposed tadpole cancellation conditions.
This is less important here where our focus is simply on the form of
the wavefunctions, 
but for a complete and realistic model it will
be necessary to ensure that the full brane configuration is
consistent. 

\item
It would be instructive to understand the connection between the geometric Yukawa couplings and the Yukawa couplings in the singular
limit, along the lines of \cite{08022809}. In the singular limit anti-branes are supersymmetric objects and the Yukawa couplings can be non-vanishing for
supersymmetric brane configurations. In the geometric limit the Yukawa couplings automatically vanish for 
supersymmetric brane configurations. It would be nice to understand
better the interpolation between these two regimes.

\item 
It would be interesting to generalise the computations in this paper to more complex surfaces, such as del Pezzos. In some cases
analytic metrics exist (e.g. see \cite{hepth0605129}) and it may be possible to explicitly solve the Dirac equation and compute the 
zero mode wavefunctions.

\item
The techniques developed here may be extended to the case of Euclidean
D3-branes wrapping the corresponding 4-cycle. This could be
interesting to study instanton induced effective couplings in the
four-dimensional effective action (see \cite{ralph} for a recent
review on these techniques).

\item
The 4-surface metrics, for example the use of the Fubini-Study metric for $\mbb{P}^2$, are 
those appropriate for the case that the surface is embedded in a non-compact Calabi-Yau. In this case the
local metric contains isometries that act as flavour symmetries of the Dirac equation. It would be interesting to 
obtain the form of the wavefunctions for surfaces embedded in Calabi-Yaus that are compact but of very large volume.
In this case the largeness of the bulk provides a small breaking parameter for the local flavour symmetry \cite{BCHKMQ}.
By obtaining the wavefunctions on such perturbed spaces, possibly
using numerical methods such as \cite{hepth0506129, hepth0612075,
  hepth0703057, 07123563},
it will be possible to study explicitly the small breaking of flavour symmetry.

\end{enumerate}

\acknowledgments{} 
For advice, encouragement and information we would like to thank Chris Beasley, Michele Cicoli, Diego
Correa, Brian Dolan, Matt Dolan, Nick Dorey, Gary Gibbons, Sven
Krippendorf, Harvey Reall, James Sparks, David Tong and Toby Wiseman.
JC is funded by Trinity College, Cambridge and would like to thank the
University of Texas at Austin for hospitality while part of this work
was carried out. This work was supported in part by the National
Science Foundation under NSF Grant PHY-0455649. AM is funded by STFC.
Part of this work was carried out during his participation in the
Tata Monsoon workshop of the International Center for Theoretical 
Studies of TIFR,
he thanks the organizers for hospitality and support.
FQ is funded by STFC and a Royal Society Wolfson Merit Award. He
thanks
the Mitchell family and the organizers of the Cook Branch's meeting 2008 for
hospitality.

\appendix

\section{Kaluza-Klein vector modes}

In this appendix
we briefly note that scalar eigenfunctions of the Laplace equation automatically generate vector
eigenfunctions, allowing the spectrum and profiles of KK vector modes to be derived from those of
scalar vector modes. We use here the notation of section \ref{sec2}.

To do so we suppose that
$\Phi^{j,ab} = X^{jk} \tilde{D}_k \Omega^{ab}$,
where $\Omega^{ab}$ is some scalar mode, and
$X^{jk}$ is a covariantly constant tensor. In particular $X^{jk}$ may be either $J^{jk}$ or $g^{jk}$, where $J_{jk}$
is the almost complex structure.
$\tilde{D}_k = \nabla_k - i \langle A \rangle_k^{ab}$ is the gauge covariant derivative acting on scalars.
Then
\bea
\tilde{D}_i \tilde{D}^i \Phi^{j,ab} & = & \tilde{D}_i \tilde{D}^i X^{jk} \tilde{D}_k \Omega^{ab} \nonumber \\
& = & \tilde{D}_i \left( X^{jk} [\tilde{D}^i, \tilde{D}_k ] +
X^{jk} \tilde{D}_k \tilde{D}^i \right)  \Omega^{ab}.
\eea
Acting on a scalar, $[\tilde{D}_i, \tilde{D}_k] \, \Omega^{ab} =
-i \langle G \rangle_{ik}^{ab}\, \Omega^{ab}$, and so we obtain
\be
\tilde{D}_i \tilde{D}^i \Phi^{j,ab} = \tilde{D}_i \left( -i X^{jk}
 \langle G \rangle^{i\phantom{k},ab}_{\phantom{i}k}\Omega^{ab}\right)
+ \tilde{D}_i (X^{jk} \tilde{D}_k \tilde{D}^i ) \Omega^{ab}.
\ee
Now, the flux is such that $\langle G \rangle_{ij} \sim \epsilon_{ij}$ and so is covariantly constant under $\nabla$. Therefore,
$$
\tilde{D}_i \langle G \rangle_{ik}^{ab} = \langle G \rangle_{ik}^{ab} \tilde{D}_i.
$$
We then obtain
\bea
\tilde{D}_i \tilde{D}^i \Phi^{j,ab} & = & -i X^{jk} \langle G \rangle^{i\phantom{k}ab}_{\phantom{i}k} \tilde{D}_i \Omega^{ab}
+ X^{jk} \tilde{D}_i \tilde{D}_k \tilde{D}^i \Omega^{ab} \nonumber \\
& = & -i X^{jk} \langle G \rangle^{i\phantom{k}ab}_{\phantom{i}k} \tilde{D}_i \Omega^{ab}
+ X^{jk} [\tilde{D}_i, \tilde{D}_k] (\tilde{D}^i \Omega^{ab}) + X^{jk} \tilde{D}_k (\tilde{D}_i \tilde{D}^i \Omega^{ab}).
\eea
Now as before $[\tilde{D}_i, \tilde{D}_k]$ on a vector gives $[\nabla_i, \nabla_k] - i \langle G \rangle_{ik}^{ab}$, and so
we get
\bea
-\frac{1}{2g^2} \tilde{D}_i \Phi_j^{ba} \tilde{D}^i \Phi^{j,ab} & = & \frac{\Phi_j^{ba}}{2g^2} \left(
-2 i X^{jk} \langle G \rangle_{ik} \tilde{D}^i \Omega^{ab} +
\right. \nonumber \\
& & \left. X^{jk} [\nabla_i, \nabla_k] \left( \tilde{D}^i \Omega^{ab} \right)
+ X^{jk} \tilde{D}_k \left( \tilde{D}_i \tilde{D}^i \Omega^{ab} \right) \right). 
\eea
$\Phi_j^{ba} = X_j^k \tilde{D}_k \Omega^{ba}$, giving\footnote{We supress the $ab$ indices in $\langle G\rangle$ to
  simplify the notation.}
\bea
-\frac{1}{2g^2} \tilde{D}_i \Phi_j^{ba} \tilde{D}^i \Phi^{j,ab} & = & \frac{\tilde{D}_p \Omega^{ba}}{2g^2} \left(
-2 i X_j^p X^{jk} \langle G \rangle_{ik} \tilde{D}^i \Omega^{ab} \right) \\
& & + \frac{\tilde{D}_p \Omega^{ba}}{2g^2}
X_j^p X^{jk} [\nabla_i, \nabla_k] \left( \tilde{D}^i \Omega^{ab} \right)
+ \frac{\tilde{D}_p \Omega^{ba}}{2g^2} X_j^p X^{jk} \tilde{D}_k \left( \tilde{D}_i \tilde{D}^i \Omega^{ab} \right) . \nonumber
\eea
We now take $X_{jk} = g_{jk}$, and so $X_j^p X^{jk} = g^{pk}$. In this case
\bea
-\frac{1}{2g^2} \tilde{D}_i \Phi_j^{ba} \tilde{D}^i \Phi^{j,ab} & = &
\frac{1}{2g^2} [ (\tilde{D}_p \Omega)^{ba} (-2i \langle G \rangle _i^p) \tilde{D}^i \Omega^{ab}
+  \\
& & ( \tilde{D}_k \Omega)^{ba} [\nabla_i, \nabla_k] (\tilde{D}^i \Omega^{ab}) + (\tilde{D}^k \Omega^{ba})\tilde{D}_k
(\tilde{D}_l \tilde{D}^l \Omega^{ab}) ] \nonumber \\
& = & \frac{1}{2g^2} [ (\Phi_p)^{ba} (-2i \langle G \rangle _i^p) \Phi_i^{ab}
+  \nonumber \\
& & (\Phi_k)^{ba} [\nabla_i, \nabla_k] (\Phi^{ab}) + (\Phi^{k,ba})\tilde{D}_k
(\tilde{D}_l \tilde{D}^l \Omega^{ab}) ]
\label{dd}
\eea
Combining (\ref{dd}) with (\ref{extra}) we obtain a total result of
$$
\frac{1}{2g^2} (D^k \Omega^{ba}) \tilde{D}_k \left( \tilde{D}_l
\tilde{D}^l \Omega^{ab} \right) = -\frac{m^2}{2g^2} \left( D^k \Omega^{ba} \right)
\left( D_k \Omega_{ab} \right).
$$
In particular, all the flux and curvature contributions have cancelled, and if $\Omega$ is a scalar eigenfunction
of $\tilde{D}_i \tilde{D}^i$ with eigenvalue $-m^2$, then $\tilde{D}_k \Omega$ is also a vector eigenfunction with
eigenvalue $-m^2$.

A similar result is obtained for $X_{jk} = J_{jk}$: the vector mode $\Phi^{j,ab} = X^{jk} \tilde{D}_k \Omega^{ab}$
is a vector eigenfunction with mass $-m^2$. On their own neither of these two modes satisfy the gauge-fixing condition
$\tilde{D}_i \Phi^{i,ab} = 0$. However the gauge-fixing condition can be satisfied by writing
\be
\Phi^{j,ab} = \alpha \, g^{jk} \tilde{D}_k \Omega^{ab} + \beta \, J^{jk} \tilde{D}_k \Omega^{ab},
\ee
for appropriate constants $\alpha$ and $\beta$.

\section{Patches}

In this section we give some details as to the explicit patch transition functions for the fermionic
wavefunctions on $\mbb{P}^1$.
We start with coordinates $z, \bar{z}$ with
\be
ds^2 = \frac{4 dz d \bar{z}}{(1 + z \bar{z})^2}.
\ee
These coordinates are good for describing all of $\mbb{P}^1$ except the point at $\infty$. We denote this patch by $\mc{A}$.
To describe the point at infinity, we must change patches to the patch $\mc{B}$ coordinatised by $u=-1/z$. The functional form of the metric
is unaltered,
\be
ds^2 = \frac{4 du d \bar{u}}{(1 + u \bar{u})^2}.
\ee
Assoicated with the patches $\mc{A}$ and $\mc{B}$ there are two separate vielbeins that
are both valid in the overlap region $\mc{A} \int \mc{B}$, $e_{\mc{A}}^1, e_{\mc{A}}^2$ and $e_{\mc{B}}^1, e_{\mc{B}}^2$. We have
\bea
e^1_{\mc{A},z} = e^1_{\mc{A}, \bar{z}} = \frac{1}{1 + z \bar{z}}, & \qquad & e^2_{\mc{A}, z} = -e^2_{\mc{A}, \bar{z}} = \frac{i}{1+ z \bar{z}}. \nonumber \\
e^1_{\mc{B},u} = e^1_{\mc{B},\bar{u}} = \frac{1}{1 + u \bar{u}}, & \qquad & e^2_{\mc{B},u} = -e^2_{\mc{B}, \bar{u}} = \frac{i}{1+ u \bar{u}}.
\eea
Using the coordinate relation $u = - z^{-1}$ that holds on the overlap region, we find that the two vielbeins are related by
$$
e^i_{\mc{B},z} = \left( \frac{\bar{z}}{z} \right) e^i_{\mc{A},z}, \qquad  e^i_{\mc{B}, \bar{z}} = \left( \frac{z}{\bar{z}} \right) e^i_{\mc{A}, \bar{z}}.
$$
In terms of real vectors $e^1$ and $e^2$, we can write
\be
\left( \begin{array}{c} e^1_{\mc{B}} \\ e^2_{\mc{B}} \end{array} \right) = \left( \begin{array}{cc} \cos(2 \theta) & - \sin (2 \theta)
\\ \sin (2 \theta) & \cos (2 \theta) \end{array} \right) \left( \begin{array}{c} e^1_{\mc{A}} \\ e^2_{\mc{A}} \end{array} \right).
\ee
Lifting this to an action on spinors using $\tilde{\Lambda} = \exp \left( \frac{1}{4} \omega^{\alpha \beta} \Lambda_{\alpha \beta} \right)$,
we obtain $\psi_B = \mc{O}_{\mc{AB},spin} \psi_A$, with
\be
\mc{O}_{\mc{AB},spin} = \left( \begin{array}{cc} \left( \frac{\bar{z}}{z} \right)^{1/2} & 0 \\
0 & \left( \frac{z}{\bar{z}} \right)^{1/2} \end{array} \right).
\ee
In the presence of a nonvanishing gauge background there is an additional
gauge contribution to the transition functions. The gauge fields on the two patches are
\bea
A_{\mc{A}} & = & \frac{ i M \bar{z} dz}{2 (1 + z \bar{z})} - \frac{i M z d \bar{z}}{2 (1 + z \bar{z})}, \nonumber \\
A_{\mc{B}} & = & \frac{i M \bar{u} du}{2 ( 1+ u \bar{u})} - \frac{i M u d \bar{u}}{2 ( 1 + u \bar{u})}.
\eea
We can verify that on the overlap region these are related by
\be
A_{\mc{B}} = A_{\mc{A}} + d \left( i \ln \left( \frac{\bar{z}}{z} \right)^{M/2} \right) \equiv A_1 + d \lambda,
\ee
and so the gauge transition function is
$$
\mc{O}_{\mc{A}\mc{B},gauge} = e^{i \lambda} = e^{-\ln \left( \frac{\bar{z}}{z} \right)^{M/2}} = \left( \frac{z}{\bar{z}} \right)^{M/2}.
$$
The overall patch transition function is then
\be
\mc{O}_{\mc{AB}} = \mc{O}_{\mc{AB},spin} \mc{O}_{\mc{AB},gauge} = \left( \begin{array}{cc} \left( \frac{\bar{z}}{z} \right)^{\frac{1-M}{2}} & 0 \\
0 & \left( \frac{z}{\bar{z}} \right)^{\frac{1+M}{2}} \end{array} \right).
\ee
Acting on zero modes of the Dirac equation with the patch transition function, we obtain
\be
\mc{O}_{\mc{AB}}  \left( \begin{array}{c} \psi_{1}(z, \bar{z}) \\ \psi_2(z, \bar{z}) \end{array} \right) =
\mc{O}_{\mc{AB}} \left( \begin{array}{c}
 f_{\mc{A}}(\bar{z})( 1 + z \bar{z} )^{\left( \frac{1-M}{2}\right)} \\
 g_{\mc{A}}({z})( 1 + z \bar{z} )^{\left( \frac{1+M}{2}\right)} \end{array} \right) =
 \left( \begin{array}{c}
 f_{\mc{B}}(\bar{u})( 1 + u \bar{u} )^{\left( \frac{1-M}{2}\right)} \\
 g_{\mc{B}}({u})( 1 + u \bar{u} )^{\left( \frac{1+M}{2}\right)} \end{array} \right),
\ee
where both $f_{\mc{A}},g_{\mc{A}}$ and $f_{\mc{B}},g_{\mc{B}}$ are analytic polynomials of maximal degree $|M|-1$.
The patch transition functions therefore indeed take normalised solutions of the Dirac equation to normalised
solutions of the Dirac equation.


\begin{thebibliography}{99}

\bibitem{hepth0502005}
  R.~Blumenhagen, M.~Cvetic, P.~Langacker and G.~Shiu,
  Ann.\ Rev.\ Nucl.\ Part.\ Sci.\  {\bf 55}, 71 (2005)
  [arXiv:hep-th/0502005].

\bibitem{hepth0610327}
  R.~Blumenhagen, B.~Kors, D.~Lust and S.~Stieberger,
  Phys.\ Rept.\  {\bf 445}, 1 (2007)
  [arXiv:hep-th/0610327].

\bibitem{hepth0702094}
  F.~Marchesano,
  Fortsch.\ Phys.\  {\bf 55} (2007) 491
  [arXiv:hep-th/0702094].

\bibitem{07112451}
  D.~Malyshev and H.~Verlinde,
  Nucl.\ Phys.\ Proc.\ Suppl.\  {\bf 171} (2007) 139
  [arXiv:0711.2451 [hep-th]].

\bibitem{08063905}
  H.~P.~Nilles, S.~Ramos-Sanchez, M.~Ratz and P.~K.~S.~Vaudrevange,
  arXiv:0806.3905 [hep-th].

\bibitem{hepth0005067}
  G.~Aldazabal, L.~E.~Ibanez, F.~Quevedo and A.~M.~Uranga,
  JHEP {\bf 0008}, 002 (2000)
  [arXiv:hep-th/0005067];
J.~F.~G.~Cascales, M.~P.~Garcia del Moral, F.~Quevedo and A.~M.~Uranga,
  JHEP {\bf 0402} (2004) 031
  [arXiv:hep-th/0312051].

\bibitem{hepth0203129}
  L.~F.~Alday and G.~Aldazabal,
  JHEP {\bf 0205}, 022 (2002)
  [arXiv:hep-th/0203129].

\bibitem{hepth0508089}
  H.~Verlinde and M.~Wijnholt,
  JHEP {\bf 0701}, 106 (2007)
  [arXiv:hep-th/0508089].

\bibitem{hepth0604208}
  J.~Gray, Y.~H.~He, V.~Jejjala and B.~D.~Nelson,
  Nucl.\ Phys.\  B {\bf 750} (2006) 1
  [arXiv:hep-th/0604208].

\bibitem{hepth0610007}
  M.~Buican, D.~Malyshev, D.~R.~Morrison, H.~Verlinde and M.~Wijnholt,
  JHEP {\bf 0701}, 107 (2007)
  [arXiv:hep-th/0610007].

\bibitem{hepth0703047}
  M.~Wijnholt,
  arXiv:hep-th/0703047.

\bibitem{08023391}
  C.~Beasley, J.~J.~Heckman and C.~Vafa,
  arXiv:0802.3391 [hep-th].

\bibitem{08052943}
  L.~Aparicio, D.~G.~Cerdeno and L.~E.~Ibanez,
  arXiv:0805.2943 [hep-ph].

\bibitem{08060102}
  C.~Beasley, J.~J.~Heckman and C.~Vafa,
  arXiv:0806.0102 [hep-th].

\bibitem{08060634}
  R.~Tatar and T.~Watari,
  arXiv:0806.0634 [hep-th].



\bibitem{hepth0502058}
  V.~Balasubramanian, P.~Berglund, J.~P.~Conlon and F.~Quevedo,
  JHEP {\bf 0503}, 007 (2005)
  [arXiv:hep-th/0502058].

\bibitem{hepth0505076}
  J.~P.~Conlon, F.~Quevedo and K.~Suruliz,
  JHEP {\bf 0508}, 007 (2005)
  [arXiv:hep-th/0505076].

\bibitem{hepph0611144}
  J.~P.~Conlon and D.~Cremades,
  Phys.\ Rev.\ Lett.\  {\bf 99}, 041803 (2007)
  [arXiv:hep-ph/0611144].

\bibitem{hepth0404229}
  D.~Cremades, L.~E.~Ibanez and F.~Marchesano,
  JHEP {\bf 0405} (2004) 079
  [arXiv:hep-th/0404229].

\bibitem{kobayashi}
H.~Abe, T.~Kobayashi and H.~Ohki,
  arXiv:0806.4748 [hep-th].

\bibitem{Pope}
  C.~N.~Pope,
  Phys.\ Lett.\  B {\bf 97}, 417 (1980).

\bibitem{hepth0203264}
  D.~Karabali and V.~P.~Nair,
  Nucl.\ Phys.\  B {\bf 641} (2002) 533
  [arXiv:hep-th/0203264].


\bibitem{hepth0207078}
  B.~P.~Dolan and C.~Nash,
  JHEP {\bf 0210} (2002) 041
  [arXiv:hep-th/0207078].

\bibitem{hepth0304037}
  B.~P.~Dolan,
  JHEP {\bf 0305} (2003) 018
  [arXiv:hep-th/0304037].

\bibitem{hepth0505107}
  R.~P.~Andrews and N.~Dorey,
  Phys.\ Lett.\ B {\bf 631} (2005) 74
  arXiv:hep-th/0505107.

\bibitem{hepth9511222}
  M.~Bershadsky, C.~Vafa and V.~Sadov,
  Nucl.\ Phys.\  B {\bf 463} (1996) 420
  [arXiv:hep-th/9511222].

\bibitem{Nielsen}
  N.~K.~Nielsen and P.~Olesen,
  Nucl.\ Phys.\  B {\bf 144}, 376 (1978).

\bibitem{BCHKMQ}
  C.~P.~Burgess, J.~P.~Conlon, L.~Y.~Hung, C.~H.~Kom, A.~Maharana and F.~Quevedo,
  arXiv:0805.4037 [hep-th].

\bibitem{GibbonsFreedman}
  D.~Z.~Freedman and G.~W.~Gibbons,
  ``New Higher Dimensional Ricci Flat Kahler Metrics,'', ITP Preprint
  ITP-SB-80-48 (unpublished).

\bibitem{Lutken}
  C.~A.~Lutken,
  J.\ Phys.\ A  {\bf 21} (1988) 1889.

\bibitem{HawkingPope}
  S.~W.~Hawking and C.~N.~Pope,
  Phys.\ Lett.\  B {\bf 73} (1978) 42.

\bibitem{hepth9907189}
  D.~S.~Freed and E.~Witten,
  arXiv:hep-th/9907189.

\bibitem{Charap}
  J.~M.~Charap and M.~J.~Duff,
  Phys.\ Lett.\  B {\bf 69}, 445 (1977).

\bibitem{GibbonsPope}
  G.~W.~Gibbons and C.~N.~Pope,
  Commun.\ Math.\ Phys.\  {\bf 61}, 239 (1978).



\bibitem{08022809}
  S.~G.~Nibbelink, D.~Klevers, F.~Ploger, M.~Trapletti and P.~K.~S.~Vaudrevange,
  JHEP {\bf 0804} (2008) 060
  [arXiv:0802.2809 [hep-th]].

\bibitem{hepth0605129}
  T.~Oota and Y.~Yasui,
  Phys.\ Lett.\  B {\bf 639}, 54 (2006)
  [arXiv:hep-th/0605129].

\bibitem{ralph}
N.~Akerblom, R.~Blumenhagen, D.~Lust and M.~Schmidt-Sommerfeld,
  arXiv:0712.1793 [hep-th].


\bibitem{hepth0506129}
  M.~Headrick and T.~Wiseman,
  Class.\ Quant.\ Grav.\  {\bf 22} (2005) 4931
  [arXiv:hep-th/0506129].

\bibitem{hepth0612075}
  M.~R.~Douglas, R.~L.~Karp, S.~Lukic and R.~Reinbacher,
  J.\ Math.\ Phys.\  {\bf 49} (2008) 032302
  [arXiv:hep-th/0612075].

\bibitem{hepth0703057}
  C.~Doran, M.~Headrick, C.~P.~Herzog, J.~Kantor and T.~Wiseman,
  arXiv:hep-th/0703057.

\bibitem{07123563}
  V.~Braun, T.~Brelidze, M.~R.~Douglas and B.~A.~Ovrut,
  JHEP {\bf 0805}, 080 (2008)
  [arXiv:0712.3563 [hep-th]].

\end{thebibliography}
\end{document}